\documentclass[longauth]{aa}

\usepackage{graphicx}
\usepackage{natbib}
\usepackage{scalerel}
\usepackage{xcolor}
\usepackage[normalem]{ulem}
\usepackage{subcaption}

\usepackage{txfonts}
\usepackage[pdfencoding=auto,psdextra]{hyperref}
\hypersetup{
    colorlinks=true,
   linkcolor=blue,
    filecolor=magenta,      
    urlcolor=blue,
    citecolor=blue,
    pdftitle={Euclid: Early Release Observations – Dwarf galaxies in the Perseus galaxy cluster},
    pdfauthor={Marleau et al.}
}
\makeatletter
\renewcommand*\aa@pageof{, page \thepage{} of \pageref*{LastPage}}
\makeatother

\usepackage{orcidlink} 
\newcommand{\orcid}[1]{\orcidlink{#1}}

\usepackage[utf8]{inputenc}

\usepackage[switch, modulo]{lineno}

\usepackage{euclid}
\usepackage{rotating}
\usepackage{pdflscape} 

\begin{document}

\title{\Euclid: Early Release Observations -- Dwarf galaxies in the Perseus galaxy cluster \thanks{This paper is published on behalf of the Euclid Consortium}}

\author{F.~R.~Marleau\orcid{0000-0002-1442-2947}\thanks{\email{francine.marleau@uibk.ac.at}}\inst{\ref{aff1}}
\and J.-C.~Cuillandre\orcid{0000-0002-3263-8645}\inst{\ref{aff2}}
\and M.~Cantiello\orcid{0000-0003-2072-384X}\inst{\ref{aff3}}
\and D.~Carollo\orcid{0000-0002-0005-5787}\inst{\ref{aff4}}
\and P.-A.~Duc\orcid{0000-0003-3343-6284}\inst{\ref{aff5}}
\and R.~Habas\orcid{0000-0002-4033-3841}\inst{\ref{aff3}}
\and L.~K.~Hunt\orcid{0000-0001-9162-2371}\inst{\ref{aff6}}
\and P.~Jablonka\orcid{0000-0002-9655-1063}\inst{\ref{aff7}}
\and M.~Mirabile\orcid{0009-0007-6055-3933}\inst{\ref{aff3},\ref{aff8}}
\and M.~Mondelin\orcid{0009-0004-5954-0930}\inst{\ref{aff2}}
\and M.~Poulain\orcid{0000-0002-7664-4510}\inst{\ref{aff9}}
\and T.~Saifollahi\orcid{0000-0002-9554-7660}\inst{\ref{aff10},\ref{aff11}}
\and R.~S\'anchez-Janssen\orcid{0000-0003-4945-0056}\inst{\ref{aff12}}
\and E.~Sola\orcid{0000-0002-2814-3578}\inst{\ref{aff13}}
\and M.~Urbano\orcid{0000-0001-5640-0650}\inst{\ref{aff10}}
\and R.~Z\"oller\orcid{0000-0002-0938-5686}\inst{\ref{aff14},\ref{aff15}}
\and M.~Bolzonella\orcid{0000-0003-3278-4607}\inst{\ref{aff16}}
\and A.~Lan\c{c}on\orcid{0000-0002-7214-8296}\inst{\ref{aff10}}
\and R.~Laureijs\inst{\ref{aff17}}
\and O.~Marchal\inst{\ref{aff10}}
\and M.~Schirmer\orcid{0000-0003-2568-9994}\inst{\ref{aff18}}
\and C.~Stone\orcid{0000-0002-9086-6398}\inst{\ref{aff19}}
\and A.~Boselli\orcid{0000-0002-9795-6433}\inst{\ref{aff20},\ref{aff21}}
\and A.~Ferr\'e-Mateu\orcid{0000-0002-6411-220X}\inst{\ref{aff22},\ref{aff23}}
\and N.~A.~Hatch\orcid{0000-0001-5600-0534}\inst{\ref{aff24}}
\and M.~Kluge\orcid{0000-0002-9618-2552}\inst{\ref{aff15}}
\and M.~Montes\orcid{0000-0001-7847-0393}\inst{\ref{aff23},\ref{aff22}}
\and J.~G.~Sorce\orcid{0000-0002-2307-2432}\inst{\ref{aff25},\ref{aff26},\ref{aff27}}
\and C.~Tortora\orcid{0000-0001-7958-6531}\inst{\ref{aff28}}
\and A.~Venhola\orcid{0000-0001-6071-4564}\inst{\ref{aff9}}
\and J.~B.~Golden-Marx\orcid{0000-0002-6394-045X}\inst{\ref{aff24}}
\and N.~Aghanim\inst{\ref{aff26}}
\and A.~Amara\inst{\ref{aff29}}
\and S.~Andreon\orcid{0000-0002-2041-8784}\inst{\ref{aff30}}
\and N.~Auricchio\orcid{0000-0003-4444-8651}\inst{\ref{aff16}}
\and M.~Baldi\orcid{0000-0003-4145-1943}\inst{\ref{aff31},\ref{aff16},\ref{aff32}}
\and A.~Balestra\orcid{0000-0002-6967-261X}\inst{\ref{aff33}}
\and S.~Bardelli\orcid{0000-0002-8900-0298}\inst{\ref{aff16}}
\and P.~Battaglia\orcid{0000-0002-7337-5909}\inst{\ref{aff16}}
\and R.~Bender\orcid{0000-0001-7179-0626}\inst{\ref{aff15},\ref{aff14}}
\and C.~Bodendorf\inst{\ref{aff15}}
\and E.~Branchini\orcid{0000-0002-0808-6908}\inst{\ref{aff34},\ref{aff35},\ref{aff30}}
\and M.~Brescia\orcid{0000-0001-9506-5680}\inst{\ref{aff36},\ref{aff28},\ref{aff37}}
\and J.~Brinchmann\orcid{0000-0003-4359-8797}\inst{\ref{aff38},\ref{aff39}}
\and S.~Camera\orcid{0000-0003-3399-3574}\inst{\ref{aff40},\ref{aff41},\ref{aff42}}
\and G.~P.~Candini\orcid{0000-0001-9481-8206}\inst{\ref{aff43}}
\and V.~Capobianco\orcid{0000-0002-3309-7692}\inst{\ref{aff42}}
\and C.~Carbone\orcid{0000-0003-0125-3563}\inst{\ref{aff44}}
\and J.~Carretero\orcid{0000-0002-3130-0204}\inst{\ref{aff45},\ref{aff46}}
\and S.~Casas\orcid{0000-0002-4751-5138}\inst{\ref{aff47}}
\and M.~Castellano\orcid{0000-0001-9875-8263}\inst{\ref{aff48}}
\and S.~Cavuoti\orcid{0000-0002-3787-4196}\inst{\ref{aff28},\ref{aff37}}
\and A.~Cimatti\inst{\ref{aff49}}
\and G.~Congedo\orcid{0000-0003-2508-0046}\inst{\ref{aff50}}
\and C.~J.~Conselice\orcid{0000-0003-1949-7638}\inst{\ref{aff51}}
\and L.~Conversi\orcid{0000-0002-6710-8476}\inst{\ref{aff52},\ref{aff53}}
\and Y.~Copin\orcid{0000-0002-5317-7518}\inst{\ref{aff54}}
\and F.~Courbin\orcid{0000-0003-0758-6510}\inst{\ref{aff7}}
\and H.~M.~Courtois\orcid{0000-0003-0509-1776}\inst{\ref{aff55}}
\and M.~Cropper\orcid{0000-0003-4571-9468}\inst{\ref{aff43}}
\and A.~Da~Silva\orcid{0000-0002-6385-1609}\inst{\ref{aff56},\ref{aff57}}
\and H.~Degaudenzi\orcid{0000-0002-5887-6799}\inst{\ref{aff58}}
\and A.~M.~Di~Giorgio\orcid{0000-0002-4767-2360}\inst{\ref{aff59}}
\and J.~Dinis\orcid{0000-0001-5075-1601}\inst{\ref{aff56},\ref{aff57}}
\and M.~Douspis\orcid{0000-0003-4203-3954}\inst{\ref{aff26}}
\and C.~A.~J.~Duncan\inst{\ref{aff51}}
\and X.~Dupac\inst{\ref{aff53}}
\and S.~Dusini\orcid{0000-0002-1128-0664}\inst{\ref{aff60}}
\and A.~Ealet\orcid{0000-0003-3070-014X}\inst{\ref{aff54}}
\and M.~Farina\orcid{0000-0002-3089-7846}\inst{\ref{aff59}}
\and S.~Farrens\orcid{0000-0002-9594-9387}\inst{\ref{aff2}}
\and S.~Ferriol\inst{\ref{aff54}}
\and P.~Fosalba\orcid{0000-0002-1510-5214}\inst{\ref{aff61},\ref{aff62}}
\and S.~Fotopoulou\orcid{0000-0002-9686-254X}\inst{\ref{aff63}}
\and M.~Frailis\orcid{0000-0002-7400-2135}\inst{\ref{aff4}}
\and E.~Franceschi\orcid{0000-0002-0585-6591}\inst{\ref{aff16}}
\and M.~Fumana\orcid{0000-0001-6787-5950}\inst{\ref{aff44}}
\and S.~Galeotta\orcid{0000-0002-3748-5115}\inst{\ref{aff4}}
\and B.~Garilli\orcid{0000-0001-7455-8750}\inst{\ref{aff44}}
\and W.~Gillard\orcid{0000-0003-4744-9748}\inst{\ref{aff64}}
\and B.~Gillis\orcid{0000-0002-4478-1270}\inst{\ref{aff50}}
\and C.~Giocoli\orcid{0000-0002-9590-7961}\inst{\ref{aff16},\ref{aff65}}
\and P.~G\'omez-Alvarez\orcid{0000-0002-8594-5358}\inst{\ref{aff66},\ref{aff53}}
\and A.~Grazian\orcid{0000-0002-5688-0663}\inst{\ref{aff33}}
\and F.~Grupp\inst{\ref{aff15},\ref{aff14}}
\and L.~Guzzo\orcid{0000-0001-8264-5192}\inst{\ref{aff67},\ref{aff30}}
\and M.~Hailey\inst{\ref{aff43}}
\and S.~V.~H.~Haugan\orcid{0000-0001-9648-7260}\inst{\ref{aff68}}
\and J.~Hoar\inst{\ref{aff53}}
\and H.~Hoekstra\orcid{0000-0002-0641-3231}\inst{\ref{aff69}}
\and W.~Holmes\inst{\ref{aff70}}
\and I.~Hook\orcid{0000-0002-2960-978X}\inst{\ref{aff71}}
\and F.~Hormuth\inst{\ref{aff72}}
\and A.~Hornstrup\orcid{0000-0002-3363-0936}\inst{\ref{aff73},\ref{aff74}}
\and D.~Hu\inst{\ref{aff43}}
\and P.~Hudelot\inst{\ref{aff75}}
\and K.~Jahnke\orcid{0000-0003-3804-2137}\inst{\ref{aff18}}
\and M.~Jhabvala\inst{\ref{aff76}}
\and E.~Keih\"anen\orcid{0000-0003-1804-7715}\inst{\ref{aff77}}
\and S.~Kermiche\orcid{0000-0002-0302-5735}\inst{\ref{aff64}}
\and A.~Kiessling\orcid{0000-0002-2590-1273}\inst{\ref{aff70}}
\and T.~Kitching\orcid{0000-0002-4061-4598}\inst{\ref{aff43}}
\and R.~Kohley\inst{\ref{aff53}}
\and B.~Kubik\orcid{0009-0006-5823-4880}\inst{\ref{aff54}}
\and K.~Kuijken\orcid{0000-0002-3827-0175}\inst{\ref{aff69}}
\and M.~K\"ummel\orcid{0000-0003-2791-2117}\inst{\ref{aff14}}
\and M.~Kunz\orcid{0000-0002-3052-7394}\inst{\ref{aff78}}
\and H.~Kurki-Suonio\orcid{0000-0002-4618-3063}\inst{\ref{aff79},\ref{aff80}}
\and O.~Lahav\orcid{0000-0002-1134-9035}\inst{\ref{aff81}}
\and D.~Le~Mignant\orcid{0000-0002-5339-5515}\inst{\ref{aff20}}
\and S.~Ligori\orcid{0000-0003-4172-4606}\inst{\ref{aff42}}
\and P.~B.~Lilje\orcid{0000-0003-4324-7794}\inst{\ref{aff68}}
\and V.~Lindholm\orcid{0000-0003-2317-5471}\inst{\ref{aff79},\ref{aff80}}
\and I.~Lloro\inst{\ref{aff82}}
\and D.~Maino\inst{\ref{aff67},\ref{aff44},\ref{aff83}}
\and E.~Maiorano\orcid{0000-0003-2593-4355}\inst{\ref{aff16}}
\and O.~Mansutti\orcid{0000-0001-5758-4658}\inst{\ref{aff4}}
\and O.~Marggraf\orcid{0000-0001-7242-3852}\inst{\ref{aff84}}
\and K.~Markovic\orcid{0000-0001-6764-073X}\inst{\ref{aff70}}
\and N.~Martinet\orcid{0000-0003-2786-7790}\inst{\ref{aff20}}
\and F.~Marulli\orcid{0000-0002-8850-0303}\inst{\ref{aff85},\ref{aff16},\ref{aff32}}
\and R.~Massey\orcid{0000-0002-6085-3780}\inst{\ref{aff86}}
\and S.~Maurogordato\inst{\ref{aff87}}
\and H.~J.~McCracken\orcid{0000-0002-9489-7765}\inst{\ref{aff75}}
\and E.~Medinaceli\orcid{0000-0002-4040-7783}\inst{\ref{aff16}}
\and S.~Mei\orcid{0000-0002-2849-559X}\inst{\ref{aff88}}
\and Y.~Mellier\inst{\ref{aff89},\ref{aff75}}
\and M.~Meneghetti\orcid{0000-0003-1225-7084}\inst{\ref{aff16},\ref{aff32}}
\and E.~Merlin\orcid{0000-0001-6870-8900}\inst{\ref{aff48}}
\and G.~Meylan\inst{\ref{aff7}}
\and M.~Moresco\orcid{0000-0002-7616-7136}\inst{\ref{aff85},\ref{aff16}}
\and L.~Moscardini\orcid{0000-0002-3473-6716}\inst{\ref{aff85},\ref{aff16},\ref{aff32}}
\and E.~Munari\orcid{0000-0002-1751-5946}\inst{\ref{aff4},\ref{aff90}}
\and R.~Nakajima\inst{\ref{aff84}}
\and R.~C.~Nichol\orcid{0000-0003-0939-6518}\inst{\ref{aff29}}
\and S.-M.~Niemi\inst{\ref{aff17}}
\and C.~Padilla\orcid{0000-0001-7951-0166}\inst{\ref{aff91}}
\and S.~Paltani\orcid{0000-0002-8108-9179}\inst{\ref{aff58}}
\and F.~Pasian\orcid{0000-0002-4869-3227}\inst{\ref{aff4}}
\and K.~Pedersen\inst{\ref{aff92}}
\and W.~J.~Percival\orcid{0000-0002-0644-5727}\inst{\ref{aff93},\ref{aff94},\ref{aff95}}
\and V.~Pettorino\inst{\ref{aff17}}
\and S.~Pires\orcid{0000-0002-0249-2104}\inst{\ref{aff2}}
\and G.~Polenta\orcid{0000-0003-4067-9196}\inst{\ref{aff96}}
\and M.~Poncet\inst{\ref{aff97}}
\and L.~A.~Popa\inst{\ref{aff98}}
\and L.~Pozzetti\orcid{0000-0001-7085-0412}\inst{\ref{aff16}}
\and F.~Raison\orcid{0000-0002-7819-6918}\inst{\ref{aff15}}
\and R.~Rebolo\inst{\ref{aff23},\ref{aff22}}
\and A.~Refregier\inst{\ref{aff99}}
\and A.~Renzi\orcid{0000-0001-9856-1970}\inst{\ref{aff100},\ref{aff60}}
\and J.~Rhodes\orcid{0000-0002-4485-8549}\inst{\ref{aff70}}
\and G.~Riccio\inst{\ref{aff28}}
\and H.-W.~Rix\orcid{0000-0003-4996-9069}\inst{\ref{aff18}}
\and E.~Romelli\orcid{0000-0003-3069-9222}\inst{\ref{aff4}}
\and M.~Roncarelli\orcid{0000-0001-9587-7822}\inst{\ref{aff16}}
\and E.~Rossetti\orcid{0000-0003-0238-4047}\inst{\ref{aff31}}
\and R.~Saglia\orcid{0000-0003-0378-7032}\inst{\ref{aff14},\ref{aff15}}
\and D.~Sapone\orcid{0000-0001-7089-4503}\inst{\ref{aff101}}
\and R.~Scaramella\orcid{0000-0003-2229-193X}\inst{\ref{aff48},\ref{aff102}}
\and P.~Schneider\orcid{0000-0001-8561-2679}\inst{\ref{aff84}}
\and A.~Secroun\orcid{0000-0003-0505-3710}\inst{\ref{aff64}}
\and G.~Seidel\orcid{0000-0003-2907-353X}\inst{\ref{aff18}}
\and M.~Seiffert\orcid{0000-0002-7536-9393}\inst{\ref{aff70}}
\and S.~Serrano\orcid{0000-0002-0211-2861}\inst{\ref{aff61},\ref{aff103},\ref{aff104}}
\and C.~Sirignano\orcid{0000-0002-0995-7146}\inst{\ref{aff100},\ref{aff60}}
\and G.~Sirri\orcid{0000-0003-2626-2853}\inst{\ref{aff32}}
\and L.~Stanco\orcid{0000-0002-9706-5104}\inst{\ref{aff60}}
\and P.~Tallada-Cresp\'{i}\orcid{0000-0002-1336-8328}\inst{\ref{aff45},\ref{aff46}}
\and A.~N.~Taylor\inst{\ref{aff50}}
\and H.~I.~Teplitz\orcid{0000-0002-7064-5424}\inst{\ref{aff105}}
\and I.~Tereno\inst{\ref{aff56},\ref{aff106}}
\and R.~Toledo-Moreo\orcid{0000-0002-2997-4859}\inst{\ref{aff107}}
\and A.~Tsyganov\inst{\ref{aff108}}
\and I.~Tutusaus\orcid{0000-0002-3199-0399}\inst{\ref{aff109}}
\and E.~A.~Valentijn\inst{\ref{aff11}}
\and L.~Valenziano\orcid{0000-0002-1170-0104}\inst{\ref{aff16},\ref{aff110}}
\and T.~Vassallo\orcid{0000-0001-6512-6358}\inst{\ref{aff14},\ref{aff4}}
\and G.~Verdoes~Kleijn\orcid{0000-0001-5803-2580}\inst{\ref{aff11}}
\and A.~Veropalumbo\orcid{0000-0003-2387-1194}\inst{\ref{aff30},\ref{aff35},\ref{aff111}}
\and Y.~Wang\orcid{0000-0002-4749-2984}\inst{\ref{aff105}}
\and J.~Weller\orcid{0000-0002-8282-2010}\inst{\ref{aff14},\ref{aff15}}
\and O.~R.~Williams\orcid{0000-0003-0274-1526}\inst{\ref{aff108}}
\and G.~Zamorani\orcid{0000-0002-2318-301X}\inst{\ref{aff16}}
\and E.~Zucca\orcid{0000-0002-5845-8132}\inst{\ref{aff16}}
\and C.~Baccigalupi\orcid{0000-0002-8211-1630}\inst{\ref{aff90},\ref{aff4},\ref{aff112},\ref{aff113}}
\and A.~Biviano\orcid{0000-0002-0857-0732}\inst{\ref{aff4},\ref{aff90}}
\and C.~Burigana\orcid{0000-0002-3005-5796}\inst{\ref{aff114},\ref{aff110}}
\and G.~De~Lucia\orcid{0000-0002-6220-9104}\inst{\ref{aff4}}
\and K.~George\orcid{0000-0002-1734-8455}\inst{\ref{aff14}}
\and V.~Scottez\inst{\ref{aff89},\ref{aff115}}
\and M.~Viel\orcid{0000-0002-2642-5707}\inst{\ref{aff90},\ref{aff4},\ref{aff113},\ref{aff112},\ref{aff116}}
\and P.~Simon\inst{\ref{aff84}}
\and A.~Mora\orcid{0000-0002-1922-8529}\inst{\ref{aff117}}
\and J.~Mart\'{i}n-Fleitas\orcid{0000-0002-8594-569X}\inst{\ref{aff117}}
\and D.~Scott\orcid{0000-0002-6878-9840}\inst{\ref{aff118}}}
										   
\institute{Universit\"at Innsbruck, Institut f\"ur Astro- und Teilchenphysik, Technikerstr. 25/8, 6020 Innsbruck, Austria\label{aff1}
\and
Universit\'e Paris-Saclay, Universit\'e Paris Cit\'e, CEA, CNRS, AIM, 91191, Gif-sur-Yvette, France\label{aff2}
\and
INAF - Osservatorio Astronomico d'Abruzzo, Via Maggini, 64100, Teramo, Italy\label{aff3}
\and
INAF-Osservatorio Astronomico di Trieste, Via G. B. Tiepolo 11, 34143 Trieste, Italy\label{aff4}
\and
Universit\'e de Strasbourg, CNRS, Observatoire astronomique de Strasbourg, UMR 7550, 67000 Strasbourg, France\label{aff5}
\and
INAF-Osservatorio Astrofisico di Arcetri, Largo E. Fermi 5, 50125, Firenze, Italy\label{aff6}
\and
Institute of Physics, Laboratory of Astrophysics, Ecole Polytechnique F\'ed\'erale de Lausanne (EPFL), Observatoire de Sauverny, 1290 Versoix, Switzerland\label{aff7}
\and
Gran Sasso Science Institute (GSSI), Viale F. Crispi 7, L'Aquila (AQ), 67100, Italy\label{aff8}
\and
Space physics and astronomy research unit, University of Oulu, Pentti Kaiteran katu 1, FI-90014 Oulu, Finland\label{aff9}
\and
Observatoire Astronomique de Strasbourg (ObAS), Universit\'e de Strasbourg - CNRS, UMR 7550, Strasbourg, France\label{aff10}
\and
Kapteyn Astronomical Institute, University of Groningen, PO Box 800, 9700 AV Groningen, The Netherlands\label{aff11}
\and
UK Astronomy Technology Centre, Royal Observatory, Blackford Hill, Edinburgh EH9 3HJ, UK\label{aff12}
\and
Institute of Astronomy, University of Cambridge, Madingley Road, Cambridge CB3 0HA, UK\label{aff13}
\and
Universit\"ats-Sternwarte M\"unchen, Fakult\"at f\"ur Physik, Ludwig-Maximilians-Universit\"at M\"unchen, Scheinerstrasse 1, 81679 M\"unchen, Germany\label{aff14}
\and
Max Planck Institute for Extraterrestrial Physics, Giessenbachstr. 1, 85748 Garching, Germany\label{aff15}
\and
INAF-Osservatorio di Astrofisica e Scienza dello Spazio di Bologna, Via Piero Gobetti 93/3, 40129 Bologna, Italy\label{aff16}
\and
European Space Agency/ESTEC, Keplerlaan 1, 2201 AZ Noordwijk, The Netherlands\label{aff17}
\and
Max-Planck-Institut f\"ur Astronomie, K\"onigstuhl 17, 69117 Heidelberg, Germany\label{aff18}
\and
Department of Physics, Universit\'{e} de Montr\'{e}al, 2900 Edouard Montpetit Blvd, Montr\'{e}al, Qu\'{e}bec H3T 1J4, Canada\label{aff19}
\and
Aix-Marseille Universit\'e, CNRS, CNES, LAM, Marseille, France\label{aff20}
\and
INAF - Osservatorio Astronomico di Cagliari, Via della Scienza 5, 09047 Selargius (CA), Italy\label{aff21}
\and
Departamento de Astrof\'isica, Universidad de La Laguna, 38206, La Laguna, Tenerife, Spain\label{aff22}
\and
Instituto de Astrof\'isica de Canarias, Calle V\'ia L\'actea s/n, 38204, San Crist\'obal de La Laguna, Tenerife, Spain\label{aff23}
\and
School of Physics and Astronomy, University of Nottingham, University Park, Nottingham NG7 2RD, UK\label{aff24}
\and
Univ. Lille, CNRS, Centrale Lille, UMR 9189 CRIStAL, 59000 Lille, France\label{aff25}
\and
Universit\'e Paris-Saclay, CNRS, Institut d'astrophysique spatiale, 91405, Orsay, France\label{aff26}
\and
Leibniz-Institut f\"{u}r Astrophysik (AIP), An der Sternwarte 16, 14482 Potsdam, Germany\label{aff27}
\and
INAF-Osservatorio Astronomico di Capodimonte, Via Moiariello 16, 80131 Napoli, Italy\label{aff28}
\and
School of Mathematics and Physics, University of Surrey, Guildford, Surrey, GU2 7XH, UK\label{aff29}
\and
INAF-Osservatorio Astronomico di Brera, Via Brera 28, 20122 Milano, Italy\label{aff30}
\and
Dipartimento di Fisica e Astronomia, Universit\`a di Bologna, Via Gobetti 93/2, 40129 Bologna, Italy\label{aff31}
\and
INFN-Sezione di Bologna, Viale Berti Pichat 6/2, 40127 Bologna, Italy\label{aff32}
\and
INAF-Osservatorio Astronomico di Padova, Via dell'Osservatorio 5, 35122 Padova, Italy\label{aff33}
\and
Dipartimento di Fisica, Universit\`a di Genova, Via Dodecaneso 33, 16146, Genova, Italy\label{aff34}
\and
INFN-Sezione di Genova, Via Dodecaneso 33, 16146, Genova, Italy\label{aff35}
\and
Department of Physics "E. Pancini", University Federico II, Via Cinthia 6, 80126, Napoli, Italy\label{aff36}
\and
INFN section of Naples, Via Cinthia 6, 80126, Napoli, Italy\label{aff37}
\and
Instituto de Astrof\'isica e Ci\^encias do Espa\c{c}o, Universidade do Porto, CAUP, Rua das Estrelas, PT4150-762 Porto, Portugal\label{aff38}
\and
Faculdade de Ci\^encias da Universidade do Porto, Rua do Campo de Alegre, 4150-007 Porto, Portugal\label{aff39}
\and
Dipartimento di Fisica, Universit\`a degli Studi di Torino, Via P. Giuria 1, 10125 Torino, Italy\label{aff40}
\and
INFN-Sezione di Torino, Via P. Giuria 1, 10125 Torino, Italy\label{aff41}
\and
INAF-Osservatorio Astrofisico di Torino, Via Osservatorio 20, 10025 Pino Torinese (TO), Italy\label{aff42}
\and
Mullard Space Science Laboratory, University College London, Holmbury St Mary, Dorking, Surrey RH5 6NT, UK\label{aff43}
\and
INAF-IASF Milano, Via Alfonso Corti 12, 20133 Milano, Italy\label{aff44}
\and
Centro de Investigaciones Energ\'eticas, Medioambientales y Tecnol\'ogicas (CIEMAT), Avenida Complutense 40, 28040 Madrid, Spain\label{aff45}
\and
Port d'Informaci\'{o} Cient\'{i}fica, Campus UAB, C. Albareda s/n, 08193 Bellaterra (Barcelona), Spain\label{aff46}
\and
Institute for Theoretical Particle Physics and Cosmology (TTK), RWTH Aachen University, 52056 Aachen, Germany\label{aff47}
\and
INAF-Osservatorio Astronomico di Roma, Via Frascati 33, 00078 Monteporzio Catone, Italy\label{aff48}
\and
Dipartimento di Fisica e Astronomia "Augusto Righi" - Alma Mater Studiorum Universit\`a di Bologna, Viale Berti Pichat 6/2, 40127 Bologna, Italy\label{aff49}
\and
Institute for Astronomy, University of Edinburgh, Royal Observatory, Blackford Hill, Edinburgh EH9 3HJ, UK\label{aff50}
\and
Jodrell Bank Centre for Astrophysics, Department of Physics and Astronomy, University of Manchester, Oxford Road, Manchester M13 9PL, UK\label{aff51}
\and
European Space Agency/ESRIN, Largo Galileo Galilei 1, 00044 Frascati, Roma, Italy\label{aff52}
\and
ESAC/ESA, Camino Bajo del Castillo, s/n., Urb. Villafranca del Castillo, 28692 Villanueva de la Ca\~nada, Madrid, Spain\label{aff53}
\and
Universit\'e Claude Bernard Lyon 1, CNRS/IN2P3, IP2I Lyon, UMR 5822, Villeurbanne, F-69100, France\label{aff54}
\and
UCB Lyon 1, CNRS/IN2P3, IUF, IP2I Lyon, 4 rue Enrico Fermi, 69622 Villeurbanne, France\label{aff55}
\and
Departamento de F\'isica, Faculdade de Ci\^encias, Universidade de Lisboa, Edif\'icio C8, Campo Grande, PT1749-016 Lisboa, Portugal\label{aff56}
\and
Instituto de Astrof\'isica e Ci\^encias do Espa\c{c}o, Faculdade de Ci\^encias, Universidade de Lisboa, Campo Grande, 1749-016 Lisboa, Portugal\label{aff57}
\and
Department of Astronomy, University of Geneva, ch. d'Ecogia 16, 1290 Versoix, Switzerland\label{aff58}
\and
INAF-Istituto di Astrofisica e Planetologia Spaziali, via del Fosso del Cavaliere, 100, 00100 Roma, Italy\label{aff59}
\and
INFN-Padova, Via Marzolo 8, 35131 Padova, Italy\label{aff60}
\and
Institut d'Estudis Espacials de Catalunya (IEEC),  Edifici RDIT, Campus UPC, 08860 Castelldefels, Barcelona, Spain\label{aff61}
\and
Institut de Ciencies de l'Espai (IEEC-CSIC), Campus UAB, Carrer de Can Magrans, s/n Cerdanyola del Vall\'es, 08193 Barcelona, Spain\label{aff62}
\and
School of Physics, HH Wills Physics Laboratory, University of Bristol, Tyndall Avenue, Bristol, BS8 1TL, UK\label{aff63}
\and
Aix-Marseille Universit\'e, CNRS/IN2P3, CPPM, Marseille, France\label{aff64}
\and
Istituto Nazionale di Fisica Nucleare, Sezione di Bologna, Via Irnerio 46, 40126 Bologna, Italy\label{aff65}
\and
FRACTAL S.L.N.E., calle Tulip\'an 2, Portal 13 1A, 28231, Las Rozas de Madrid, Spain\label{aff66}
\and
Dipartimento di Fisica "Aldo Pontremoli", Universit\`a degli Studi di Milano, Via Celoria 16, 20133 Milano, Italy\label{aff67}
\and
Institute of Theoretical Astrophysics, University of Oslo, P.O. Box 1029 Blindern, 0315 Oslo, Norway\label{aff68}
\and
Leiden Observatory, Leiden University, Einsteinweg 55, 2333 CC Leiden, The Netherlands\label{aff69}
\and
Jet Propulsion Laboratory, California Institute of Technology, 4800 Oak Grove Drive, Pasadena, CA, 91109, USA\label{aff70}
\and
Department of Physics, Lancaster University, Lancaster, LA1 4YB, UK\label{aff71}
\and
Felix Hormuth Engineering, Goethestr. 17, 69181 Leimen, Germany\label{aff72}
\and
Technical University of Denmark, Elektrovej 327, 2800 Kgs. Lyngby, Denmark\label{aff73}
\and
Cosmic Dawn Center (DAWN), Denmark\label{aff74}
\and
Institut d'Astrophysique de Paris, UMR 7095, CNRS, and Sorbonne Universit\'e, 98 bis boulevard Arago, 75014 Paris, France\label{aff75}
\and
NASA Goddard Space Flight Center, Greenbelt, MD 20771, USA\label{aff76}
\and
Department of Physics and Helsinki Institute of Physics, Gustaf H\"allstr\"omin katu 2, 00014 University of Helsinki, Finland\label{aff77}
\and
Universit\'e de Gen\`eve, D\'epartement de Physique Th\'eorique and Centre for Astroparticle Physics, 24 quai Ernest-Ansermet, CH-1211 Gen\`eve 4, Switzerland\label{aff78}
\and
Department of Physics, P.O. Box 64, 00014 University of Helsinki, Finland\label{aff79}
\and
Helsinki Institute of Physics, Gustaf H{\"a}llstr{\"o}min katu 2, University of Helsinki, Helsinki, Finland\label{aff80}
\and
Department of Physics and Astronomy, University College London, Gower Street, London WC1E 6BT, UK\label{aff81}
\and
NOVA optical infrared instrumentation group at ASTRON, Oude Hoogeveensedijk 4, 7991PD, Dwingeloo, The Netherlands\label{aff82}
\and
INFN-Sezione di Milano, Via Celoria 16, 20133 Milano, Italy\label{aff83}
\and
Universit\"at Bonn, Argelander-Institut f\"ur Astronomie, Auf dem H\"ugel 71, 53121 Bonn, Germany\label{aff84}
\and
Dipartimento di Fisica e Astronomia "Augusto Righi" - Alma Mater Studiorum Universit\`a di Bologna, via Piero Gobetti 93/2, 40129 Bologna, Italy\label{aff85}
\and
Department of Physics, Institute for Computational Cosmology, Durham University, South Road, DH1 3LE, UK\label{aff86}
\and
Universit\'e C\^{o}te d'Azur, Observatoire de la C\^{o}te d'Azur, CNRS, Laboratoire Lagrange, Bd de l'Observatoire, CS 34229, 06304 Nice cedex 4, France\label{aff87}
\and
Universit\'e Paris Cit\'e, CNRS, Astroparticule et Cosmologie, 75013 Paris, France\label{aff88}
\and
Institut d'Astrophysique de Paris, 98bis Boulevard Arago, 75014, Paris, France\label{aff89}
\and
IFPU, Institute for Fundamental Physics of the Universe, via Beirut 2, 34151 Trieste, Italy\label{aff90}
\and
Institut de F\'{i}sica d'Altes Energies (IFAE), The Barcelona Institute of Science and Technology, Campus UAB, 08193 Bellaterra (Barcelona), Spain\label{aff91}
\and
Department of Physics and Astronomy, University of Aarhus, Ny Munkegade 120, DK-8000 Aarhus C, Denmark\label{aff92}
\and
Waterloo Centre for Astrophysics, University of Waterloo, Waterloo, Ontario N2L 3G1, Canada\label{aff93}
\and
Department of Physics and Astronomy, University of Waterloo, Waterloo, Ontario N2L 3G1, Canada\label{aff94}
\and
Perimeter Institute for Theoretical Physics, Waterloo, Ontario N2L 2Y5, Canada\label{aff95}
\and
Space Science Data Center, Italian Space Agency, via del Politecnico snc, 00133 Roma, Italy\label{aff96}
\and
Centre National d'Etudes Spatiales -- Centre spatial de Toulouse, 18 avenue Edouard Belin, 31401 Toulouse Cedex 9, France\label{aff97}
\and
Institute of Space Science, Str. Atomistilor, nr. 409 M\u{a}gurele, Ilfov, 077125, Romania\label{aff98}
\and
Institute for Particle Physics and Astrophysics, Dept. of Physics, ETH Zurich, Wolfgang-Pauli-Strasse 27, 8093 Zurich, Switzerland\label{aff99}
\and
Dipartimento di Fisica e Astronomia "G. Galilei", Universit\`a di Padova, Via Marzolo 8, 35131 Padova, Italy\label{aff100}
\and
Departamento de F\'isica, FCFM, Universidad de Chile, Blanco Encalada 2008, Santiago, Chile\label{aff101}
\and
INFN-Sezione di Roma, Piazzale Aldo Moro, 2 - c/o Dipartimento di Fisica, Edificio G. Marconi, 00185 Roma, Italy\label{aff102}
\and
Satlantis, University Science Park, Sede Bld 48940, Leioa-Bilbao, Spain\label{aff103}
\and
Institute of Space Sciences (ICE, CSIC), Campus UAB, Carrer de Can Magrans, s/n, 08193 Barcelona, Spain\label{aff104}
\and
Infrared Processing and Analysis Center, California Institute of Technology, Pasadena, CA 91125, USA\label{aff105}
\and
Instituto de Astrof\'isica e Ci\^encias do Espa\c{c}o, Faculdade de Ci\^encias, Universidade de Lisboa, Tapada da Ajuda, 1349-018 Lisboa, Portugal\label{aff106}
\and
Universidad Polit\'ecnica de Cartagena, Departamento de Electr\'onica y Tecnolog\'ia de Computadoras,  Plaza del Hospital 1, 30202 Cartagena, Spain\label{aff107}
\and
Centre for Information Technology, University of Groningen, P.O. Box 11044, 9700 CA Groningen, The Netherlands\label{aff108}
\and
Institut de Recherche en Astrophysique et Plan\'etologie (IRAP), Universit\'e de Toulouse, CNRS, UPS, CNES, 14 Av. Edouard Belin, 31400 Toulouse, France\label{aff109}
\and
INFN-Bologna, Via Irnerio 46, 40126 Bologna, Italy\label{aff110}
\and
Dipartimento di Fisica, Universit\`a degli studi di Genova, and INFN-Sezione di Genova, via Dodecaneso 33, 16146, Genova, Italy\label{aff111}
\and
INFN, Sezione di Trieste, Via Valerio 2, 34127 Trieste TS, Italy\label{aff112}
\and
SISSA, International School for Advanced Studies, Via Bonomea 265, 34136 Trieste TS, Italy\label{aff113}
\and
INAF, Istituto di Radioastronomia, Via Piero Gobetti 101, 40129 Bologna, Italy\label{aff114}
\and
Junia, EPA department, 41 Bd Vauban, 59800 Lille, France\label{aff115}
\and
ICSC - Centro Nazionale di Ricerca in High Performance Computing, Big Data e Quantum Computing, Via Magnanelli 2, Bologna, Italy\label{aff116}
\and
Aurora Technology for European Space Agency (ESA), Camino bajo del Castillo, s/n, Urbanizacion Villafranca del Castillo, Villanueva de la Ca\~nada, 28692 Madrid, Spain\label{aff117}
\and
Department of Physics and Astronomy, University of British Columbia, Vancouver, BC V6T 1Z1, Canada\label{aff118}}

\abstract{We make use of the unprecedented depth, spatial resolution, and field of view of the \Euclid Early Release Observations (EROs) of the Perseus galaxy cluster to detect and characterise the dwarf galaxy population in this massive system. Using a dedicated annotation tool, the \Euclid high resolution VIS and combined VIS+NIR colour images were visually inspected and dwarf galaxy candidates were identified. Their morphologies, the presence of nuclei, and their globular cluster (GC) richness were visually assessed, complementing an automatic detection of the GC candidates. Structural and photometric parameters, including \Euclid filter colours, were extracted from 2-dimensional fitting. Based on this analysis, a total of 1100 dwarf candidates were found across the image, with 638 appearing to be new identifications. The majority (96\%) are classified as dwarf ellipticals, 53\% are nucleated, 26\% are GC-rich, and 6\% show disturbed morphologies. A relatively high fraction of galaxies, 8\%, are categorised as ultra-diffuse galaxies. The majority of the dwarfs follow the expected scaling relations of galaxies. Globally, the GC specific frequency, $S_{{N}}$, of the Perseus dwarf candidates is intermediate between those measured in the Virgo and Coma clusters. While the dwarf candidates with the largest GC counts are found throughout the \Euclid field of view, the dwarfs located around the east-west strip, where most of the brightest cluster members are found, exhibit larger $S_{{N}}$ values, on average. The spatial distribution of the dwarfs, GCs, and intracluster light show a main iso-density/isophotal centre displaced to the west of the bright galaxy light distribution. The ERO imaging of the Perseus cluster demonstrates the unique capability of \Euclid to concurrently detect and characterise large samples of dwarf galaxies, their nuclei, and their GC systems, allowing us to construct a detailed picture of the formation and evolution of galaxies over a wide range of mass scales and environments.}

    
\titlerunning{\Euclid: ERO -- Perseus cluster dwarfs}
\authorrunning{Marleau et al.}
 \maketitle

\section{\label{sc:Intro} Introduction}

The Perseus cluster, also known as Abell\,426, is a nearby, massive, and rich galaxy cluster at a distance of $72\,{\rm Mpc} \pm 3\,{\rm Mpc}$ \citep{Tully2009}, with a virial radius ($r_{200}$) of $2.2\, \rm {Mpc}$ and mass ($M_{200}$) of $1.2 \times 10^{15}\,{\rm M_{\odot}}$ \citep{Aguerri2020}, and member of the Perseus-Pisces supercluster. Its central core lacks late-type galaxies (LTGs) and is dominated by early-type galaxies (ETGs; \citealt{Kent1983}). The cluster is very bright in the X-ray regime; the peak of this emission coincides with NGC\,1275, although there is an offset between the centroids in X-ray and visible light \citep{Nulsen1980,Ulmer1992}. Similarly to the Virgo cluster, there are indications that Perseus is a dynamically young cluster with ongoing assembly processes \citep{Andreon1994}. In addition to the strong X-ray emission, several other properties characterise the cluster, such as: the peculiarity of the central galaxy NGC\,1275 \citep{Conselice2003}; the presence of structures in the intracluster medium (ICM) -- bubbles, ripples, and a weak shock front -- which are likely related to the AGN in NGC\,1275 \citep{Mathews2006, Fabian2011}; and the high cluster velocity dispersion ($\sigma = 1240\,\kms$; \citealt{Aguerri2020}). 

In the Perseus cluster, like in all galaxy clusters, dwarf galaxies are the dominant population, although they have a small contribution in terms of total luminosity and mass content of the cluster. Dwarf galaxies are generally defined as having $M_{\ast}\,\leq 10^{9}\,{\rm M}_{\odot}$, or as galaxies that have absolute visual magnitude $M_{V}\,\geq\,-17$ and are more spatially extended than globular clusters (GCs; \citealt{Tammann1994}). This limit in magnitude can be more conservative, such that $M_{V}$\,$\geq$\,$-18$ \citep{Grebel2003}. The average effective radius of ultra-faint dwarfs \citep{Simon2019} is on average a factor of 10 larger than the one for GCs ($R_{\rm e} \simeq 3$\,pc), although some smaller dwarfs and large GCs, such as Crater-I \citep{Torrealba2016}, are more similar in size and therefore are distinguished by their dark matter (DM) content. 

Dwarf galaxies are usually classified on the basis of their morphology. Late-type, which can be split into dwarf irregular (dI) and blue compact dwarfs (BCD); early-type, split into dwarf spheroidals (dSph), dwarf ellipticals (dE), and ultra-faint dwarfs (UFDs); nucleated or non-nucleated), and on their star-formation (SF) activity (star-forming or quenched). The main difference between dI and BCD galaxies is in their star formation rate, while the difference between dSphs and dEs is not clearly defined, and is mainly based on the total mass of the system (dEs are typically more massive than dSphs).\footnote{In this paper the term dEs refers to all quenched dwarfs.}

Star-forming dwarf galaxies (SFDs) are characterised by low mass, low chemical abundances, along with high gas and DM content, while quenched dwarf galaxies are non-star-forming, gas-poor objects \citep{Ferguson1994} and DM dominated. They have also been divided into subclasses based on their effective radii ($R_{\rm e}$): from the most compact dwarfs such as the ultra-compact dwarfs (UCDs; \citealt{Hilker1999,Drinkwater2000}) with 10\,$<$\,$R_{\rm e}$ $<$\,100\,pc, to the most extended ones, i.e.\ the ultra diffuse galaxies (UDGs; \citealt{vanDokkum2015,Marleau2021,Zoeller_2024}), with $R_{\rm e}$\,$\geq$\,1.5\,kpc. 

Early-type dwarf galaxies (dEs) were thought to exist mostly in group and cluster environments \citep{Binggeli1987, Geha2012}. This is due to various mechanisms that take place in galaxy clusters mainly causing quenching of dwarf galaxies \citep{Boselli2008,Kormendy2009,Boselli2022}. Such quenching mechanisms include ram-pressure stripping in which interactions between hot cluster gas and the star-forming gas of an infalling galaxy eventually removes the reservoir of the star-forming galaxy \citep{Gunn1972}, tidal harassment/stripping where gas is removed via tidal deformations \citep{Moore1996,Mayer2001,Smith2010}, and starvation wherein a galaxy is depleted of star-forming gas in a cluster environment \citep{Larson1980}. However, a recent census of dwarf galaxies has identified dEs in large numbers in low density environments \citep{Habas2020} which suggests that there is likely more than one pathway to their formation.

Several properties of dEs are now well known since the pioneering work of Binggeli in the 1990s \citep{Binggeli1991}. It was first noticed by \citet{Binggeli1998} and \citet{Gavazzi2005} that the S\'ersic index $n$ decreases from $\simeq$\,4 in massive ellipticals down to $ n \simeq 1$ in dEs \citep{Poulain2021}. Kinematic studies of dEs show that these systems are not pressure supported but rather rotationally supported \citep{Toloba2011, Toloba2015}. From an analysis of dEs in the Virgo cluster, \citet{Lisker2007} showed that these systems can also be divided into multiple sub-populations that differ significantly in their morphology and clustering properties. From dEs with disc features like spiral arms or bars to dEs with central star formation, or to ordinary, bright dEs that have no or only a weak nucleus. These systems are distributed like spiral and irregular galaxies, while the ordinary nucleated dwarf elliptical galaxies, are centrally clustered. It was also found that the frequency of nuclei decreases with decreasing luminosity of galaxies. These properties can provide important clues on the formation and evolution of such systems. The proposed scenario is that the majority of nucleated dEs or their progenitors should have experienced infall in the earliest phases of the cluster, or they could have formed in dark matter halos along with the cluster itself. All the other dE sub-classes are unrelaxed populations, implying that they have formed later than the the nucleated dEs, probably from (continuous) infall of progenitor galaxies \citep{Lisker2006, Lisker2007, Lisker2009, Su2022}. 

\begin{figure*}[ht!]
\centerline{
\includegraphics[width=1.0\linewidth]{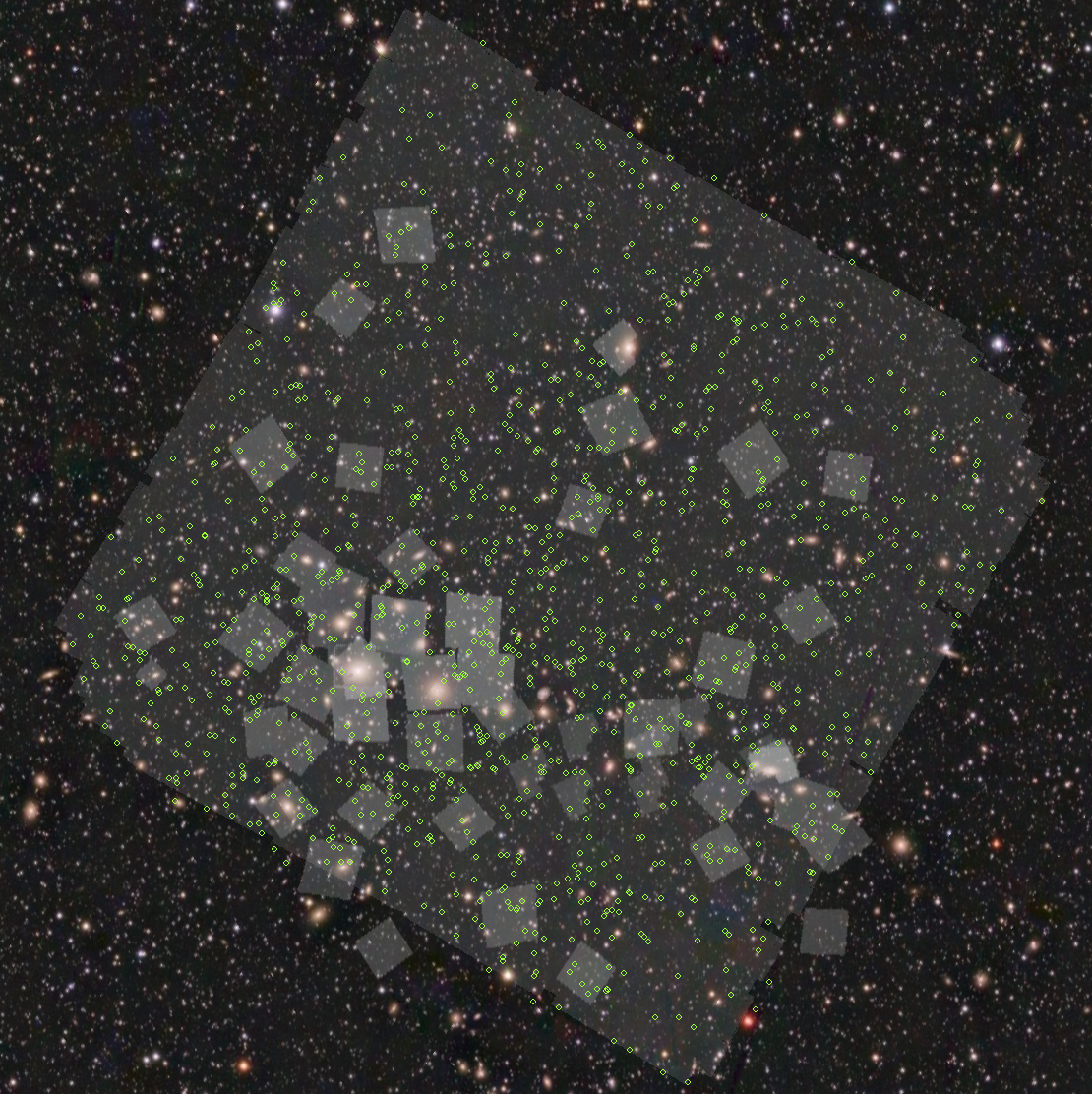}
}
\caption{VIS image of the Perseus galaxy cluster covering \ang{0.84;;} $\times$ \ang{0.84;;} and displayed on top of the PanSTARRS DR1 $g$\,+$r$\,+$i$ colour image with FoV $\sim$\,\ang{1.17;;} $\times$ \ang{1.17;;}, with north up and east left. The {green circles} show the location of the dwarf galaxy candidates identified in the ERO data, while the grey patches represent the footprint of HST/ACS observations (F850LP/SDSSz, F814w, F791w, F785LP, F775w, F625w, F622w filters). The {green circles} that fall just outside of the \Euclid footprint were identified in an earlier data product with a slightly larger FoV.
\label{fig:Perseus}}
\end{figure*}

The two most-extensively studied galaxy clusters are Virgo and Fornax, which are the closest to us, at distances of 20 and 16.5 Mpc, respectively \citep{Blakeslee2009}. Recently, these clusters were surveyed with wide field of view (FoV) cameras and in new deep multi-band images (e.g., \citealt{Ferrarese2012,Ferrarese2020}, \citealt{Durrell2014}, \citealt{Boselli2018} and \citealt{Lim2020} for Virgo; \citealt{Munoz2015}, \citealt{Iodice2016} and \citealt{Venhola2018, Venhola2019} for Fornax) that map out cluster galaxies down to low stellar masses and surface brightnesses.

In the case of the Perseus cluster, photometric surveys have been conducted mainly in the central core region. \citet{Conselice2002,Conselice2003} observed an area of 170 arcmin$^{2}$ with the WIYN 3.5-m telescope in $U$, $B$, and $R$, with reliable photometry up to $B=24$. These observations led to the identification of 53 dwarf galaxy candidates in the cluster core. In this sample, the galaxies are spheroidal or elliptical, 17 of which are nucleated. \citet{Wittmann2017} carried out a deep wide-field imaging survey with the 4.2-m William Herschel Telescope (WHT) reaching a $V$ band depth of about 27~mag arcsec$^{-2}$, focusing on low surface brightness (LSB; $\mu_{g,0}$ $\geq $ 24\,mag\,arcsec$^{-2}$) galaxies and likely UDGs. Their catalogue consists of 89 LSB dwarf galaxy candidates in the Perseus cluster core. In a subsequent paper, \citet{Wittmann2019} detected an additional 496 early-type cluster members, including dwarf galaxies, 182 of which are nucleated. The possibility to detect additional dwarf galaxies in the Perseus cluster, other LSB objects such as features like tidal tails and streams, as well as simultaneously detecting GCs, can only be achieved with very high spatial resolution, not realisable from the ground, combined with a large FoV. This is now possible with the launch of {\Euclid} in July 2023, which was commissioned in August/September 2023 and has already begun its survey observations.

The \Euclid space mission \citep{Laureijs11,EuclidSkyOverview} is planned to observe close to one-third of the sky across four photometric bands. The telescope is equipped with two instruments: VIS \citep{EuclidSkyVIS} for imaging at red optical wavelengths with a broad bandwidth (one filter: $\IE$); and NISP \citep{Schirmer-EP18,EuclidSkyNISP} for imaging in the near infrared (three filters: $\YE$, $\JE$, and $\HE$) as well as low-resolution near-infrared (NIR) spectroscopy. The imaging data of the Euclid Wide Survey (EWS; \citealt{Scaramella-EP1}) will reach a depth of $\IE=26.2$, and $\YE=24.3$, $\JE=24.5$, $\HE=24.4$ (5\,$\sigma$ detection for point-like sources). Because of its unprecedented surface brightness sensitivity (Wide [Deep] Survey: $\simeq 29.8$\,[31.8]\,mag\,arcsec$^{-2}$), spatial resolution (VIS: $\simeq \ang{;;0.16}$, NISP: $\simeq \ang{;;0.48}$) and large survey area ($\sim$\,14\,000\,deg$^{2}$; \citealt{Scaramella-EP1}), it will enable the simultaneous detection and characterisation of hundreds of thousands of new dwarf galaxies -- including thousands of new UCDs, dEs, and UDGs with their GC systems, in a wide range of environments. The latter is particularly interesting given the recent observations of GC-rich UDGs and the ongoing debate on their formation models compared to the general population of dwarf galaxies (\citealp{Lim2018,Forbes2020,Marleau2021,Gannon2022,Saifollahi2022,Marleau2024}). This new era of wide-field imaging surveys is set to be transformational in our understanding of galaxy formation and evolution at the low-mass end of the galaxy mass function. Examples of these exquisite capabilities are shown in other Early Release Observation (ERO; \citealt{EROcite})  papers describing the observation and characterisation of GCs in the Fornax cluster (\citealt{EROFornaxGCs}, Euclid Collaboration: Voggel, K., et al.\ 2024, in prep.), and the ERO project on nearby galaxies \citep{ERONearbyGals}. 

In this paper, we present the detection and characterisation of the dwarf galaxy population in the central region of the Perseus cluster using the \Euclid ERO images. The FoV of the \Euclid ERO observations of the Perseus cluster is shown in Fig.\,\ref{fig:Perseus} (VIS image) and compared to previous space-based HST observations in the same region. This FoV was selected for the \Euclid ERO as it contains many dwarf galaxies, intracluster light (ICL; \citealt{EROPerseusICL}), and more. The wide field coverage of the \Euclid observations, as compared to the footprint of the HST surveys in the same region of the sky, illustrates the gain in survey capability of \Euclid for the study of dwarf galaxies.

The paper is organised as follows. In Sect.~\ref{sc:Data} of the paper, \Euclid data and the complementary archival data are described. Section~\ref{sc:Methods} describes the methodology to analyse the data and search for dwarf galaxy candidates. Section~\ref{sc:Catalogs} provides the comparison with previous catalogues, while Sect.~\ref{sc:Phot-Str} describes the methodology for the measurement of the photometric and structural parameters of the galaxies. Section~\ref{sc:Results} studies the properties of the dwarfs using the output of the analysis. Finally, Sect.~\ref{sc:Conclusion} discusses and summarises the findings. 

\section{\label{sc:Data} Data} 

The ERO data of the Perseus galaxy cluster were obtained during \Euclid's performance verification (PV) phase at the beginning of September in 2023 \citep{EROData}. The FoV of size 0.7 deg$^{2}$ ($\ang{0.84;;} \times \ang{0.84;;}$) is centred at RA~$= 3^{\rm h}\,18^{\rm m}\,33^{\rm s}.12$ and ${\rm Dec} = \ang{41;39;03.60}$ and is rotated clockwise by \ang{30;;} from north. The data were obtained in a dithered observation sequence where an image in \IE\ is taken simultaneously with slitless grism spectra in the NIR, followed by NIR images taken through \JE, \HE\, and then \YE. The telescope is then dithered and the sequence is repeated again. This observation sequence is similar to the Reference Observation Sequence (ROS) that is being used to observe the EWS. Four ROSs were acquired for a total of 16 exposures per band, resulting in images of the Perseus cluster having a maximum exposure time of 9056.0\,s in the \IE\ filter and 1395.2\,s in the \YE, \JE, and \HE\ filters (see Table\,1 in \citealt{EROPerseusOverview}). In comparison, the expected depth should therefore be 0.75\,magnitude deeper than the EWS that is being observed in one standard ROS with four dithered images per band for a total exposure time of 70.2\,minutes, combining four repetitions of 566\,s for VIS and 112\,s for each NISP band.

\begin{figure}[ht!]
\centerline{
\includegraphics[width=\linewidth]{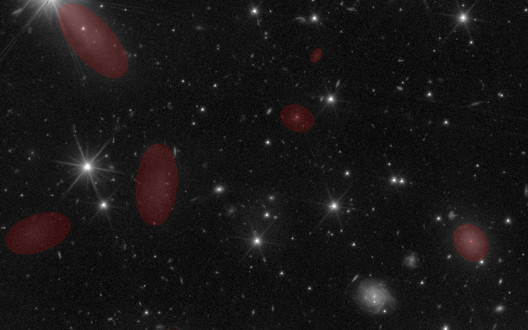}}
\caption{Illustrations of annotated dwarf candidates (drawn with ellipses) and some nuclei (drawn with circles).}
\label{fig:annotated_dwarfs}
\end{figure}

Note that the limiting surface brightness of the Perseus ERO field is fainter than that expected for the EWS, as a result of the longer total exposure time. Therefore the dwarf galaxies studied in this paper have a higher signal-to-noise ratio (SNR), and a higher accuracy for the derived global properties than the dwarfs to be found in the EWS at the same distance. A comprehensive analysis of the ERO surface brightness limits can be found in \citet{EROData}. As presented in this paper, the LSB performance for the ERO Perseus field is constrained by the non-uniformity of the background attributed to Galactic cirrus. The radial profiles of galaxies go down to $\mu_{\IE}=30.1$\,mag\,arcsec$^{-2}$ \citep{EROPerseusOverview} when integrating light at increasing radii, by combining over 360\,degrees the signal of many 100\,arcsec$^{-2}$ areas, each at the SNR$\sim$2 level. The ICL reaches down to $\mu_{\IE}=29.4$\,mag\,arcsec$^{-2}$ at an SNR of 1 by integrating the signal over very large areas \citep{EROPerseusICL}. With respect to the dwarfs, the faintest dwarf galaxies in the new ERO Perseus cluster catalogue present a typical effective radius of \ang{;;1} and reach down to an average effective surface brightness of $\langle \mu_{\IE, \rm{e}} \rangle=26.3$\,mag\,arcsec$^{-2}$, and a surface brightness at the effective radius of $\mu_{\IE, \rm{e}}=28.7$\,mag\,arcsec$^{-2}$ (see also discussion in Sect.~\ref{sc:Results}), at a total SNR within the effective radius high enough to enable derivation of physical parameters (SNR\,$\gtrsim$\,12, the performance being limited by photon statistics at this scale).

The pixel sizes for the VIS and NIR images are 0\farcs1 and 0\farcs3, respectively, which means that for both instruments the point spread function (PSF) is slightly undersampled. The final ERO stacked frames have a median PSF full-width-at-half-maximum (FWHM) of 0\farcs 16, 0\farcs 48, 0\farcs 49, and 0\farcs 50 (1.6, 1.6, 1.63, 1.67 pixels) in \IE, \YE, \JE, and \HE, respectively \citep{EROData}. The uncertainty of the photometric calibration for the EROs was required to be $\la$\,10\%, and subsequent checks show that the nominal zero point (ZP) of ${\rm ZP} = 30\,{\rm AB\,mag}$ satisfies this requirement for NISP, and of ${\rm ZP} = 30.13 \,{\rm AB\,mag}$ for \IE. The details of the data reduction are described in \citet{EROData}. Hereafter, we refer to AB magnitudes as simply magnitudes.

\begin{figure}[ht!]
\centerline{
\includegraphics[width=\linewidth]{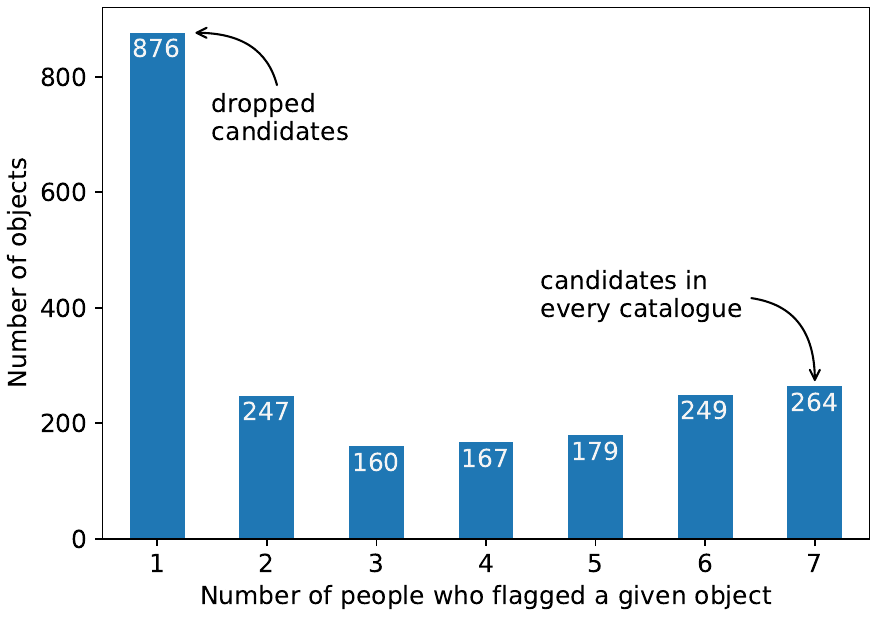}}
\caption{Agreement between the initial catalogues of the seven classifiers. The catalogues were cross-matched within a \ang{;;2} radius, and for each object we counted how many people annotated it as a dwarf galaxy.}
\label{fig:num_flagged}
\end{figure}

\begin{figure*}[ht!]
\centerline{
\includegraphics[width=0.9\linewidth]{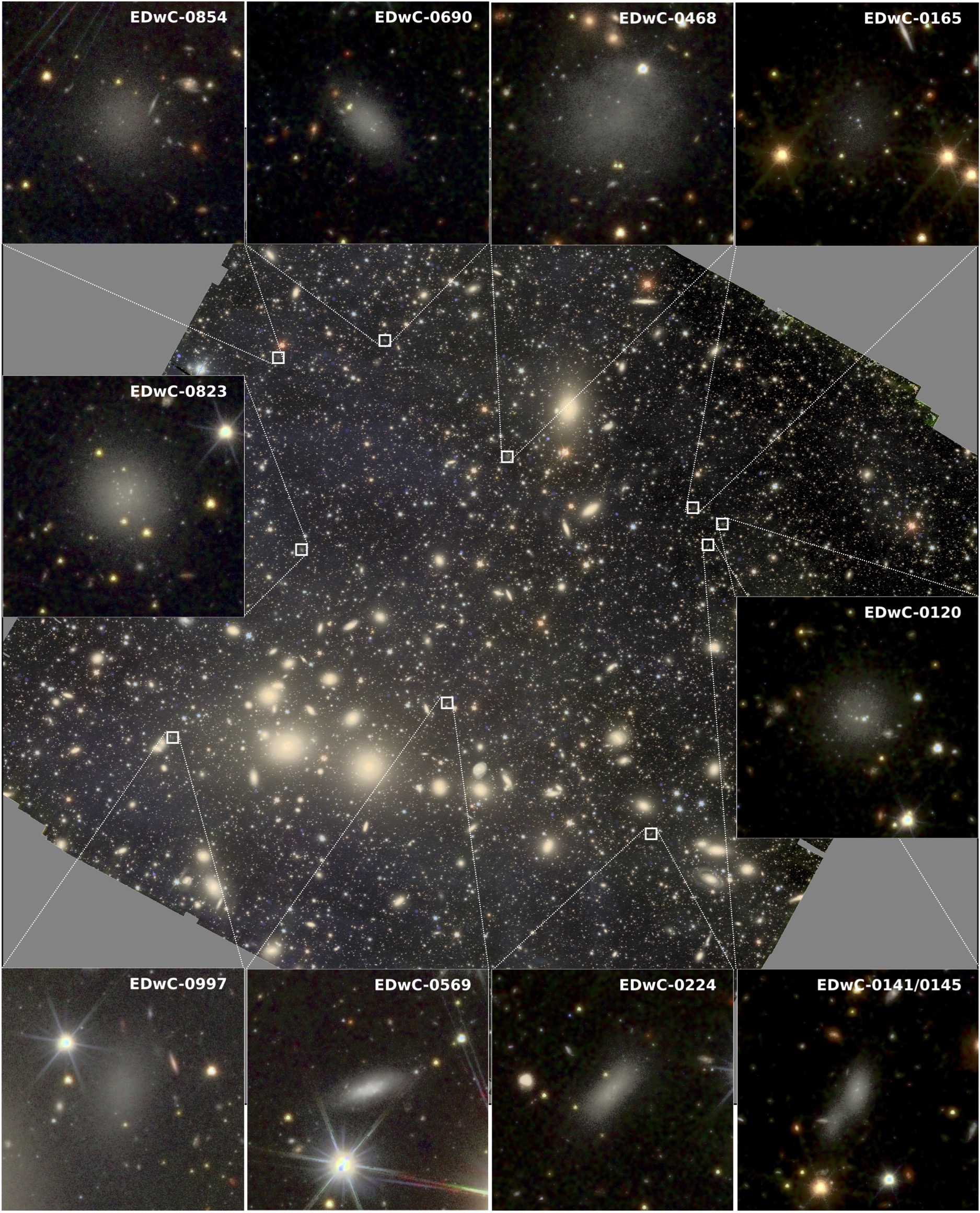}
}
\caption{VIS-NISP colour image of the Perseus galaxy cluster. The full FoV of \ang{0.84;;} $\times$ \ang{0.84;;} is shown, with north up and east to the left. Examples of dwarf galaxy candidates are shown in the individual cutouts of size \ang{;;40} $\times$ \ang{;;40}. The galaxies EDwC-0854, EDwC-0468, EDwC-0165, and EDwC-0997 are newly identified UDG candidates. The system composed of EDwC-0141 and EDwC-0145 ({\it bottom right cutout}) show one example of two newly identified dwarfs that appear to be interacting.
\label{fig:Perseuscolour}}
\end{figure*}

\begin{figure*}[ht!]
\centerline{\includegraphics[width=\linewidth]{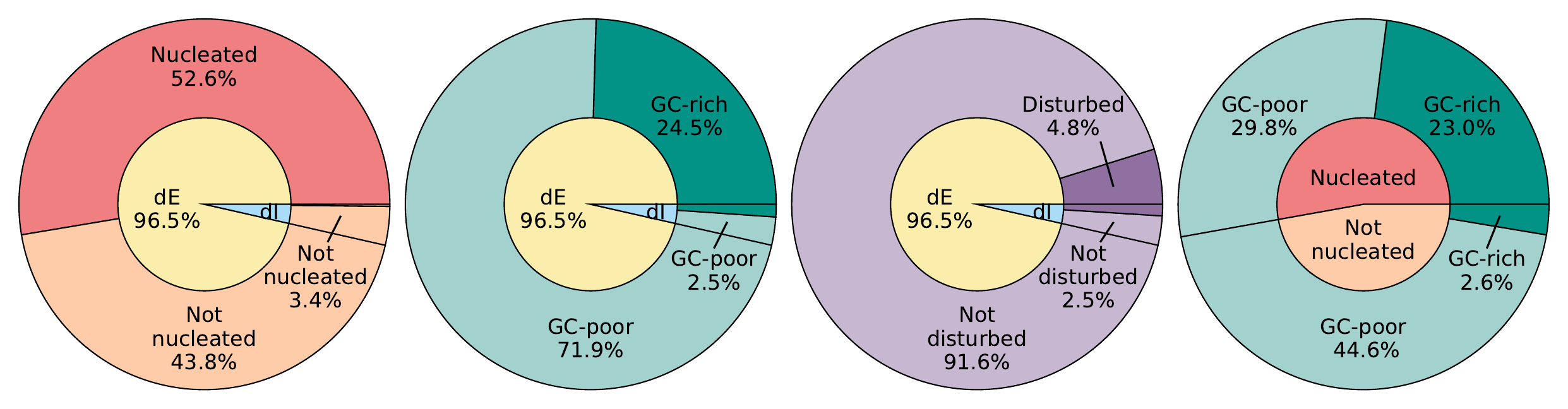}}
\caption{Classification of the final catalogue of 1100 dwarf galaxy candidates. Darker shades on the {outer circle} correspond to the presence of the feature of interest. \textit{Left}: Nucleated fraction ({outer circle}) as a function of morphology ({inner circle}). \textit{Middle,left}: GC richness ({outer circle}) as a function of morphology ({inner circle}). \textit{Middle,right}: Signs of disturbed morphology ({outer circle}) as a function of morphology ({inner circle}). \textit{Right}: GC richness ({outer circle}) as a function of the nucleated fraction. A total of 1061 (96\%) dwarf galaxy candidates are classified as dE (dI: 39 or 4\%), 581 (53\%) are nucleated, 282 (26\%) are GC-rich, and 64 (6\%) have a disturbed morphology. }
\label{fig:pie-chart-morphology}
\end{figure*}

\section{\label{sc:Methods} Dwarf catalogue and classification}

In this section we present the methods used to produce the catalogue of dwarf galaxy candidates in the ERO Perseus field. The detection is carried out visually by inspecting the high resolution \IE\ image and lower resolution combined \IE, \JE,\ and \HE\ colour image. Several steps were taken to clean the list of objects before constructing the final catalogue.

\subsection{Visual inspection}

Our dwarf galaxy candidates were visually identified by seven contributors using \texttt{Jafar}, an online annotation tool described in detail in \citet{Sola2022}. In brief, \texttt{Jafar} allows one to display and navigate around images, zoom in and out, while offering various drawing tools that enable users to precisely delineate the shapes of any objects of interest. To annotate an object, contributors select the most appropriate shape from several options (e.g., ellipse, circle, or polygon), position, resize, and rotate the shape as necessary to trace the contours of the object, and attach a label to classify its type (e.g., ETG, LTG, dwarf galaxy, nucleus, stream, or image artefact). The coordinates of the annotated object are taken from the calculated contours of the drawn shape. All relevant parameters are then stored in an online database. To assist with the classifications, \texttt{Jafar} also has the capability to overlay multiple images and adjust their dynamics, contrast, and brightness. 

Upon login, the classifiers were presented with an arcsinh stretched \IE\ image and a combined \IE+ \JE+ \HE\ colour image. The high spatial resolution of the VIS image, combined with the arcsinh stretch that enhances the faintest structures, enabled the classifier to identify substructures within the galaxies, for example spiral arms, and hence avoid as much as possible background galaxy contaminants. The presence of any nuclei and/or a large number of GCs was very useful in providing further confirmation of the presence of a dwarf galaxy candidate in the image and sometimes were even used to identify them. Therefore, looking for high concentrations of GCs in the \Euclid surveys promises to be a good tool for the detection of GC-rich dwarfs and UDGs. 

The colour image was also helpful to distinguish dwarf galaxies from background objects and artefacts. In particular, there are many round ghost halos that resemble dwarf galaxies in the VIS image, but are less prominent in the NIR images and thus appear to have blue colours in the \IE+ \JE+ \HE\ colour image. 

Dwarf candidates were annotated with ellipses, while potential nuclei were delineated with circles. Examples of such annotations are presented in Fig.\,\ref{fig:annotated_dwarfs}. Ghost halos were delineated by several contributors with circles and classified as such. All seven classifiers individually inspected the full ERO image, producing seven unique catalogues of candidates with their coordinates, centres, radii, and areas. In total, the classifiers produced 6157 annotations of objects that were flagged as dwarf galaxies. 

\subsection{Merging the individual catalogues}

Before merging the individual catalogues into a single list of unique dwarf candidates, we first crossmatched each catalogue against itself. Ideally, the smallest separation between nearest neighbours amongst all seven catalogues would determine the maximum search radius that could be applied to avoid merging neighbouring dwarfs into a single detection, assuming each catalogue is relatively complete and contains no duplicates. However, this test revealed a small number of objects in almost every catalogue with separations $\lesssim\ang{;;1}$. A visual inspection of all objects with separations $<\ang{;;3}$ found that 11 objects with the smallest separations were actually nuclei that had been mislabelled as dwarf galaxies, and we also had multiple dwarf-dwarf pairs with a separations $\geqslant$\,\ang{;;2.6}. These nuclei were removed from the sample, and we adopted a search radius of \ang{;;2} to retain the dwarf-dwarf pairs. The catalogues were then crossmatched using the \texttt{Astropy} \citep{astropy:2013, astropy:2018, astropy:2022} \texttt{SkyCoords} module, generating a list of 2140 dwarf candidates. This number still includes many duplicate objects (40) that were identified in subsequent visual inspections (described below). For this preliminary catalogue, we adopted the average right ascension and declination of the matched objects; the average positional uncertainty for the sample is $\pm \ang{;;0.38}$. 

We then investigated the initial agreement between classifiers, as shown in Fig.\,\ref{fig:num_flagged}. Care should be taken when interpreting this plot, however. The classifiers were not asked to annotate every object in the image, and therefore, it is unclear how many discrepancies are due to actual disagreements on the classifications of individual objects, and how many objects of interest were simply missed by one or more classifiers. Ideally, the latter case is minimised through the participation of more classifiers, but having more opinions means that it is also difficult to reach full agreement on the classification of any single object. In general, we consider the agreement to be good; 859 objects were flagged as a potential dwarf by a majority ($ \geq 4$) of the participants, while 264 objects were flagged by all seven classifiers.  At the other end, some 876 objects were annotated by a single contributor, and were subsequently removed from further analysis. Most of these are very small objects for which any classification is difficult, background galaxies including lenticulars or spirals, and dwarf irregular galaxies at unknown distances. This cleaning led to an intermediate catalogue of 1260 candidates identified by at least two contributors. To ensure the consistency and robustness of this intermediate list, a final validation step was performed.

\begin{figure*}[ht!]
\centerline{\includegraphics[width=0.9\linewidth]{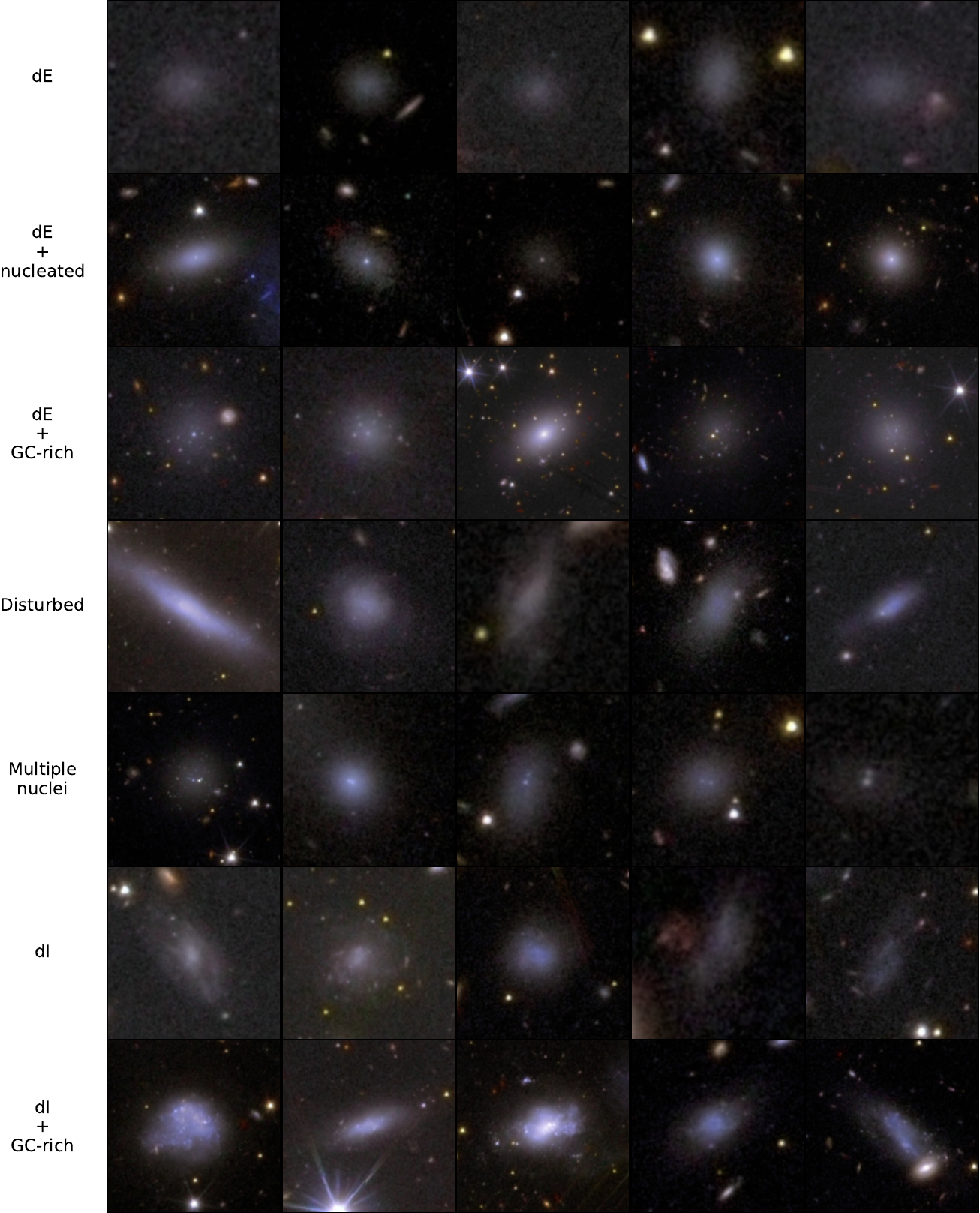}} 
\caption{Cutouts of some dwarf candidates, taken from the VIS-NISP colour image created using the \IE\ band in blue, the \YE\ band in green, and the \HE\ band in red. The colours are projected onto the high-resolution \IE\ band to best reflect their appearance as detected. From top to bottom: dE; nucleated dE; GC-rich dE; disturbed morphologies; multiple nuclei; dI; and GC-rich dI. The sizes of the cutouts are proportional to twice the area determined from the annotation of classifiers, with north up and east to the left.}
\label{fig:cutouts-examples}
\end{figure*}

The 1260 dwarfs in the intermediate catalogue were validated using an on-line interface. It simultaneously displayed cutouts of the \IE\ and \IE+\,\JE+\,\HE\ colour image, with a fixed FoV \ang{;;0.5} across, centred on the object. The same seven contributors re-evaluated the status of each dwarf candidate, assessed their morphology, GC richness, and counted the number of nuclei. The possible responses were: (a) presence of a dwarf, ${\rm `yes'} = 1$, ${\rm `unsure'} = 0.5$, or ${\rm `no'} = 0$; (b) morphology, `dE' -- regular early-type galaxy, `dI' -- star-forming dwarf irregular galaxy, `Disturbed' -- tidally disrupted or interacting object, either dE or dI; (c) number of nuclei, from zero to three; and (d) GC richness, where we define a galaxy as GC-rich if at least two GC candidates were visible within the optical body. A score assessing the presence of a dwarf was computed by taking the mean of all votes from (a).

A consensus list was determined by looking at the histograms of the individual scores. The majority of objects (874) had a score higher than 0.9 and only a few (84) had a score lower than 0.5. To avoid keeping objects that were likely contaminants, we set a threshold at 0.7 (corresponding to a score of 5/7 classifiers), above which the object is securely considered to be a dwarf. This leads to a final catalogue of 1100 dwarf galaxy candidates. Unless explicitly stated, the rest of our analysis is based on these 1100 galaxies. Figure~\ref{fig:Perseuscolour} shows examples of dwarf galaxy candidates identified in the ERO Perseus field. Table\,\ref{appendix:final-dwarf-catalogue} summarises the properties of the candidates in our final catalogue, while Table\,\ref{appendix:table-rejected-dwarfs} list the 160 objects with a score lower than 0.7 that were rejected from the final catalogue.

Completeness is not an issue for this study, since we aim to characterise the properties of secure cluster members. Nevertheless, it is important in the context of the analysis of the ERO Perseus cluster luminosity function presented in \citet{EROPerseusOverview}. In this paper, it is demonstrated that the completeness begins to rapidly decline at $M(\IE)=-12$, dropping to 50\% at $-11$. 

\subsection{Morphology, GC richness and nuclei}
\label{sec:morpho_gc_nsc}
To determine the final classification of the dwarf candidates, we assigned scores by averaging the number of votes for the morphology (dE/dI, disturbed), GC richness, and presence of any nuclei. Based on the histograms of the scores and a final inspection of the images of each object, we determined the threshold above which the candidate falls into one of the categories. There were generally few cases with scores between 2/7 and 5/7, as most of the histograms were distributed either towards low or high scores or were bimodal around 0 and 1. We set a threshold of 2 votes out of 7 ($ {\rm score} = 0.286$) to consider the dwarf candidate as a dE, and the same threshold was used to classify it as `disturbed', GC-rich, and nucleated (i.e. presence of one or more nucleus). This threshold was chosen to include as many galaxies with nuclei, GCs, and signs of tidal disturbances as possible, while still being confirmed by at least two of the seven contributors.

Figure~\ref{fig:pie-chart-morphology} presents the classification break down of the 1100 dwarf candidates in our catalogue. A total of 1061 (96\%) are classified as dE (dI: 39 or 4\%), 581 (53\%) are nucleated, 282 (26\%) are GC-rich, and 64 (6\%) have a disturbed morphology. Examples of these galaxies are displayed in in Fig.\,\ref{fig:cutouts-examples}, where we show cutouts of select dwarfs classified as: dE, dI, disturbed morphologies, nucleated, GC-rich, or a combination of these features.

\begin{table*}[ht!]
\caption[]{Previous surveys of the Perseus cluster that extend into the dwarf regime. The second and third columns give the number of catalogue entries that fall within the \Euclid FoV and the number matched sources within \ang{;;3.5}, respectively. Columns (4) -- (7) list properties of the surveys: the spatial coverage, the limiting surface brightness, the primary telescope used for the survey, and the filters used. The \citet{Meusinger2020} and \citet{Wittmann2019} samples are further separated into dwarf and non-dwarf members (see text) for a better comparison. }
\label{table:cat_numbs}
\begin{centering}
\begin{tabular}{lrrcccc}
\hline
\hline
\noalign{\smallskip}
Reference & $N_{\rm FoV}$ & $N_{\rm matched}$ & Coverage & $\mu_{\rm lim}$ & Telescope & Bands \\
 & & & [deg$^{2}$] & [mag arcsec$^{-2}$] & & \\
\noalign{\smallskip}
\hline
\noalign{\smallskip}
\citet{Meusinger2020} & 211 & 60 & 10 &  \dots & Alfred-Jensch\tablefootmark{a} & $BR$ 6670\tablefootmark{a} \\
\hspace{6mm} {\textit{dwarf members}} & 56 & 32 & & \\
\hline \noalign{\smallskip}
\citet{Wittmann2019} & 5339 & 426 & 0.27 & 27 & WHT\tablefootmark{b} & $V$ \\
\hspace{6mm} {\textit{cluster members}} & 484 & 389 & & & & \\
\hspace{14mm} $M_{\rm V} \leq - 18$ & 20 & 1& & & & \\
\hspace{14mm} $M_{\rm V} > - 18$ & 464 & 388 & & & \\
\hspace{6mm} {\textit{non-cluster members}} & 4855 & 37 & & & & \\
\hline
\noalign{\smallskip}
\citet{Wittmann2017} & 85 & 72 & 0.27&27 & WHT& $V$ \\
\hline
\noalign{\smallskip}
\citet{Conselice2003} & 53 & 25 & 173 arcmin$^{2}$ & 24 & WIYN & $BRU$
\tablefootmark{c} \\
\hline
\noalign{\smallskip}
\citet{Brunzendorf1999} & 128 & 4 & 10 & 27 & Alfred-Jensch & $B$ \\
\noalign{\smallskip}
\hline
\end{tabular}
\tablefoot{
\tablefoottext{a}{\citet{Meusinger2020} also utilize several auxiliary data sources, including: spectra and $u$, $g$, $r$, $i$, and $z$ imaging from SDSS; $B$,$V$, and $R$ imaging and optical spectra from the CAHA telescope at Calar Alto; NED; WISE; and H$\alpha$ data from the literature.
\tablefoottext{b}{In addition to the WHT imaging, \citet{Wittmann2019} used archival Subaru HSC data in $g$, $r$, and $z$ bands as auxiliary data sets to obtain colours and the \ang{;;0.5} seeing conditions for the morphological classifications. The HSC $g$ band reached a limiting surface brightness of 28.8\,mag\,arcsec$^{-2}$.}
\tablefoottext{c}{Only partial $U$ band coverage is available. \hfill}
}
}
\end{centering}
\end{table*}

\begin{figure}[ht!]
\centerline{
\includegraphics[width=0.45\textwidth]{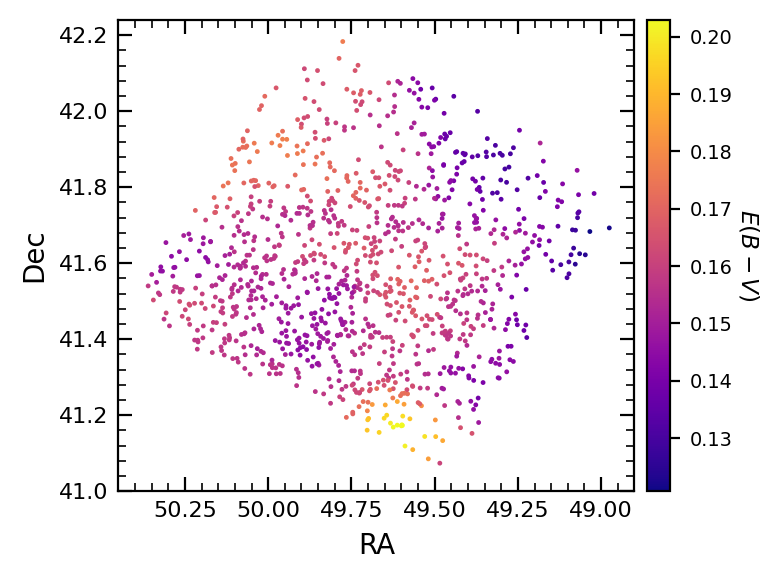}
}
\centerline{
\includegraphics[width=0.45\textwidth]{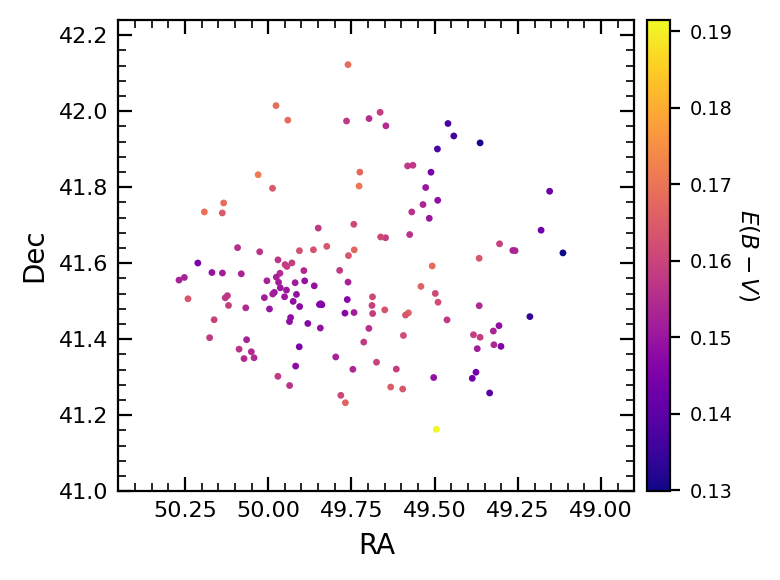}
}
\caption{Map of the colour excess $E(B-V)$ for the dwarf galaxy sample ({\it top}) and the bright galaxy sample ({\it bottom}) presented in \citet{EROPerseusOverview}.
\label{fig:extinction}}
\end{figure}

\begin{figure}[ht!]
    \centering
    \includegraphics[width=0.45\textwidth]{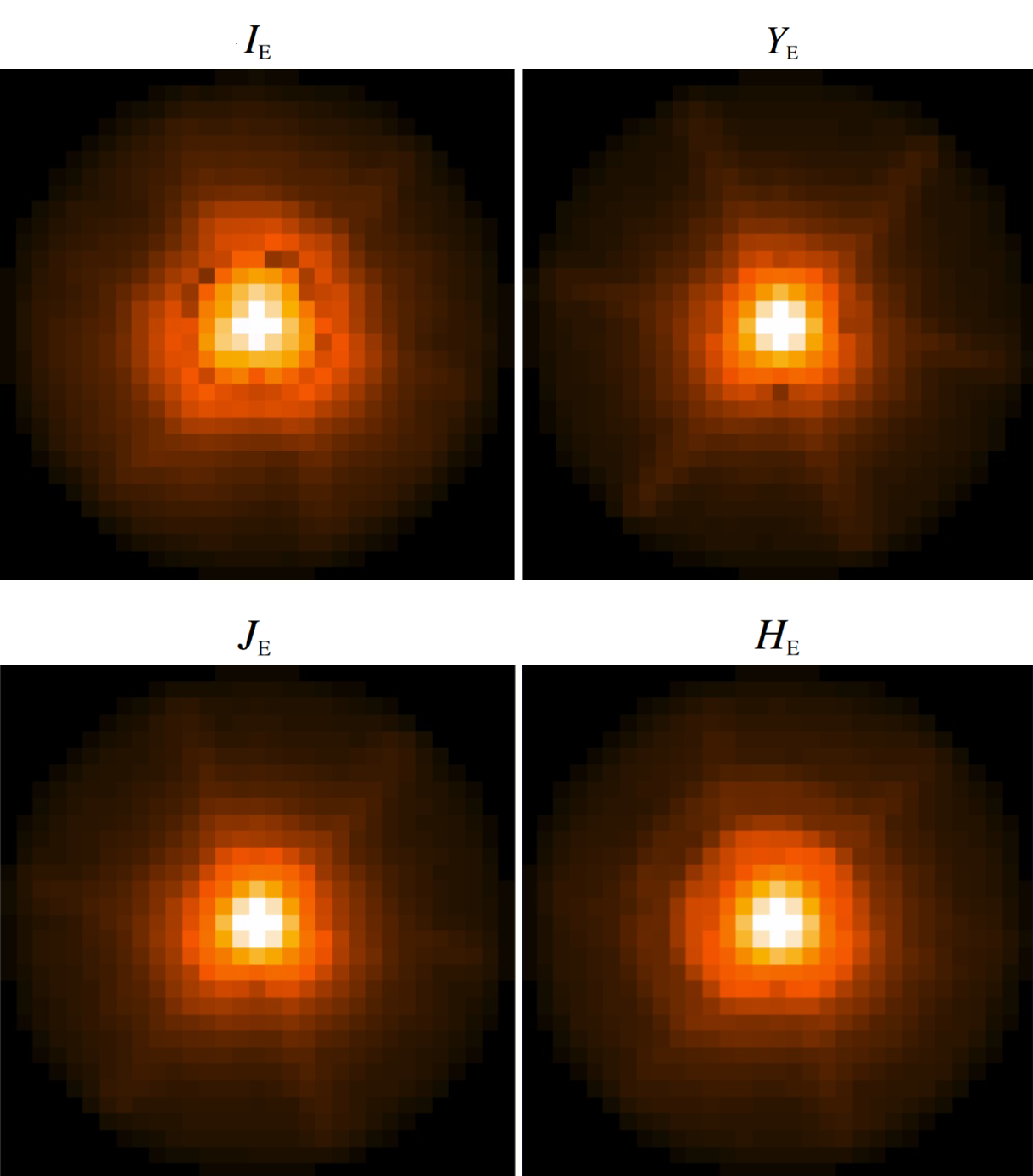}
    \caption{PSF models with the native pixel scale of $\IE$, $\YE$, $\JE$, and $\HE$ in the centre of the ERO Perseus images. The displayed PSF models have a sidelength of \ang{;;3.1} for \IE and \ang{;;9.3} for $\YE$, $\JE$, and $\HE$. The models are displayed on a logarithmic scale.}
    \label{fig:PSF}
\end{figure}

\section{\label{sc:Catalogs} Comparison with previous dwarf catalogues}

A number of studies in the literature have attempted to identify membership in the Perseus cluster, with several studies extending into the dwarf regime (e.g., \citealt{Brunzendorf1999,Conselice2003,Penny2011,Penny2014,Wittmann2017,Wittmann2019, Meusinger2020,Gannon2022}). The methods to determine cluster membership differ; some galaxies in these catalogues have been confirmed as members with measured distances, while others are identified as likely members through their morphology and colour.

We first confirmed which galaxies from each source fall within the \Euclid ERO image using Multi-Order Coverage (MOC) maps and the {\texttt{Aladin}} \citep{Bonnarel2000} `filtering by MOC' feature. MOC maps provide an International Virtual Observatory Alliance (IVOA) standard file format to define arbitrary sky regions (e.g., a FoV) in spherical geometry, using the \texttt{HEALPix} \citep{Gorski2005} tessellation technique; this format allows for the rapid and accurate comparison of two such regions, or regions and catalogs. The filtered catalogues were then cross-matched against our dwarf candidate list, using a \ang{;;3.5} search radius, roughly the average effective radius of our dwarf sample (see Sect.\,\ref{sc:Phot-Str}). It should be noted that the vast majority of our matches have much smaller separations. We summarise the matches between our catalogue and previous surveys in Table\,\ref{table:cat_numbs}. Finally, we also cross-matched the literature catalogues against each other to create a master list of unique members. 

\begin{figure*}[ht!]
\centerline{
\includegraphics[width=0.95\linewidth]{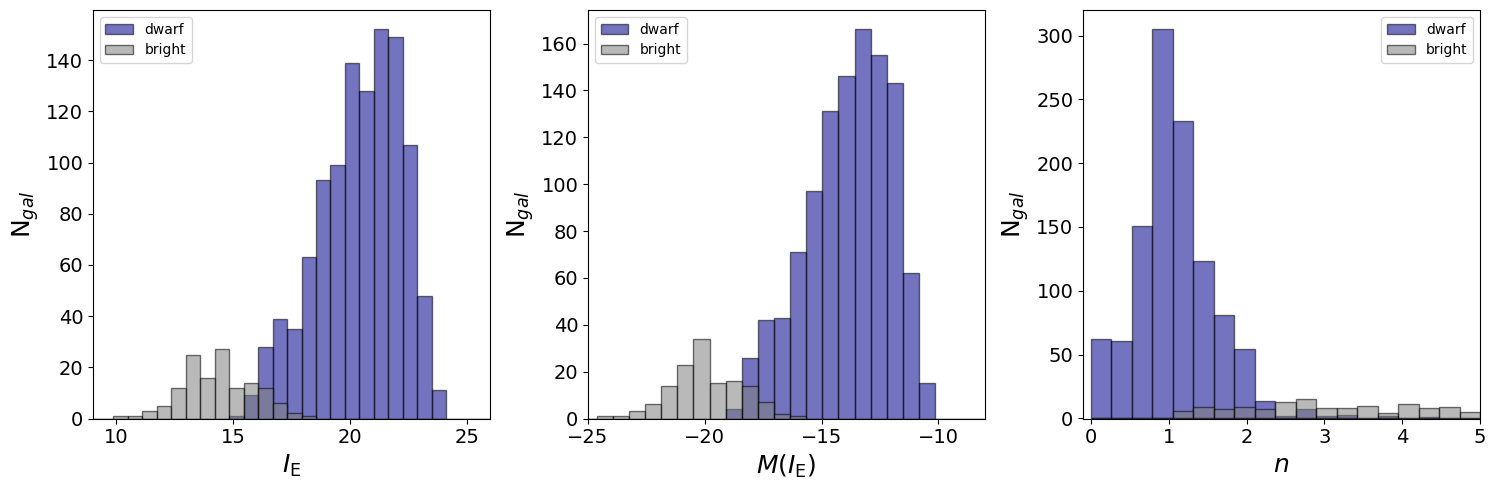}
}
\centerline{
\includegraphics[width=0.95\linewidth]{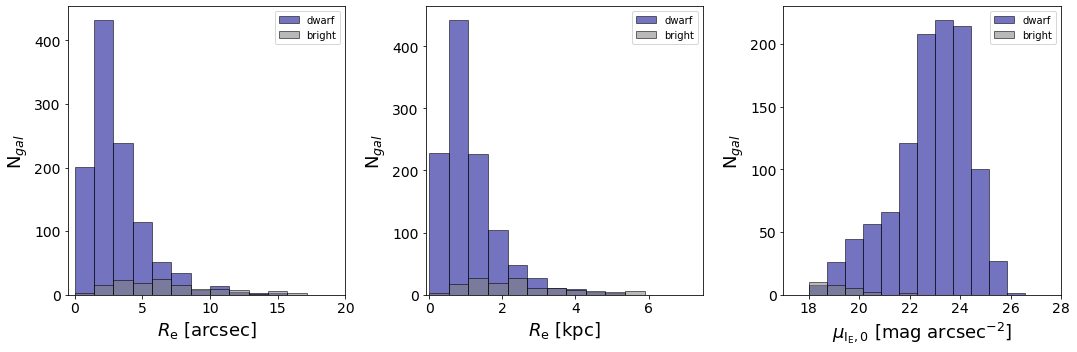}
}
\caption{Photometric and structural parameters derived from the surface profile fitting of the 1100 dwarf galaxy candidates and the bright galaxy sample from \citet{EROPerseusOverview} in the ERO Perseus field shown in {violet-blue} and {light-grey colours}, respectively. All magnitudes and surface brightness values were corrected for extinction, as described in Sect.~\ref{sec:extcorr}.
\label{fig:struct}}
\end{figure*}

Two prior surveys attempted to separate cluster members from background objects \citep{Meusinger2020,Wittmann2019}. This is particularly important because the Perseus cluster, a member of the Perseus-Pisces supercluster, is not truly isolated. Additionally, structures have been identified behind the cluster at $z\approx 0.03$ and $z\simeq 0.06$ (see Fig.\,8 in \citealt{Meusinger2020}, and \citealt{Aguerri2020}) that could be a source of confusion.

\citet{Wittmann2019} present a catalogue of 5437 galaxies in the direction of the Perseus cluster core, nearly all of which fall within the \Euclid ERO FoV. They use morphologies and colour to separate cluster members from background galaxies, and classify each galaxy into one of eight categories: dE/ETG cluster members, candidate cluster members, background ETGs, LTGs, edge-on disk galaxies, galaxies with weak substructure, merging background galaxies, and excluded sources (e.g., image artefacts or Galactic cirrus). For this comparison, we adopt the Wittmann et al.\ nomenclature, where the first two categories are considered candidate cluster members. They did not attempt to separate cluster and background LTGs, disk galaxies, or galaxies with weak substructure; for simplicity, we have merged these with the remaining categories -- aside from the excluded sources -- and collectively refer to them as non-cluster members. After matching the catalogues, we find 349 cluster members in common between our samples (348 `dE/ETG cluster candidates' and 41 `candidate cluster members'). In Table\,\ref{table:cat_numbs}, we break down the number of proposed cluster members in two magnitude bins using the $M_V$ magnitudes from \citet{Wittmann2019}: massive galaxies with $M_V \leq -18$ and a dwarf sample ($M_V > -18$). Only one of our dwarf candidates falls in the massive galaxy bin, and it has $M_V = -18.1$, placing it right on the cusp of the separation. Among the dwarf candidates, our agreement is $\gtrsim$\,80\%.  Interestingly, this is also the typical level of agreement between the seven classifiers in this work with each other (80\% -- 90\%), and may reflect inherent biases in the selection of dwarf galaxies that translate into differences between catalogs.

Only 36 of our proposed dwarf members were labelled as non-cluster objects (18 background ETGs, 7 LTGs, 7 edge-on disk galaxies, and 5 galaxies with weak substructures) in the \citet{Wittmann2019} sample; none of our galaxies matched with objects identified as cirrus or image artefacts in their sample. However, Wittmann et al.\ note a few caveats in the classifications: small sources that could not be readily classified given the resolution of the WHT were classified as background ETGs, some of the LTGs may be legitimate cluster members, and the classification of galaxies with weak substructures is ambiguous. Thus, it is entirely possible that our matched objects are legitimate cluster members, although distance estimates will be required to prove or disprove their membership. In the region of overlap between the two surveys, we identify an additional 143 dwarf candidates with no matching sources within \ang{;;3.5}.

\begin{figure*}[ht!]
\centerline{
\includegraphics[width=0.75\linewidth]{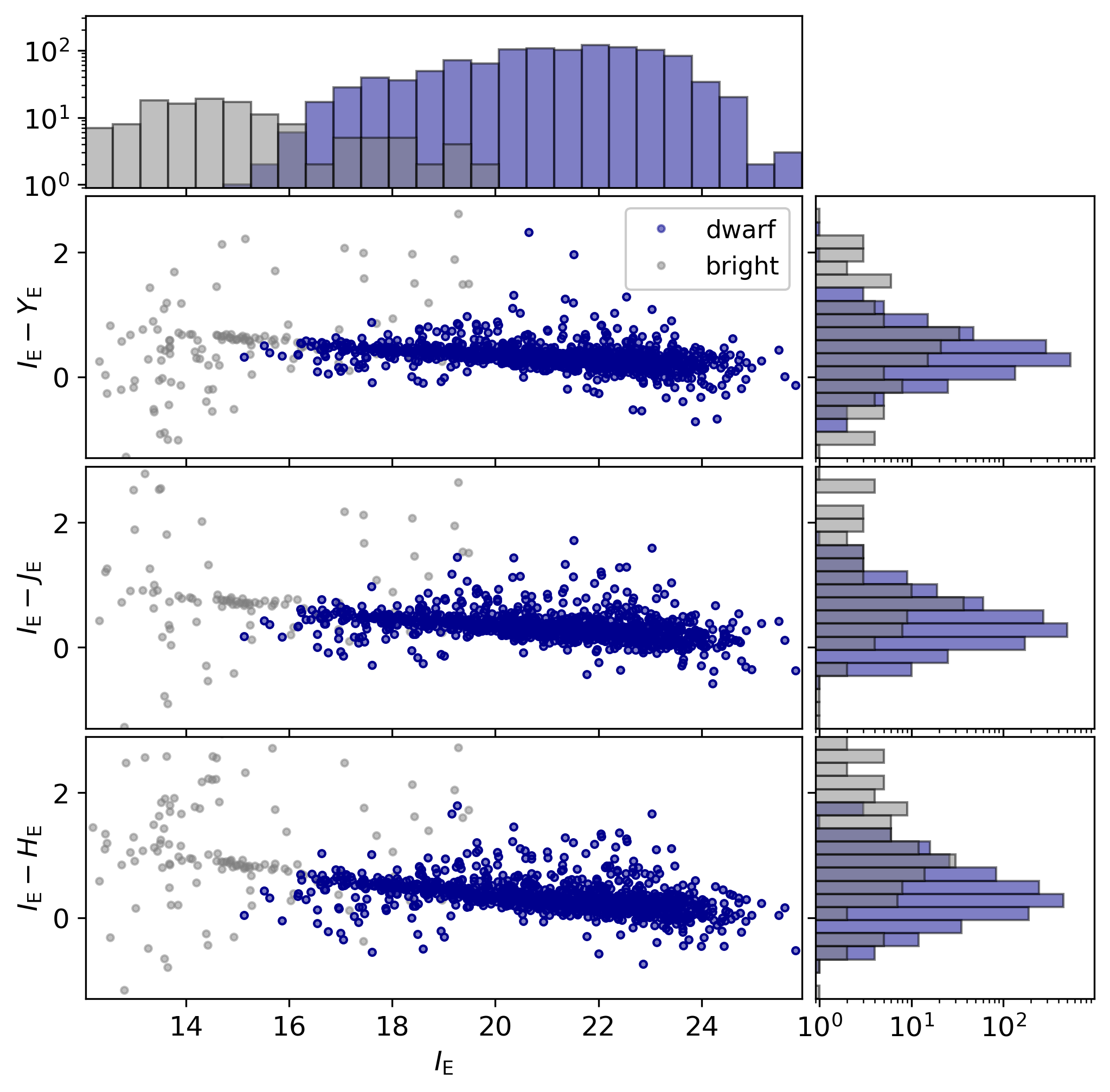}
}
\caption{\Euclid VIS-NISP colours as a function of \IE\ measured using aperture photometry within 1\,$R_{\rm e}$ of the 1100 dwarf galaxy candidates and the bright galaxy sample in the ERO Perseus field from \citet{EROPerseusOverview} shown in {violet-blue} and {light-grey colours}, respectively. The magnitudes were corrected for extinction before the colours were computed, as described in Sect.~\ref{sec:extcorr}. 
\label{fig:colours}}
\end{figure*}

The \citet{Meusinger2020} sample includes 1294 galaxies in a 10\,${\rm deg}^2$ field. Only 212 of these galaxies fall within the \Euclid ERO FoV, however, of which 56 are classified as dwarfs according to their absolute magnitudes. We have a much poorer agreement with their catalogue, matching just under 50\% of their dwarfs, and it is not immediately clear why. However, their sample was selected through a combination of a visual examination of Alfred-Jensch Telescope images and visually cleaning an automated catalog generated from \texttt{SExtractor}. It is entirely possible that the software, being able to quantify properties of the detections, allowed the authors to select dwarf candidates that would not have been identified in a visual inspection, for example: brighter or more disturbed dwarfs. It is also worth noting that the identification of dwarf galaxies in their sample is somewhat dependent on which parameter (e.g., absolute magnitude or stellar mass) that is applied, so these numbers should not be taken as absolutes. 

Three of the samples contain distance measurements that can be used to robustly test cluster membership \citep{Penny2014, Meusinger2020, Gannon2022}. Penny et al. have Keck-ESI spectra of six dEs in the cluster core. \citet{Meusinger2020} take spectroscopic redshifts from various sources (SDSS, TLS Tautenburg telescope, CAHA telescope at Calar Alto Observatory, NED, and H$\alpha$ data from \citealt{Sakai2012} and \citealt{Moss2000,Moss2006}), while \citet{Gannon2022} obtained integral field spectroscopy with the Keck Cosmic Web Imager (KCWI; \citealt{Morrissey2018}). A total of 125 of these confirmed cluster members lie within the \Euclid ERO image (all 6 from \citealt{Penny2014}, 3 from \citealt{Gannon2022}, 116 from \citealt{Meusinger2020}). Four of the galaxies from \citet{Penny2014} are in our sample; the other two are badly contaminated by intracluster light in the core, and were not flagged during our initial inspection of the images. In the case of \citet{Gannon2022}, we identified all three of their dwarfs that fall within our FoV. As discussed above, the \citet{Meusinger2020} sample includes massive and dwarf galaxies; only 9 dwarfs -- identified by their absolute $r$ magnitudes -- are confirmed cluster members. Of these, four are part of our sample.

Matching our catalogue against others specifically targeting UDGs is also useful to test our completeness at the faint end. The \citet{Wittmann2017} catalogue contains 89 UDG candidates; 85 are in the \Euclid FoV and we match 72 candidates with our catalogue. A number of other papers (e.g., \citealt{Li2022,Gannon2022}) have built upon the \citet{Wittmann2017} sample and furthered the study of select UDG candidates. \citet{Li2022} identified 11 UDG candidates based on over-densities of intergalactic GC populations, detected in the HST PIPER survey, of which nine are in our sample. We also examined their galaxy CDG-1, a potential galaxy with an over-density of GCs but no detected diffuse stellar content. All GCs are found in our GC catalogue (see Sect.~\ref{sec:GCauto}) but we do not find any diffuse emission associated with these GCs. The most likely scenario is that these are a random grouping of objects that are not associated with a galaxy. \citet{Gannon2022} obtained follow-up Subaru/Hyper Suprime-Cam imaging of five UDGs in the \citet{Wittmann2017} sample; all three of these galaxies that fall within the \Euclid ERO FoV are in our sample. Thus, we appear to be fairly complete at the faint end.

At another extreme, \citet{Penny2012} identified 84 ultra compact dwarf (UCD) candidates in the Perseus cluster core, and \citet{Penny2014a} spectroscopically confirmed 14 as members of the cluster. None of these galaxies -- neither the confirmed UCD cluster members nor the UCD candidates -- are in our sample. This is unsurprising, however, as their small sizes and relatively high surface brightnesses would have precluded their selection during the visual examination of the images.

In total, 462 of the dwarfs in our sample have already been identified in previous studies. The remaining 638 galaxies appear to be new detections. 

\begin{figure*}[ht!]
\centerline{
\includegraphics[width=\linewidth]{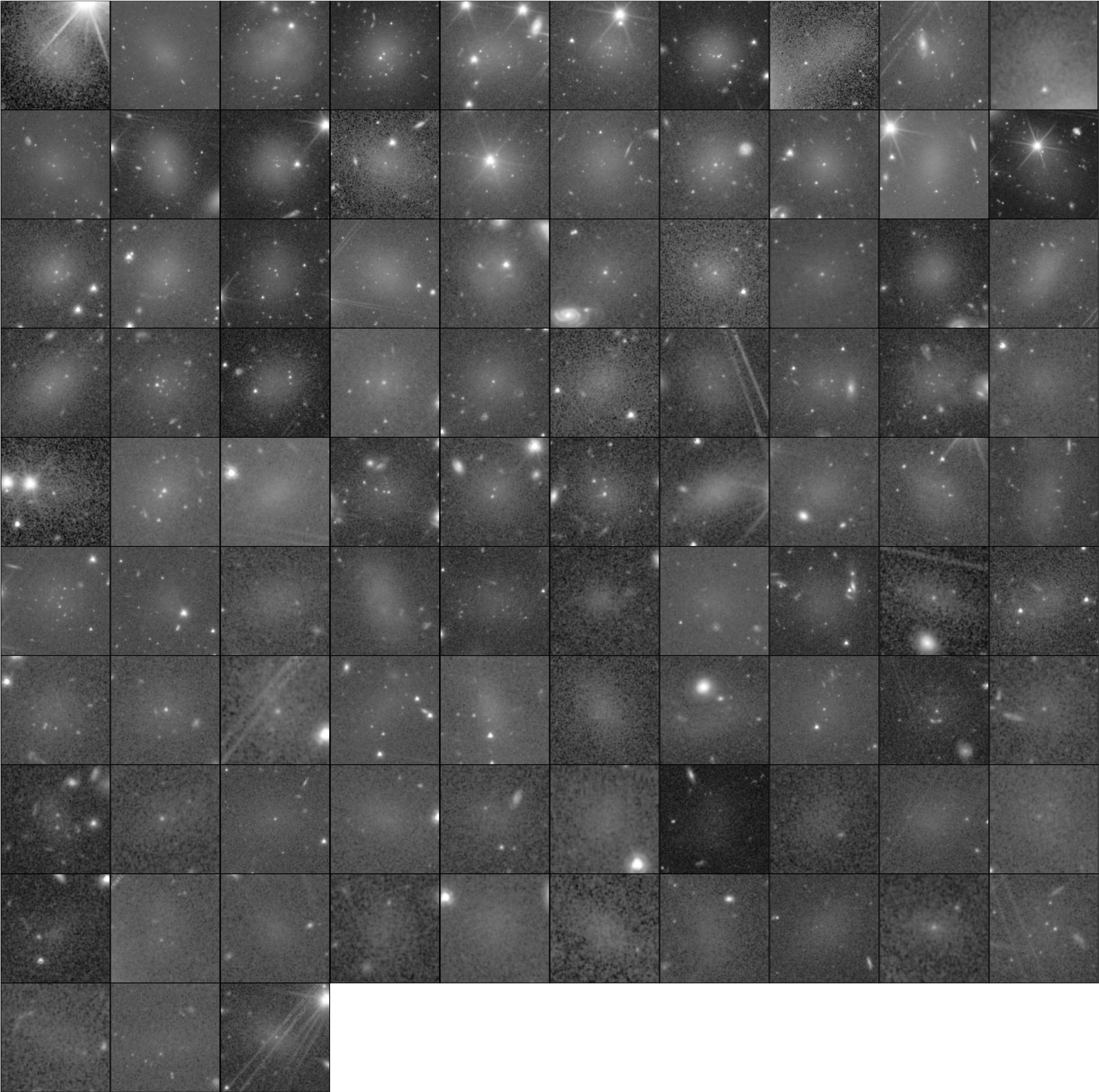}
}
\caption{Cutouts of the VIS image for the 93 UDGs in the ERO Perseus field, comprising 8\% of the dwarf sample. The UDGs were selected to have an effective radius $R_{\rm e} \geq 1.5$\,kpc and a central surface brightness in the $g$ band $\mu_{g,0} \geq 24$\,mag\,arcsec$^{-2}$. The size of the cutouts are proportional to twice the area determined from the annotation of classifiers, with north up and east to the left. The UDGs are ordered by decreasing surface brightness $\langle \mu_{\IE,\rm e} \rangle$.
\label{fig:UDG}}
\end{figure*}

\begin{figure*}[ht!]
\centerline{
\includegraphics[width=0.7\linewidth]{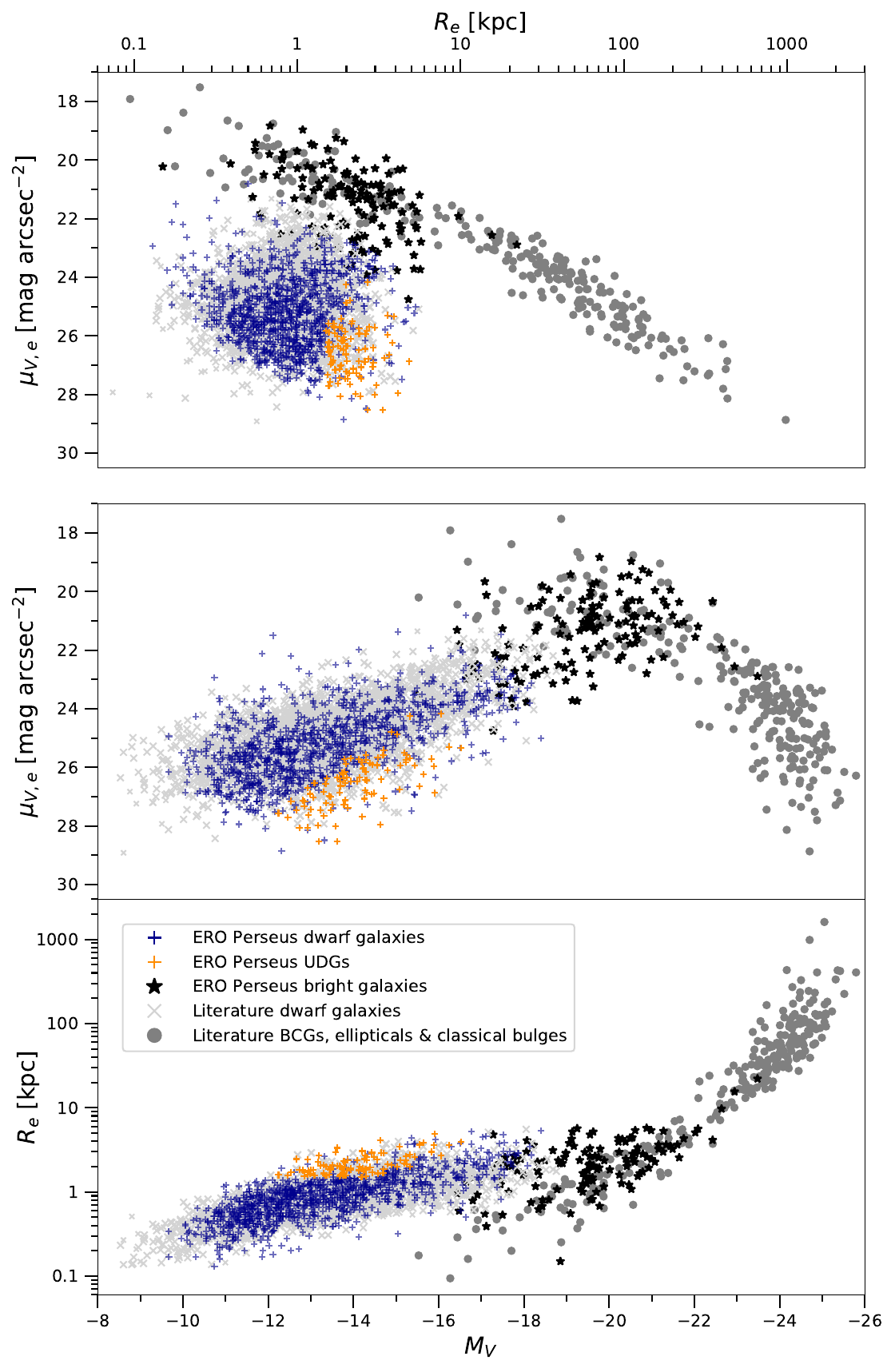}
}
\caption{Comparison between $M_{\rm V}$, $R_{\rm e}$, and $\mu_{\rm {V, e}}$ scaling relations of the ERO Perseus dwarf sample ({violet-blue}), including UDGs ({orange}) from this work, and bright galaxies ({black}) from \citet{EROPerseusOverview}. The basis for this plot is Figure\,37 in \citet{Kormendy2009} with updates from \citet{KormendyBender2012}, \citet{Bender2015}, \citet{kluge2020}, \citet{Marleau2021} and \citet{Zoeller_2024}, including brightest cluster galaxies (BCGs), ellipticals, and classical bulges ({dark grey}), as well as dwarf galaxies including UDGs ({light grey}). For the sake of clarity, we do not show errorbars. The uncertainties vary significant between the different galaxies.
\label{fig:dwarfscalingrelations}}
\end{figure*}

\begin{figure*}[ht!]
\centerline{
\includegraphics[width=0.75\linewidth]{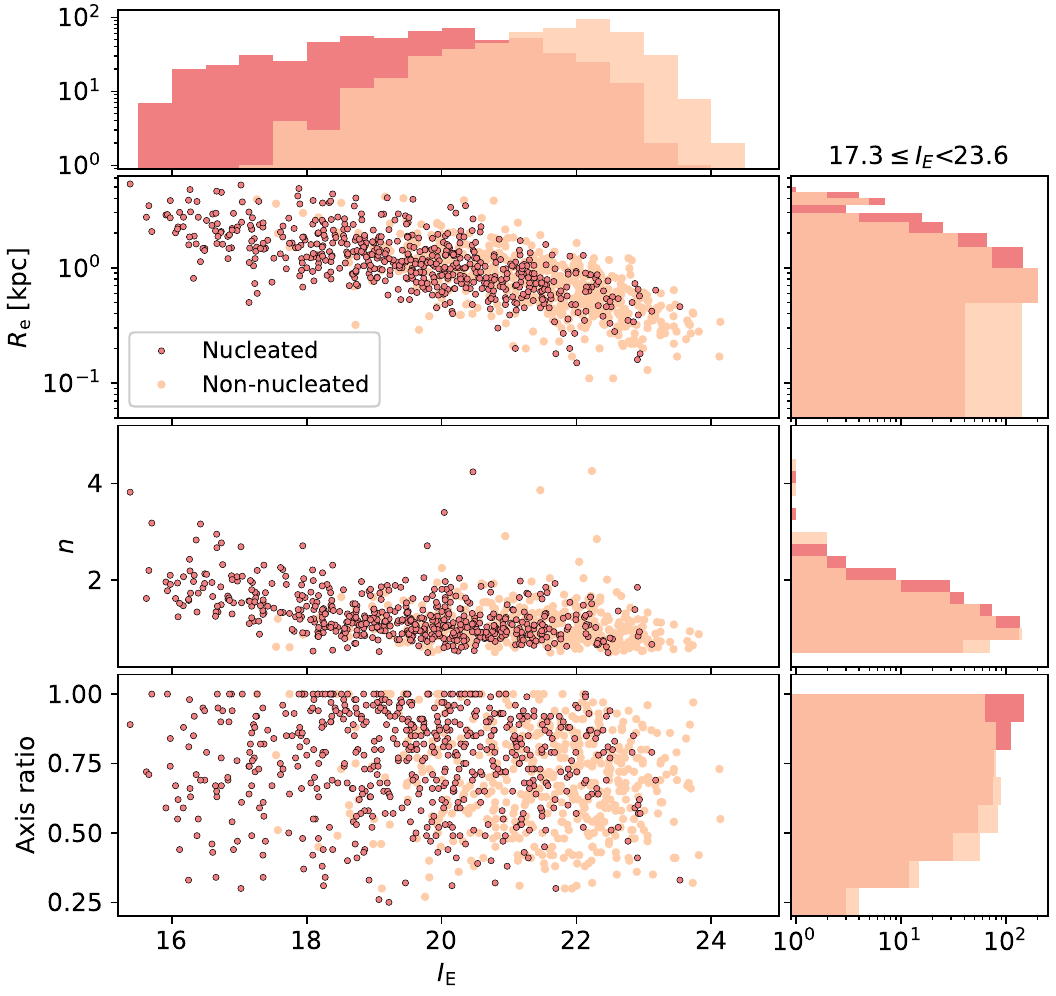}}
\caption{Scaling relations of \IE\ as a function of the $R_{\rm e}$ (\emph{top}), S\'ersic index $n$ (\emph{middle}), and axis ratio (\emph{bottom}) of the nucleated (\emph{pink}) and non-nucleated (\emph{peach}) dEs in the ERO Perseus field. We display the distribution of magnitude for the whole samples, as well as of the $R_{\rm e}$, $n$, and the axis ratio within the same magnitude range $17.3\leq \IE<23.6$.
\label{fig:nucprop}}
\end{figure*}

\section{\label{sc:Phot-Str} Photometry and structural parameters}

\subsection{Extinction correction}
\label{sec:extcorr}

To derive the correct values of the attenuation due to the Milky Way (MW) dust the following ingredients are needed: the map of the Galactic dust distribution; an extinction law; the spectral energy distribution (SED) of the extragalactic source; and the transmission functions of the relevant filters. 

The sky region of the Perseus cluster is significantly affected by extinction from MW dust, as can be seen in Fig.\,\ref{fig:extinction}. To correct for it, we adopted the \texttt{HEALPix} \citep{Gorski2005} dust opacity map with $N_{\rm side}=2048$ released by the Planck Collaboration in 2013 \citep{2014A&A...571A..11P}.\footnote{\protect\url{{https://irsa.ipac.caltech.edu/data/Planck/release_1/all-sky-maps/}}} 
This map provides information on the colour excess $E(B-V)$ -- obtained with data from which point sources were removed --, the optical depth at 353\,GHz $\tau$, and its uncertainty $\diff \tau$. Each galaxy was associated with the \texttt{HEALPix} pixel containing its coordinates, from which we derived the values of $E(B-V)$ and its uncertainty $\diff E(B-V) = E(B-V) \, \diff \tau/\tau$. 

To derive the attenuation at different wavelengths we used the extinction curve $k(\lambda) = R_V \times A(\lambda)/A_V$, where $A(\lambda)$ and $A_V$ are the magnitudes attenuated at the wavelength $\lambda$ and in the $V$ filter, from \cite{2023ApJ...950...86G}, with $R_V=3.1$ implemented through the {\tt dust-extinction} package.\footnote{\url{https://dust-extinction.readthedocs.io/}} 

The correct derivation of the attenuation in each observed bandpass depends on the SED of each object \citep[e.g.][]{2017A&A...598A..20G}. However, since the SED of each galaxy is unknown, a general approach is to assume a test case SED, e.g., a flat spectrum in frequency or a stellar SED. We employed the flux $F$ from a $5700\,\mathrm{K}$ blackbody, to represent a G7V star, which resembles a typical galaxy continuum, as a proxy.

Finally, the exact attenuation in each band, especially in the broad bands that we are considering in the present work, depends also on the shape of the throughput, including filter, optics, mirror, and detector. We used the official bandpasses $R(\lambda)$ for NISP \citep{Schirmer-EP18}\footnote{\url{https://euclid.esac.esa.int/msp/refdata/nisp/NISP-PHOTO-PASSBANDS-V1}} and VIS \citep{2016SPIE.9904E..0QC,Scaramella-EP1}. 

With the above ingredients, for each filter $x$ we derive the quantity $c_x$ as
\begin{equation}
c_x = 2.5\logten\frac{\int R_x(\lambda)\, \lambda\, F(\lambda)\, 10^{0.4\,k(\lambda)} \, \diff\lambda}{\int R_x(\lambda)\, \lambda\, F(\lambda)\, \diff\lambda}\, ,
\end{equation}
which can be used to obtain the intrinsic magnitudes as $m_{\rm obs} - c_x E(B-V)$. With this approach, we obtained $c_x = 2.122$, $1.066$, $0.726$, and $0.470$ in the $\IE$, $\YE$, $\JE$, and $\HE$ bands, respectively.

\subsection{Cutouts}

The cutouts for the photometric and structural parameter analysis were generated with sizes of 9\,$R_{\rm e}$ \citep{Poulain2021}, where the preliminary estimate of $R_{\rm e}$ was determined by using the area in ${\rm deg}^2$ defined by the users with the annotation tool and assuming a visual extent of the galaxy corresponding to approximately 2\,$R_{\rm e}$. 

Of the 1100 dwarf cutouts, 16 had to be recut to a smaller than 9\,$R_{\rm e}$ size as they fall near the edge of the image, which caused issues with the fitting algorithms. For these galaxies, we created smaller cutouts while making sure to have sufficient remaining sky level pixels in the image. Because a number of galaxies (24) fall outside the FoV of the ERO Perseus observations in one or more \Euclid NIR filters, the measurements for those galaxies were not possible and therefore the tables of photometric parameters show no entries for that particular galaxy and filter.

The cutouts were created with the same angular size for all \IE\ as well as $\YE$, $\JE$, and $\HE$ images.
The range of cutout regions provided from the annotation tool corresponds to $ \ang{;;10}$--$\ang{;;170}$ (or $5$\,kpc -- $60$\,kpc).

\begin{figure}[ht]
        \centering
         \includegraphics[width=0.45\textwidth]{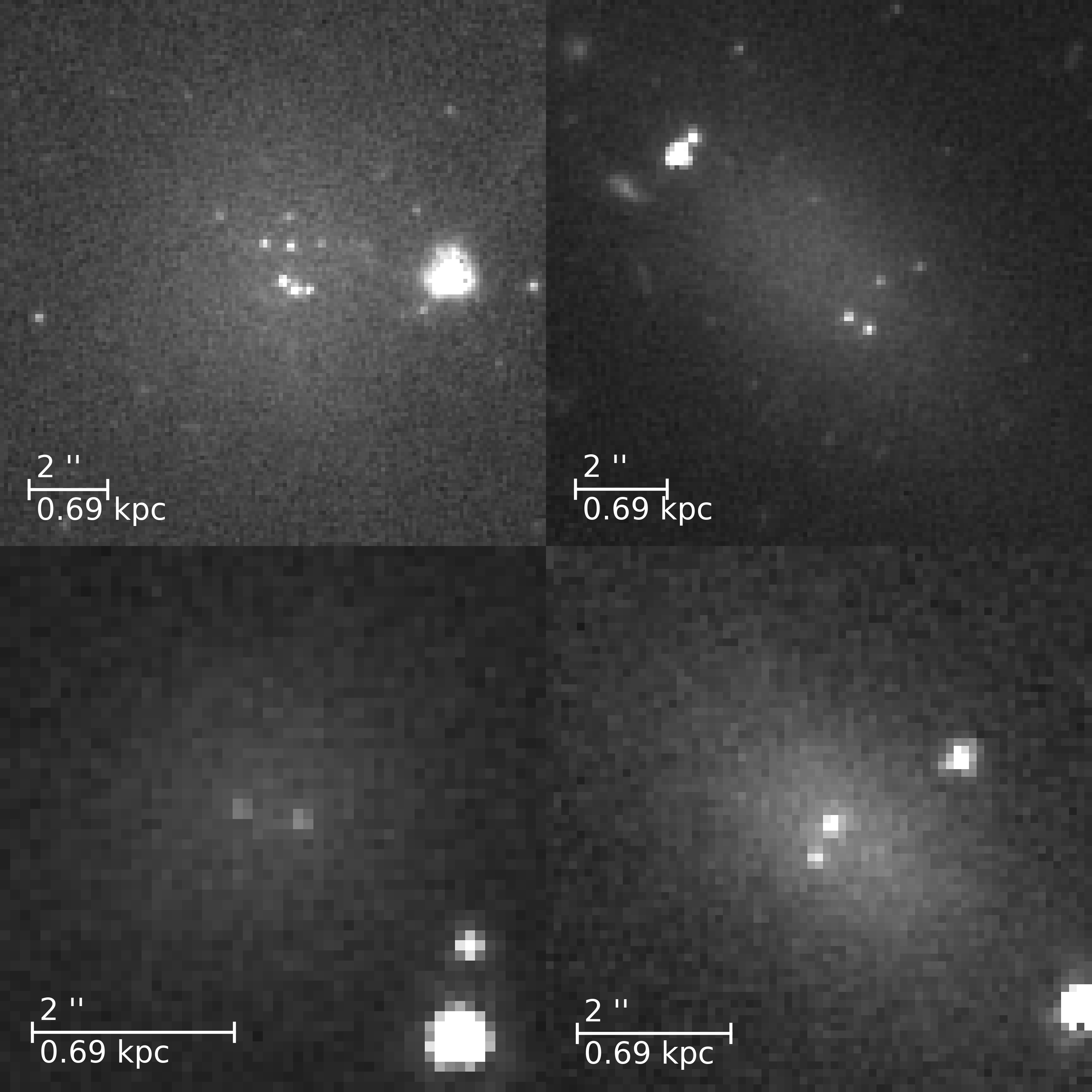}
    \caption{Examples of dwarfs that stand out due to their number of nuclei candidates.}
        \label{fig:multinuclei}
\end{figure}

\begin{figure}[ht]
    \centering
        \centering
        \includegraphics[width=0.45\textwidth]{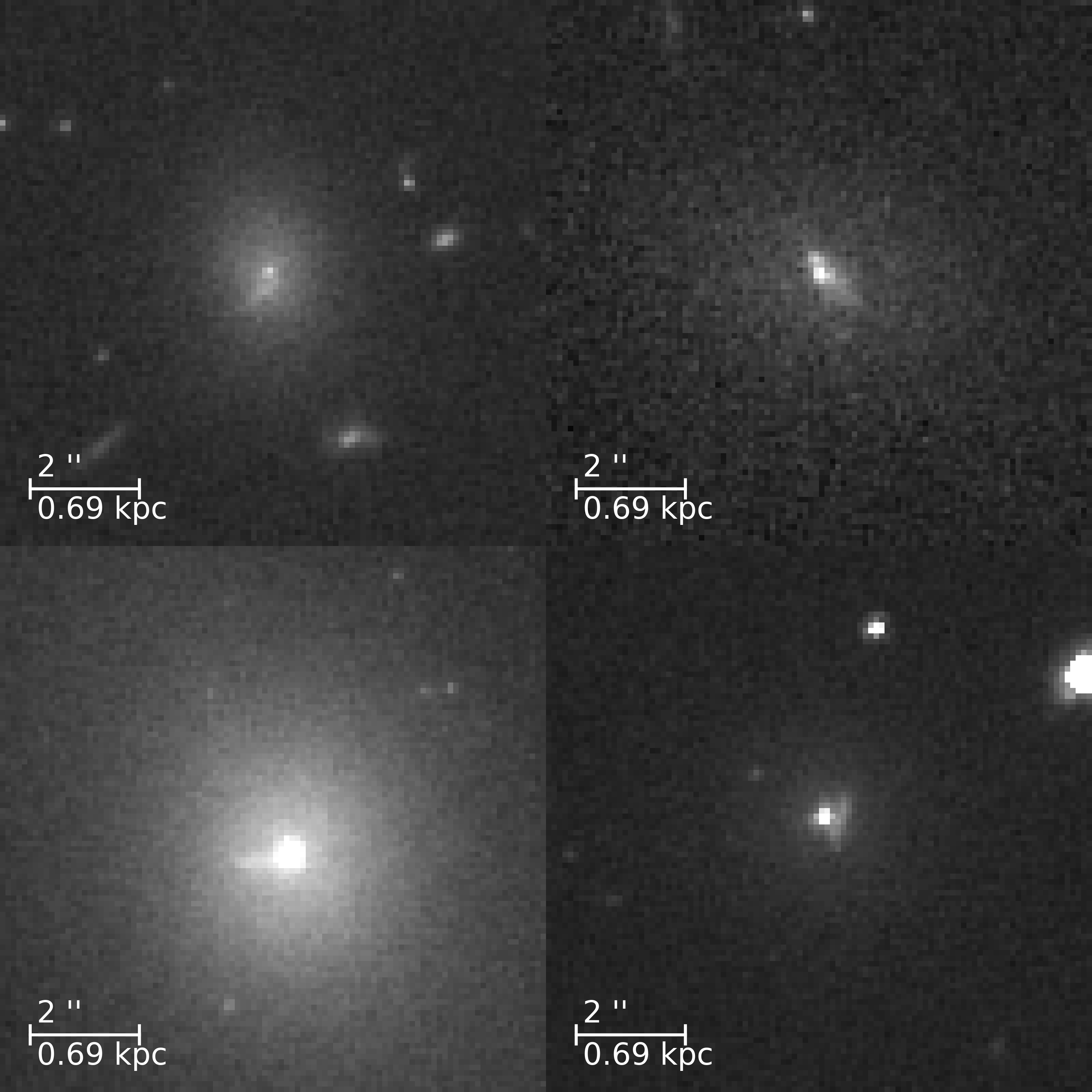}
    \caption{Examples of dwarfs that stand out due to the presence of a complex nucleus.}
        \label{fig:complexnuclei}
\end{figure}

\subsection{Masking}
\label{sec:masking}

The masking of contaminating sources in the cutouts is done via two main steps. The first step is to run \texttt{MTObjects} \citep{Teeninga2015} that produces the MTO segmentation image. The advantage of using \texttt{MTObjects} is that it is capable of locating the faint outskirts of objects. The code is run for all the \IE\ and \YE, \JE, and \HE\ image cutouts independently. In order to avoid including the dwarf galaxy in the mask, the parameter \texttt{move\_factor} is optimally selected for the ERO Perseus images and furthermore, a region of size 1\,$R_{\rm e}$, centred on the dwarf galaxy, is unmasked after this step is completed. The masks are also visually inspected. The second step involves running \texttt{SExtractor} \citep{Bertin1996}, which is good at detecting point sources -- including within the galaxy of interest -- on all of the \IE\ and \YE, \JE, and \HE\ image cutouts. We run \texttt{SExtractor} three times with different regions of the image centred on the dwarf masked, as well as different detection and analysis thresholds, minimum and maximum areas, background mesh sizes and deblending contrasts.

The MTO and \texttt{SExtractor} masks are then combined to create a final mask. To ensure that the nucleus of the dwarf has been masked by the \texttt{SExtractor} runs, the nucleus of each of the nucleated dwarfs identified by our visual inspection is masked in a final step. The exact position of the nucleus is determined by finding the maximum pixel value within a region of 15~pixels around the galaxy coordinates and a region of 4(2) pixels in radius in the \IE(\YE,\JE,\HE) images is masked in order to make sure that the nucleus does not affect the fit of the diffuse component of the dwarf galaxy.

\begin{figure*}[ht!]
\centerline{
\includegraphics[width=0.33\linewidth]{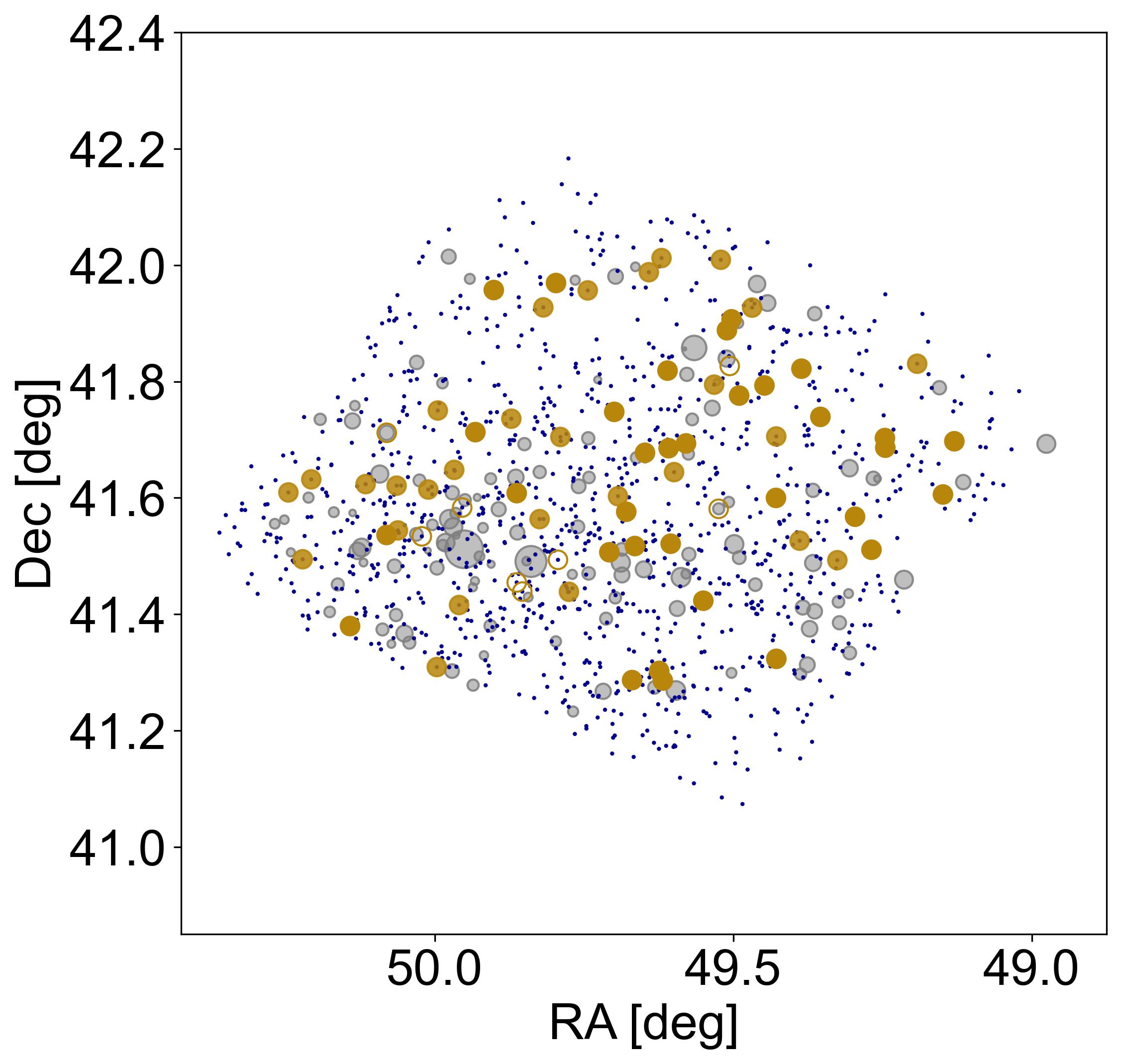}
\includegraphics[width=0.33\linewidth]{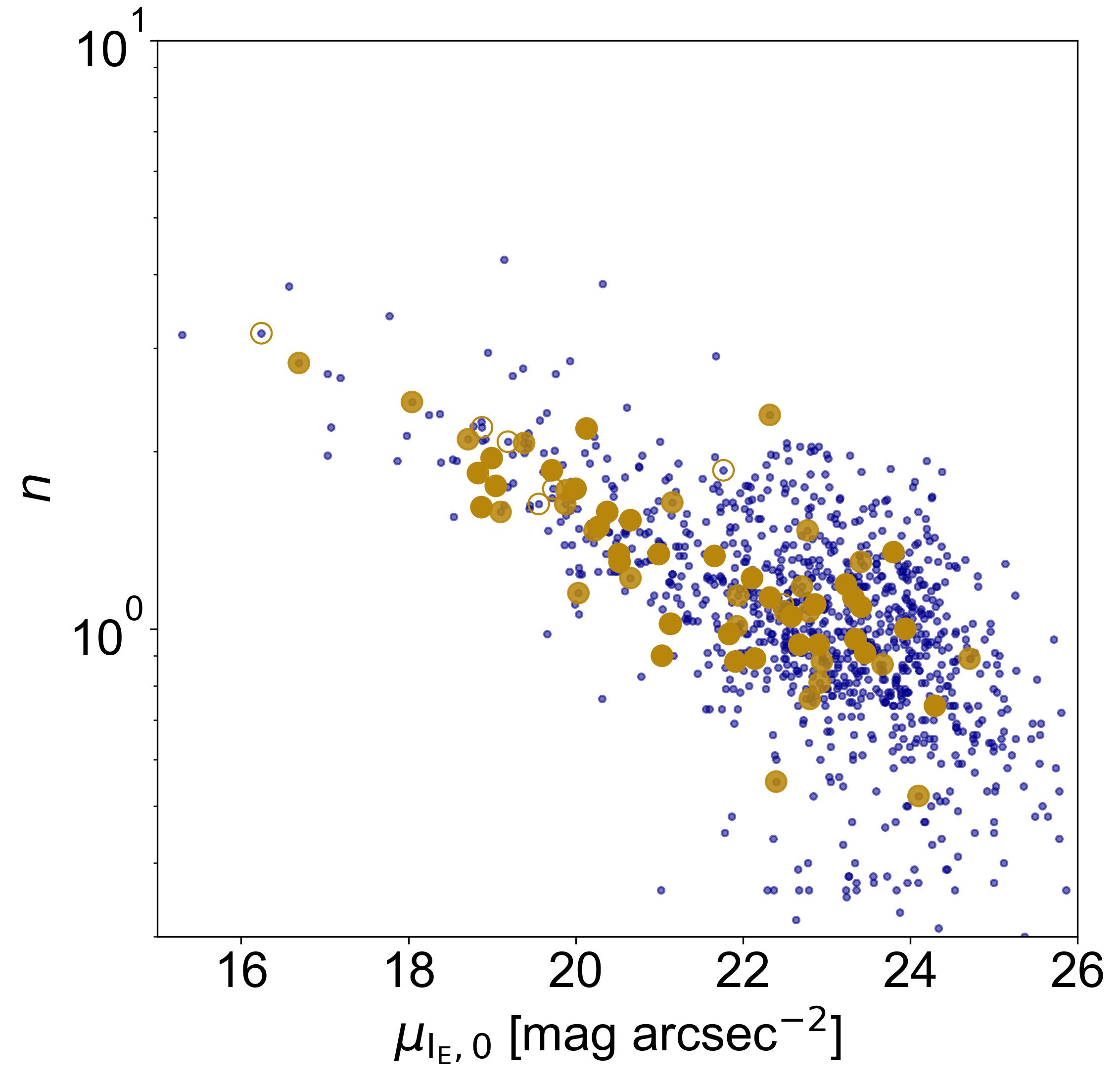}
\includegraphics[width=0.32\linewidth]{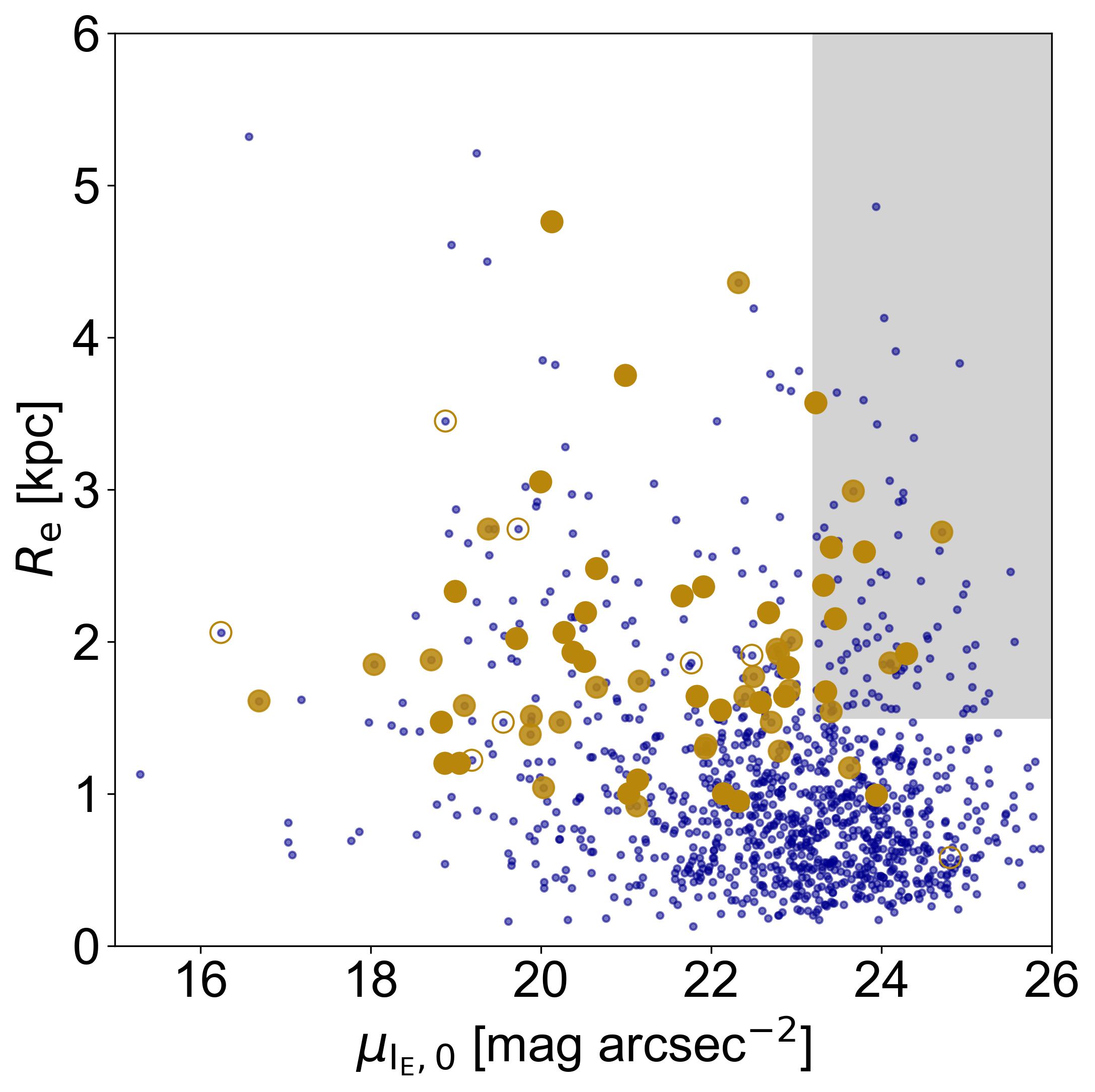}
}
\centerline{
\includegraphics[width=0.33\linewidth]{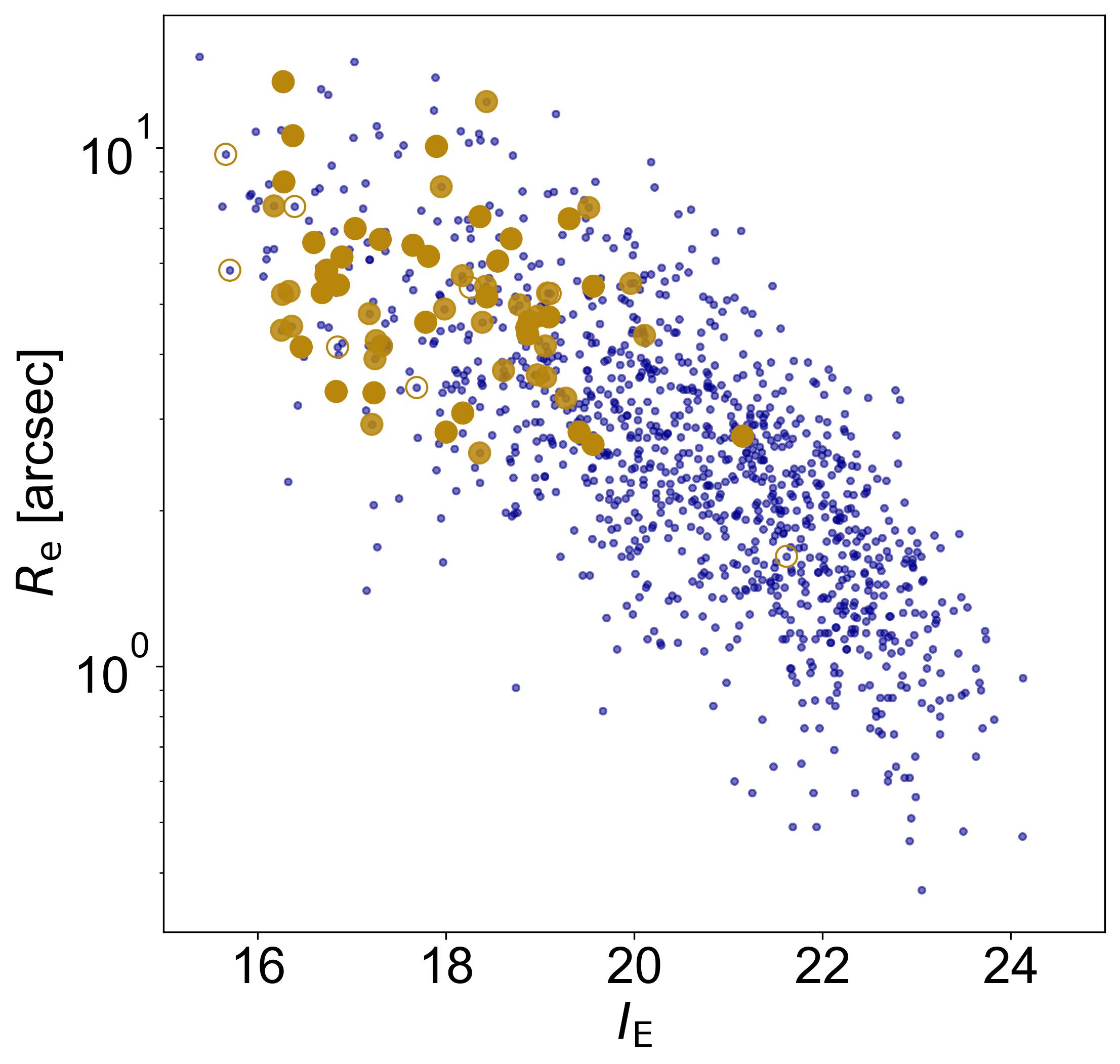}
\includegraphics[width=0.33\linewidth]{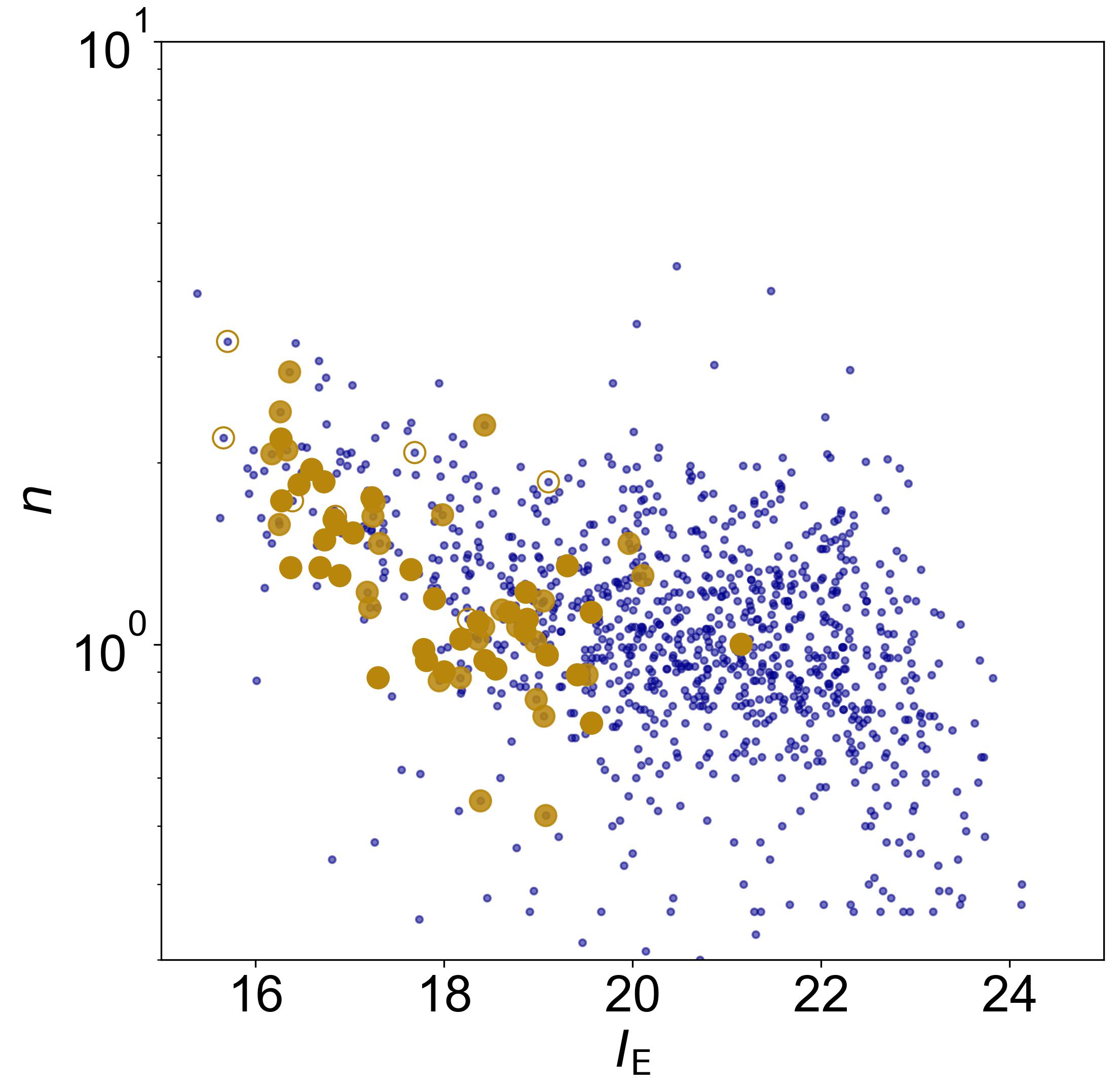}
\includegraphics[width=0.33\linewidth]{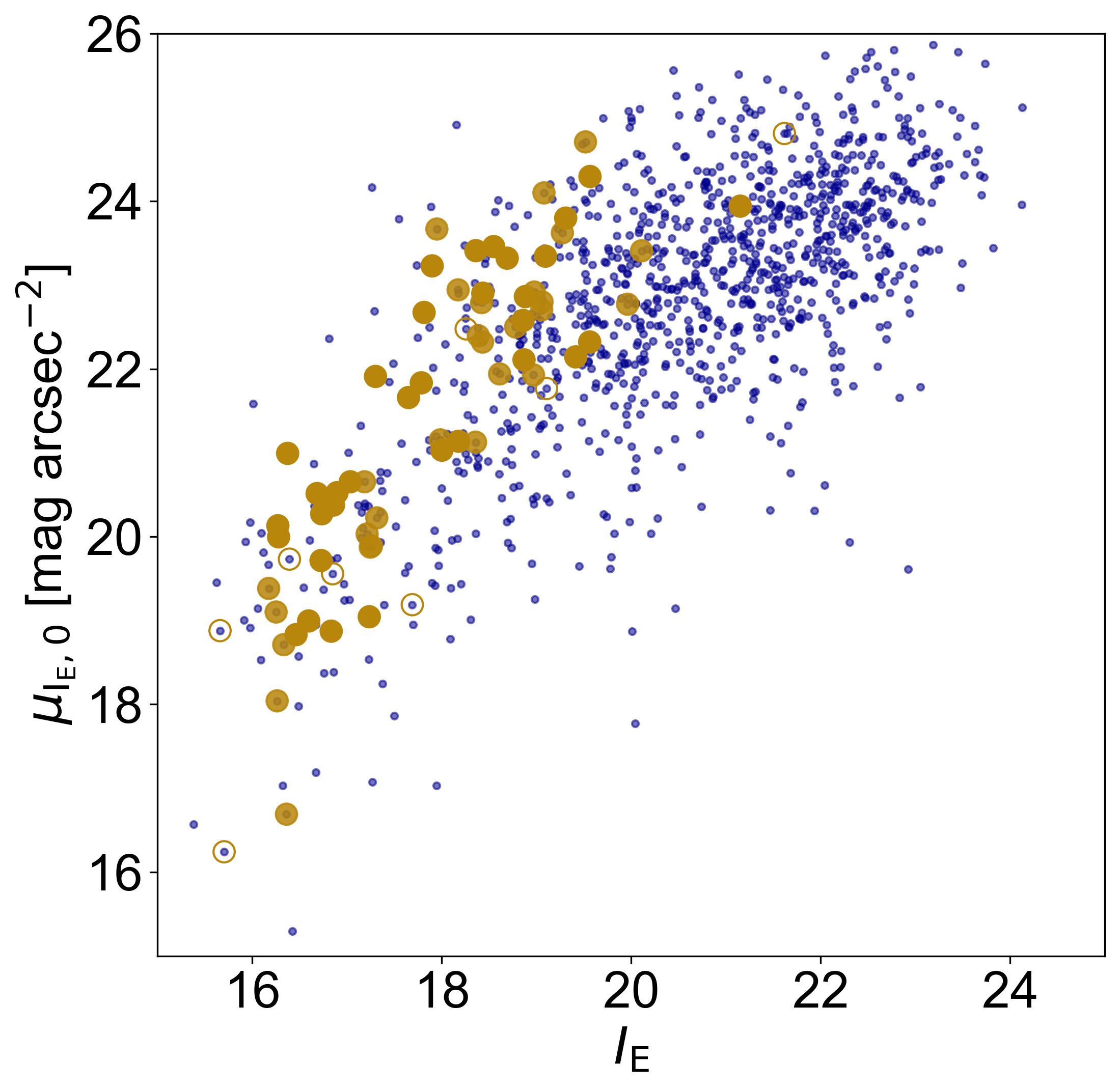}
}
\caption{Structural, photometric, and positional properties of galaxies in our sample, with visually selected GC-rich targets identified. The full sample of dwarf galaxies is shown with {violet-blue dots}. {Gray filled circles} represent galaxies brighter than $\IE=16$, with the symbol size indicating the galaxy effective radius ({\it upper left panel}). Visually identified GC-rich candidates are denoted by {filled orange} or {empty orange circles} (the latter is used for dwarfs near bright galaxies). The {grey shaded area} in the upper right panel represents the expected locus for UDGs, as defined by \citet{vanDokkum2015} and in Sect.\,\ref{sc:Results}.} 
\label{fig:visual_gcrich}
\end{figure*}

\subsection{Photometric and structural parameters}
\label{section:photo-and-struct}

The galaxy modelling was first performed on the \IE\ images due to the better spatial sampling and spatial resolution. We first applied a non-parametric galaxy image analysis method to obtain an initial guess of the structural parameters of each dwarf galaxy using \texttt{AutoProf} \citep{Stone2021}. The masks were used, but since complete isophotes cannot be masked using \texttt{AutoProf} (it ignores the mask in this case), the nuclei were not masked in this first step. A total of 872 sources finished with convergence and 228 did not converge. The following photometric and structural parameters were calculated from the \texttt{AutoProf} output: total magnitude (\IE); effective radius ($R_{\rm e}$); surface brightness (central: $\mu_{\IE,\rm 0}$, at $R_{\rm e}$: $\mu_{\IE,\rm e}$, and within $R_{\rm e}$: $\langle \mu_{\IE,\rm e} \rangle$); S\'ersic index ($n$); position angle (PA); and axis ratio (AR). For the galaxies with no convergence, we used the median value of the results from the fits that finished with convergence for the next step (see below).

These initial estimates of the structural parameters were then used as inputs to both \texttt{Galfit} \citep{Peng2010} and \texttt{AstroPhot} \citep{Stone2023}. The model used for both consists of a two-dimensional (2D) S\'ersic function \citep{Sersic1963} combined with a PSF. 

To obtain the PSF models, we first used \texttt{SExtractor} to detect high SNR sources (\texttt{DETECT\_MINAREA}\,$=5$; \texttt{DETECT\_THRESH}\,$=100$ for \IE and \texttt{DETECT\_THRESH}\,$=40$ for NISP). Then, point sources were selected and PSF models were created with \texttt{PSFEx} \citep{Bertin2011} using the pixel basis, \texttt{PSFVAR\_DEGREES}\,$=2$, and without over- nor under-sampling. The PSF models at the centres of the images were reconstructed from the \texttt{PSFEx} output as a linear combination of the PSF vectors. The PSF models for \IE, \JE, \YE, and \HE are shown in Fig.\,\ref{fig:PSF}.

The \texttt{Galfit} and \texttt{AstroPhot} codes were run in parametric mode using the above model and masks. In addition, two runs of \texttt{AstroPhot} were performed using different sky models, a flat and a plane sky model. The plane sky model was run in order to improve the fitting outcome of the dwarfs embedded in strong sky gradients due to a nearby massive (bright) galaxy. All models and residuals produced by both codes were visually inspected and the output parameters of the best fits were retained for the final catalogue. 

From these runs, a total of 854 dwarf galaxies returned a good fit and therefore robust structural parameters. The remaining 246 galaxies required manual intervention. For 123 dwarfs, it was necessary to manually edit the masks due to bright stars or other sources of contamination that fell within the central 1\,$R_{\rm e}$ of the dwarf, a region that was previously unmasked in \texttt{MTO} in order to avoid removing the dwarf itself. For the remaining 123 galaxies, patching of the image was needed because adjusting the masks was not enough. These cases were extreme and showed very strong contamination from stellar spikes, stars, bright nearby galaxies, or a combination of features. Of these, the smaller (and faint) dwarfs were the most affected and difficult to fit. These 89 dwarf galaxies have a S\'ersic index of 0.36 or below. For $n < 0.36$ the meaning of $R_{\rm e}$ and $\IE$ changes \citep{Sersic1963}; they are no longer at the half light radius. \texttt{AstroPhot} uses the fourth-order expansion of $b_n$ \citep{Sersic1963} to get down to $n=0.36$, but higher orders are needed to go further. Therefore, for the dwarfs with $n \leq 0.36$, the extracted parameters should be treated with caution and with this limit in mind. Note that some of those galaxies actually have a good fit from \texttt{Galfit}, so this caution does not apply to all 89 dwarfs.

Histograms of the extracted parameters are shown in Fig.\,\ref{fig:struct} and the values for each dwarf candidates are given in Table\,\ref{appendix:struct-param1}. The absolute magnitudes and effective radii (in kpc) were computed using the distance of 72~Mpc. 

\subsection{Colours}\label{sc:colours}

Once the photometry and structural parameters of the dwarfs were obtained, aperture photometry was performed on all VIS+NIR images to obtain colour information on the dwarfs. The aperture photometry was done using the python package \texttt{photutils} and using a circular aperture of 1\,$R_{\rm e}$, with the effective radius taken from the best fit to the dwarf in the \IE\ image. The aperture photometry was performed without using a mask and hence includes any nucleus or contaminating source. The aperture magnitudes were then corrected for extinction using the method described in Sect.~\ref{sec:extcorr} and applied at each dwarf position independently and for each \Euclid passband. 

The aperture photometry magnitudes and the extinction correction (EC) at each galaxy position and \Euclid passband are given in Table\,\ref{appendix:struct-param2}. The distribution of VIS-NISP aperture colours of the Perseus dwarfs are shown in Fig.\,\ref{fig:colours}. We compare these colours with those of the two dwarf irregular galaxies IC\,10 and Holmberg\,II presented in \citep{ERONearbyGals}. We find that their reported \IE\,$-$\,\HE colours of $-0.419\pm0.105$ and $0.029\pm0.090$, respectively, fall well within the colour distribution shown in Fig.\,\ref{fig:colours} for the ERO Perseus dwarfs. The NISP-NISP aperture colours were also computed and can be found in Fig.\,\ref{fig:colour-colour}.

\section{\label{sc:Results} Results}

\subsection{Luminosity and stellar mass function}

The luminosity and stellar mass function (LF and SMF, respectively) of the ERO Perseus galaxy population, consisting mostly of the dwarf galaxy sample presented here, is discussed in a separate paper \citep{EROPerseusOverview}. The faint end slope of the LF, $\alpha_{\rm S}$, is found to have a value of $\alpha_{\rm S} = -1.2$ to $-1.3$. The criteria used to identify all cluster members, maximize completeness and minimise the contamination of foreground and background sources, along with the interpretation of the results in terms of models, can be found in \citet{EROPerseusOverview}.

\subsection{Scaling relations}

Galaxy populations can be characterised in structural parameter spaces \citep[see, e.g.,][] {Kormendy1985, Bender1992, Binggeli1994, Kormendy2009}. In order to compare the structural parameters of our dwarf galaxy sample to scaling relations in the literature, we have to convert the \IE\ magnitudes and surface brightnesses to the $V$ band. To select the UDGs in our sample, we have to convert them to the $g$ band. 

The magnitude transformation from \IE\ to $V$ band is derived using synthetic photometry of SEDs of old stellar populations with subsolar metallicities, about ${\rm [Fe/H]} = -0.5$ \citep{EROFornaxGCs}, similar to quiescent dwarf galaxies in cluster environments: $V = $\;\IE$\,+\,0.5$. The transformation from the $V$ band to the SDSS $g$ band is taken from Lupton (2005):\footnote{\url{https://classic.sdss.org/dr4/algorithms/sdssUBVRITransform.php\#Lupton2005}} $g = V + 0.31$.

We select UDGs based on the definition of \citet{vanDokkum2015}, i.e.\ $\mu_{\rm {g,0}} \geq 24$\,mag\,arcsec$^{-2}$ and $R_{\rm e} \geq 1.5$\,kpc. This selection cut yields 93 UDGs for our 1100 sample of dwarf galaxies with structural parameters. This corresponds to $8.5\%$ of our total dwarf sample, compared to a smaller fraction of $2.7\%$ in the Mass Assembly of early Type gaLAxies with their fine Structures (MATLAS) survey \citep{Marleau2021} and $5.4\%$ reported by \citet{Zoeller_2024} for A262 and A1656. The VIS cutouts of the Perseus UDGs are shown in Fig.\,\ref{fig:UDG}. 

The higher fraction of UDGs in our sample is likely due to the greater surface brightness depth of the ERO data (by $0.5$--$1$\,mag when comparing to the MATLAS survey) and the high angular resolution of the \Euclid data that enables one to clearly distinguish faint dwarfs/UDGs from background sources. The four UDGs in our sample with the lowest central surface brightness are EDwC-0035, EDwC-0424, EDwC-0926, and EDwC-0932 ($\mu_{\IE, \rm {e}} = 26.5$--$27.3$\,mag\,arcsec$^{-2}$). Note that other dwarf candidates that are not classified as UDGs have even fainter surface brightness, with the faintest, as measured at $R_{\rm e}$, being the dwarf candidate EDwC-0239 with $\mu_{\IE, \rm {e}} = 28.7$\,mag\,arcsec$^{-2}$ (not extinction corrected).

We investigate which regions the dwarf galaxies in our sample populate in the $M_{\rm V}$--$R_{\rm e}$, $M_{\rm V}$--$\mu_{\rm {V,e}}$, and $\mu_{\rm {V,e}}$--$R_{\rm e}$ parameter spaces and compare them to scaling relations from the literature in Fig.\,\ref{fig:dwarfscalingrelations}. The structural parameters of ellipticals are taken from \citet{Bender1992} and \citet{Kormendy2009}, while those of classical bulges are from \citet{Fisher2008}, \citet{Kormendy2009}, and \citet{KormendyBender2012}. The structural parameters of BCGs are from \citet{kluge2020}. The literature data points of dwarf galaxies contain Local Group dwarf spheroidals from \citet{Mateo1998} and \citet{McConnachieIrwin2006}, Virgo dwarf spheroidals from \citet{Ferrarese2006}, \citet{Gavazzi2005}, and \citet{Kormendy2009}, dwarf galaxies (including UDGs) from the MATLAS survey \citep{Poulain2021,Marleau2021}, and dwarf spheroidals (including UDGs) in A262 and A1656 (Coma cluster) from \citet{Zoeller_2024}. Within the data set from the literature, BCGs, ellipticals, and classical bulges are represented in dark grey, and dwarf galaxies (including UDGs) are represented in light grey. Bright galaxies from the ERO Perseus luminosity function project \citet{EROPerseusOverview} are depicted in black. Our dwarf sample is split up into UDGs (orange) and and non-UDGs (violet-blue). 

The majority of our data points follow the scaling relations of dwarf galaxies from the literature. In the $M_{\rm V}$--$\mu_{\rm {V,e}}$ and $\mu_{\rm {V,e}}$--$R_{\rm e}$ parameter spaces, the dwarf distribution is slightly extended towards fainter $\mu_{\rm {V,e}}$. Furthermore, the diffuse end of the dwarf parameter relations are more densely populated than in previous studies. This indicates higher completeness of diffuse galaxies, presumably due to a combination of the high depth and high spatial resolution delivered by \Euclid that provides a better separation from contaminating objects. The increased density within the diffuse part of the parameter spaces is particularly evident in the $M_{\rm V}$--$\mu_{\rm {V,e}}$ relation, which significantly deviates from the scaling relations reported in \citet{Kormendy2009}. In that paper, a narrow $M_{\rm V}$--$\mu_{\rm {V,e}}$ scaling relation is reported for the dwarf spheroidals, whereas the distribution of our dwarf galaxy sample is nearly twice as broad. When comparing with datasets from more recent dwarf surveys, such as the MATLAS survey \citep{Marleau2021}, the Next Generation Virgo cluster Survey (NGVS; \citealt{Ferrarese2012}) and the Next Generation Fornax Survey (NGFS; \citealt{Munoz2015}), we find good agreement with the spread of the distributions. Examples of galaxies with structural parameters similar to the most extreme galaxies found in this study were also reported by \citet{Zoeller_2024}. We also find a few galaxies above the $M_{\rm V}$--$\mu_{\rm {V,e}}$ and $\mu_{\rm {V,e}}$--$R_{\rm e}$ or below the $M_{\rm V}$--$R_{\rm e}$ scaling relations of dwarf galaxies; we cannot exclude the possibility that these could be interloping background galaxies.

Furthermore, we note that while UDGs populate the diffuse extension with respect to the scaling relations of spheroidals from \citet{Kormendy2009}, there are also some galaxies populating those regions that do not fulfill the UDG definition. We conclude that the UDG definition does not cover all of the most extreme diffuse galaxies with respect to these structural parameter relations, which is in agreement with the findings from \citet{Marleau2021} and \citet{Zoeller_2024}. 

\begin{figure*}[ht!]
\centerline{
\includegraphics[width=\linewidth]{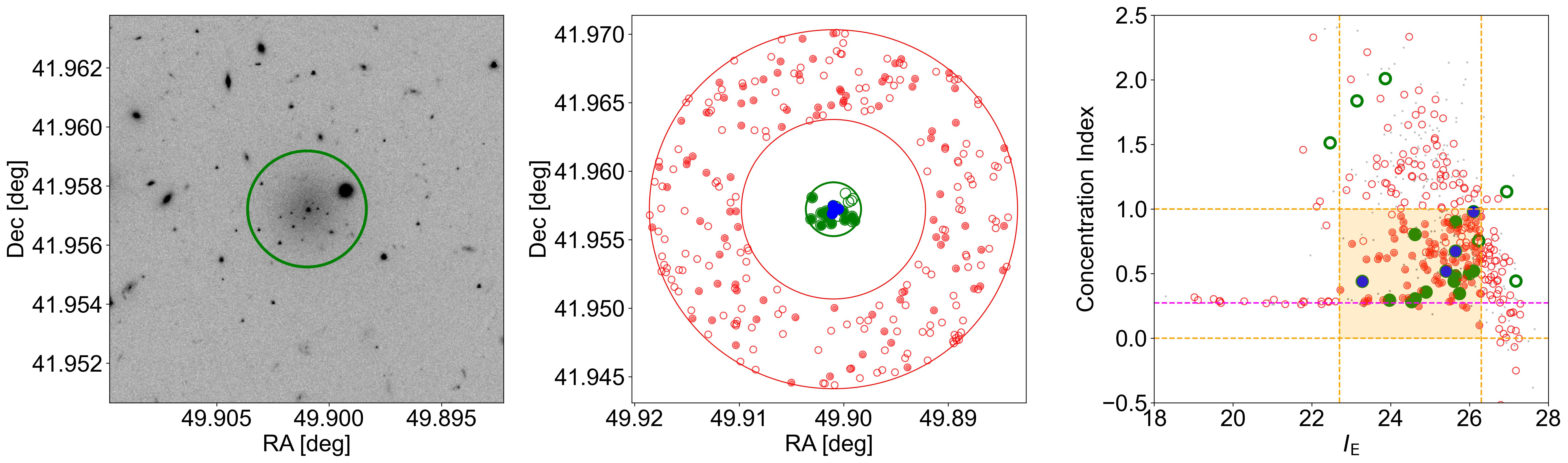}
}
\caption{GC selection procedure. {\it Left panel}: $\IE$ image of the dwarf EDwC-0791. The green circle shows the 1.5\,$R_{\rm e}$ search radius adopted to identify GC candidates around the galaxy. {\it Middle panel}: on-galaxy selection region ({large central green circle}) and the first inner background annulus used for background correction ({red circles}, 5~--~10\,$R_{\rm e}$ in size). All sources detected on the galaxy and in the background region are marked with {green} (on-galaxy) and {red} (off-galaxy) {empty circles}. The sources that fulfill all selection criteria adopted for identifying GC candidates are marked with {filled red (green) symbols} in the outer (inner) area. The {blue symbols} in the centre identify GC candidates within $R_{\rm e}/3$, which are possible nuclear star clusters. {\it Right panel}: concentration index for sources on the galaxy and in the first background annulus. Symbols and colours are the same as in the { middle panel}. The {dashed orange lines} and the {pale orange shaded area} mark the ranges adopted for identifying GC candidates in this plane. The {magenta line} is the median concentration index measured on bright stars.
}
\label{fig:GCautodetect}
\end{figure*}

\subsection{Nucleated dwarf properties}

The morphological classification of the dwarfs led to a sample of 582 nucleated galaxies, comprising 52.9\% of the dwarf sample. A thorough analysis of the effect of the galaxy luminosity and the environment on the nucleated fraction will be presented elsewhere (Euclid Collaboration: S\'anchez-Janssen, R., et al.\ 2024, in prep.). We study as a first step the dependence of the nucleated fraction on the luminosity and morphology of the dwarfs. In Table\,\ref{tab:nucfraction}, we report the nucleated fraction for the whole sample, as well as for the subpopulations of dEs and dIs. We divide the samples into two bins, namely faint and bright dwarfs, by cutting at the mean apparent magnitude of the sample $\IE=20.4$ [corresponding to $M(\IE)=-13.9$]. Studies across a wide range of environments have reported that nucleated dwarfs tend to be brighter than non-nucleated dwarfs, regardless of the morphological type, such that brighter dwarfs have higher nucleated fractions (e.g., \citealt{Cote2006,Ordenes-Briceno2018,Sanchez-Janssen2019,Habas2020,Hoyer2021,Zanatta2021,Carlsten2022}). In agreement with the literature, we find a higher nucleated fraction for the bright dwarfs as compared to the faint ones, with 81.3\% (8.3\%) of the bright dEs (dIs) being nucleated, against 32.5\% of the faint dEs and none of the faint dIs. We find overall a larger nucleation fraction for the dEs as compared to the dIs, which is consistent with the observations of \citet{Habas2020}, but different from the expectations from \citet{Neumayer2020}, where the fraction of nucleated dEs and dIs are similar. The difference between the two fractions can be explained by the low statistics of the dI sample, with two of the 39 galaxies being nucleated, but also from the fact that it is more difficult to identify nuclei in irregular galaxies due to the presence of star-forming regions and dust obscuration. 

\begin{table}
    \caption{The fraction of nucleated dwarfs as a function of the magnitude and morphology of the host dwarf.}
    \label{tab:nucfraction}
    \centering
    \begin{tabular}{lrrr}
    \hline
    \hline
            & dE & dI & All\\
    \hline
    \noalign{\smallskip}
     Bright ($\IE<20.4$) & 81.3\% & 8.3\% & 77.8\%\\
     Faint  ($\IE\geq20.4$)& 32.5\% & 0\% & 31.7\%\\
     All    & 54.6\% & 5.1\% & 52.9\%\\
    \hline
    \end{tabular}
\end{table}

\begin{figure}[ht!]
\centerline{
\includegraphics[width=0.9\linewidth]{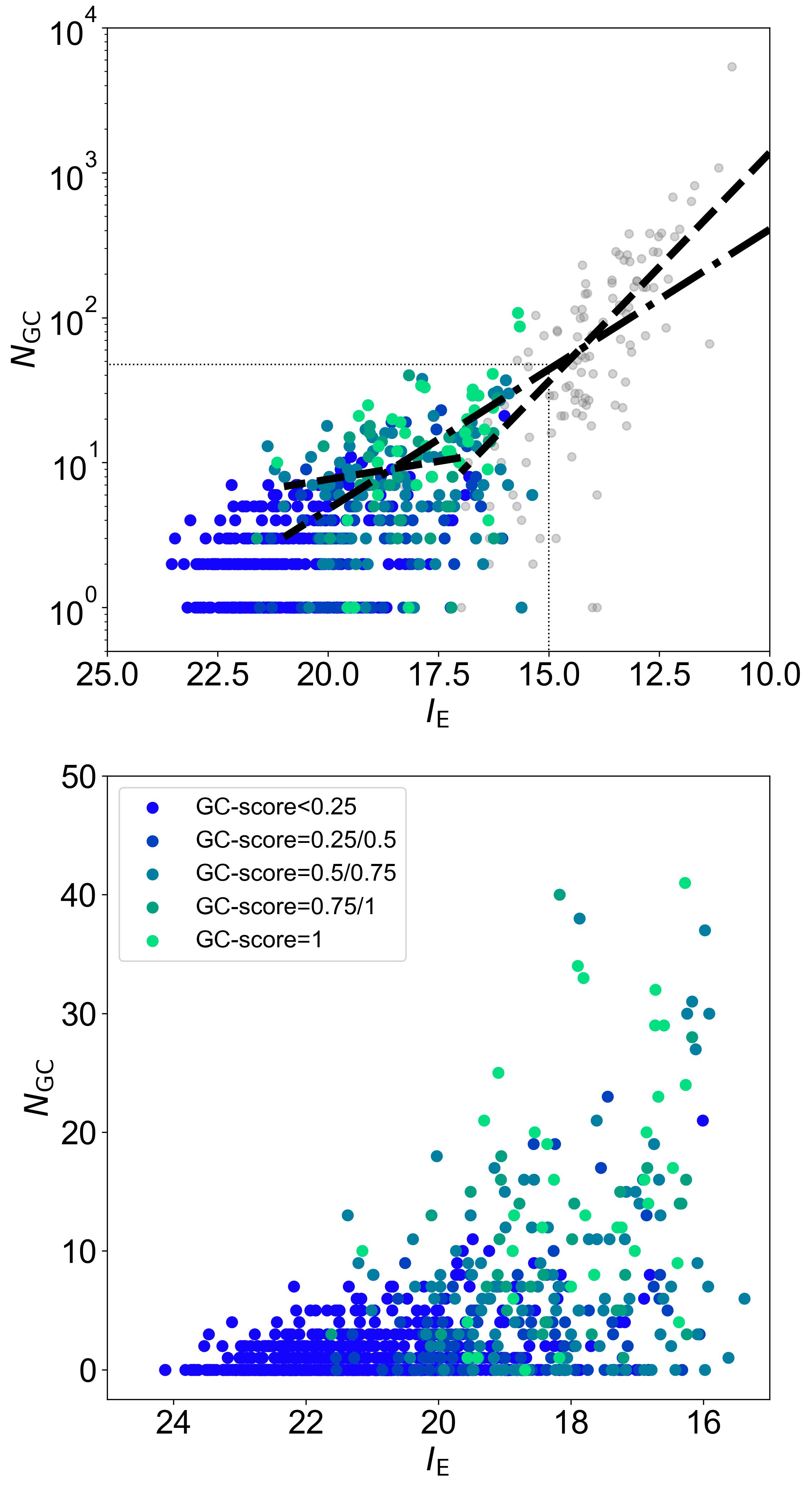}
}
\caption{Total GC population versus galaxy $\IE$ magnitude for the full sample of galaxies ({\it upper panel}, log scale on the $y$-axis) and for the sample limited to faint $\IE<15$ ones ({\it lower panel}, linear $y$-scale). The full sample is represented by {gray filled circles} in the {\it upper panel}. Dwarf galaxies are symbol and colour-coded based on their GC richness score, as labelled. In the {\it upper panel}, we also report fits to the data using the sample of galaxies with $10 \leq \IE \leq 21$ and $N_{\rm GC} \geq 5$ ({dashed-dotted line}), as well as a broken linear fit ({dashed line}) obtained adopting a separation limit at $\IE=17$.
The dotted box in the {\it upper panel} shows the position of the box in the {lower panel}.}
\label{fig:ngc_ie}
\end{figure}

\begin{figure*}
\centerline{
\includegraphics[width=0.33\linewidth]{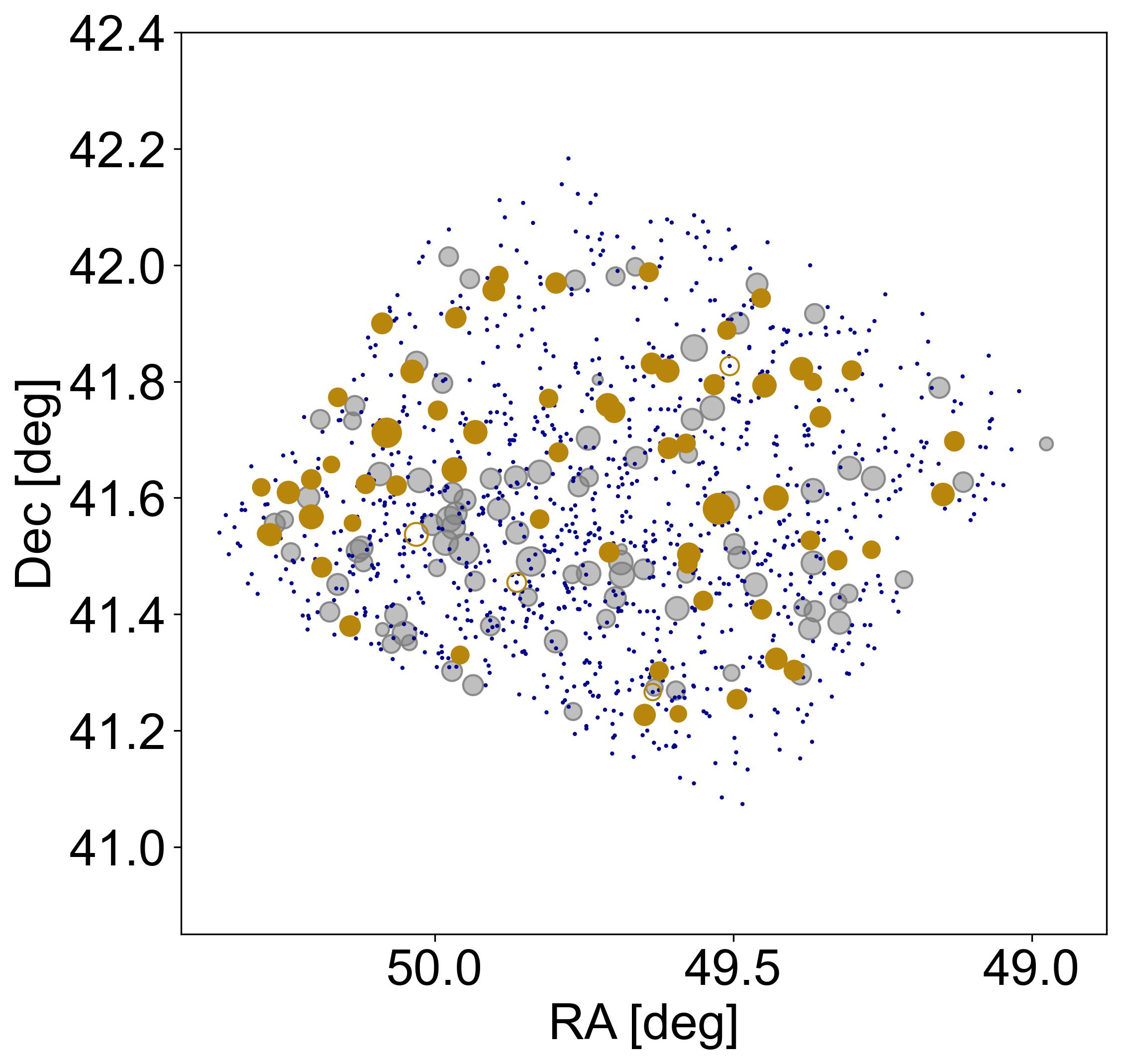}
\includegraphics[width=0.33\linewidth]{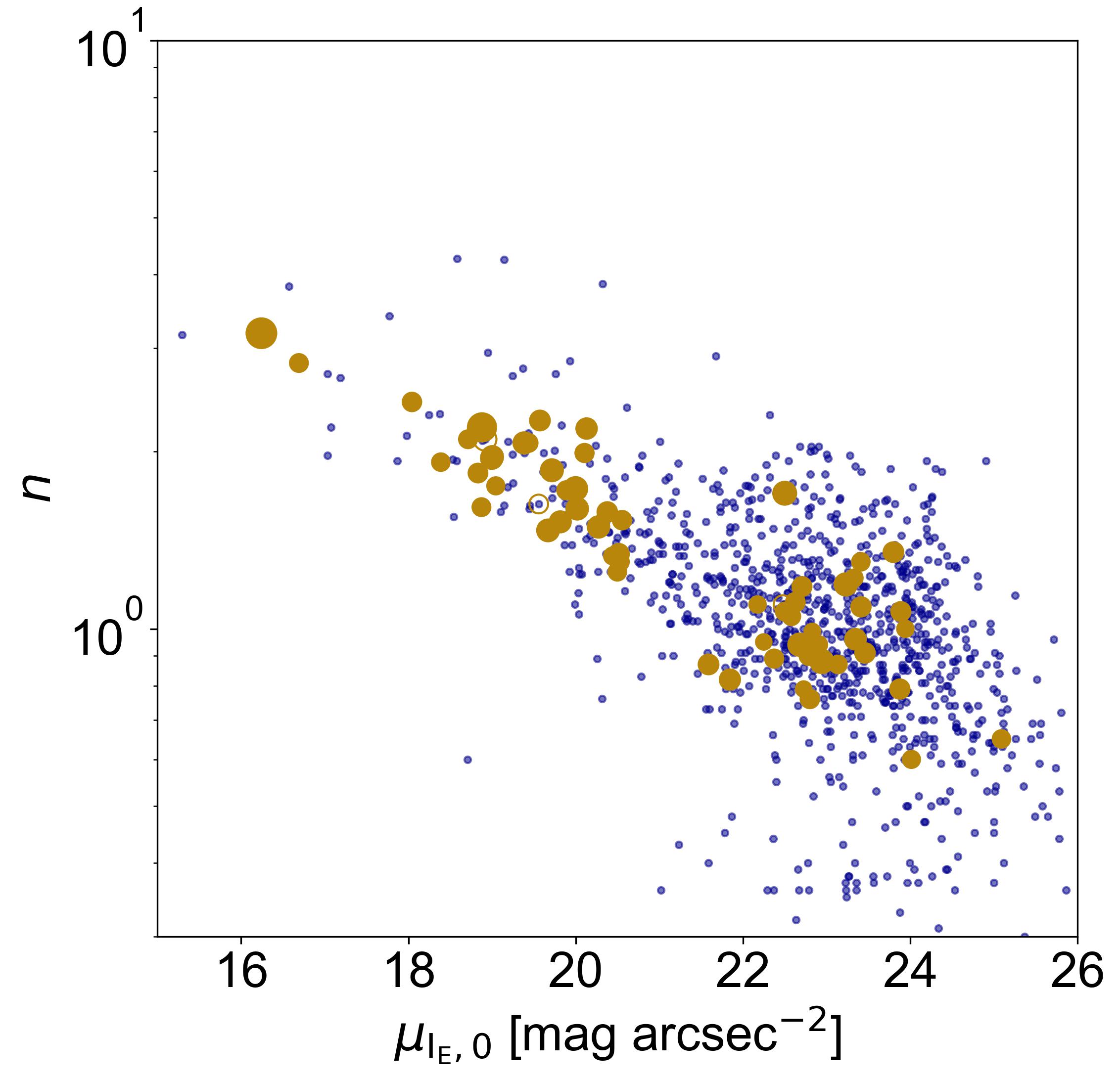}
\includegraphics[width=0.32\linewidth]{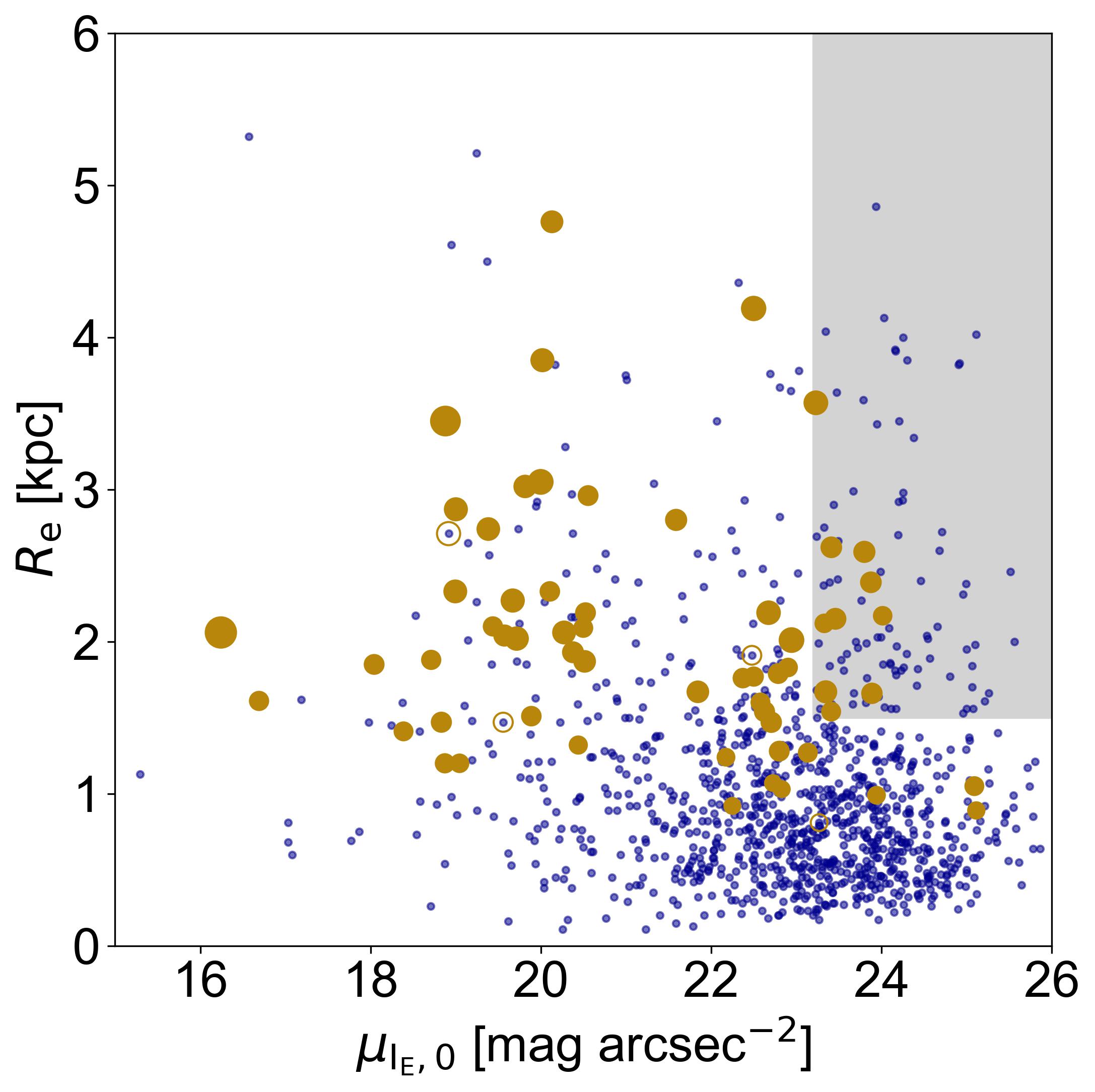}
}
\centerline{
\includegraphics[width=0.33\linewidth]{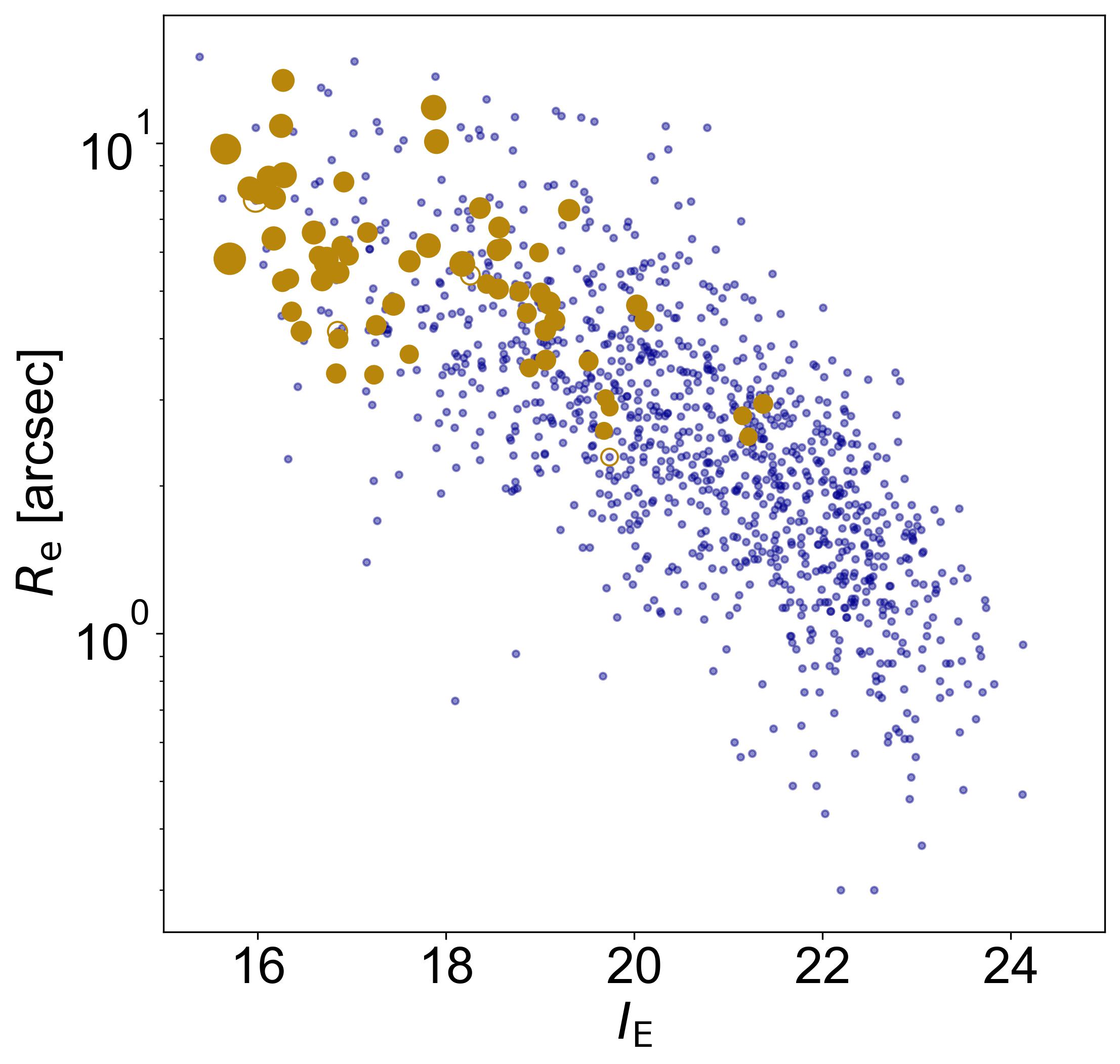}
\includegraphics[width=0.33\linewidth]{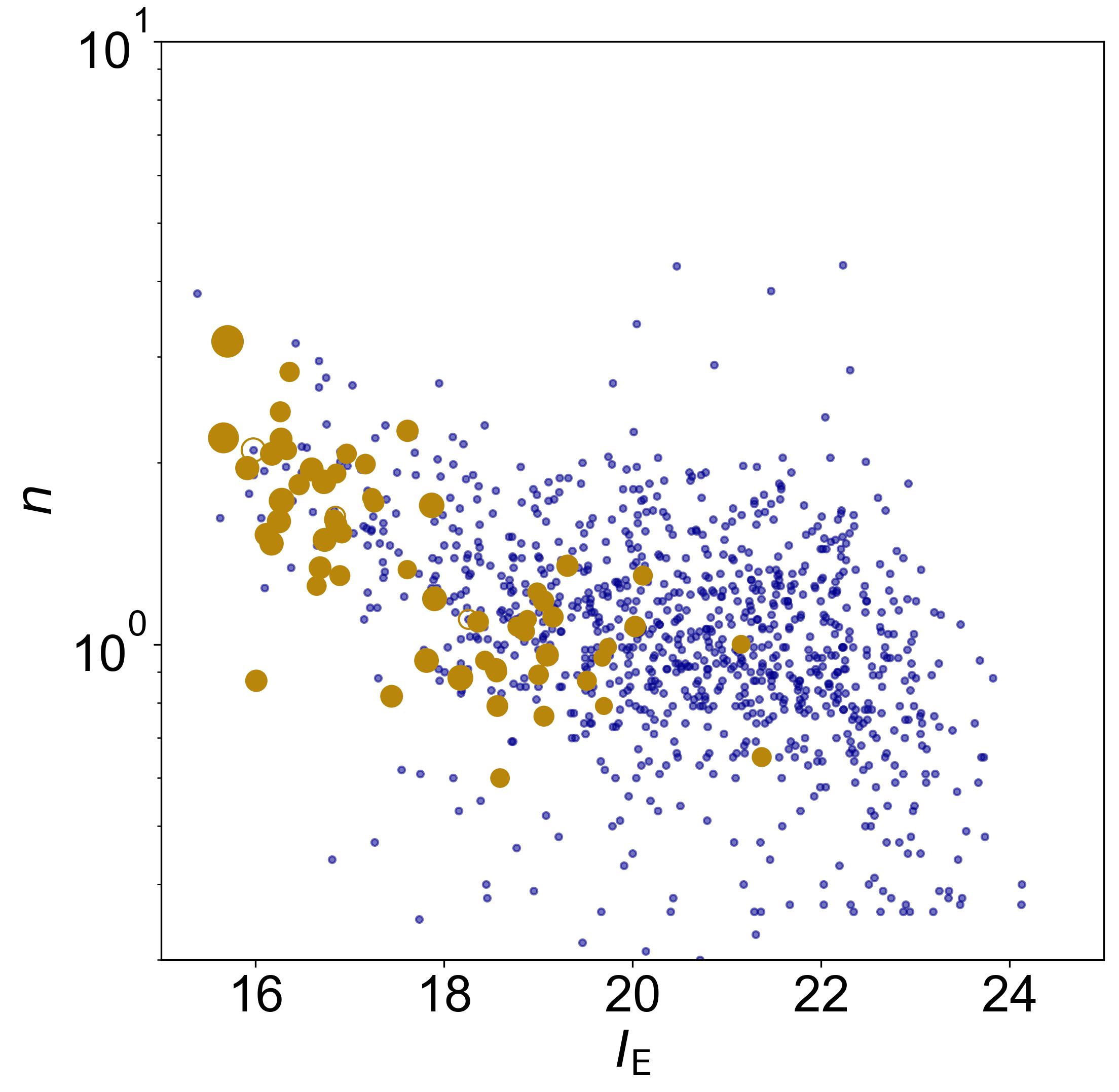}
\includegraphics[width=0.33\linewidth]{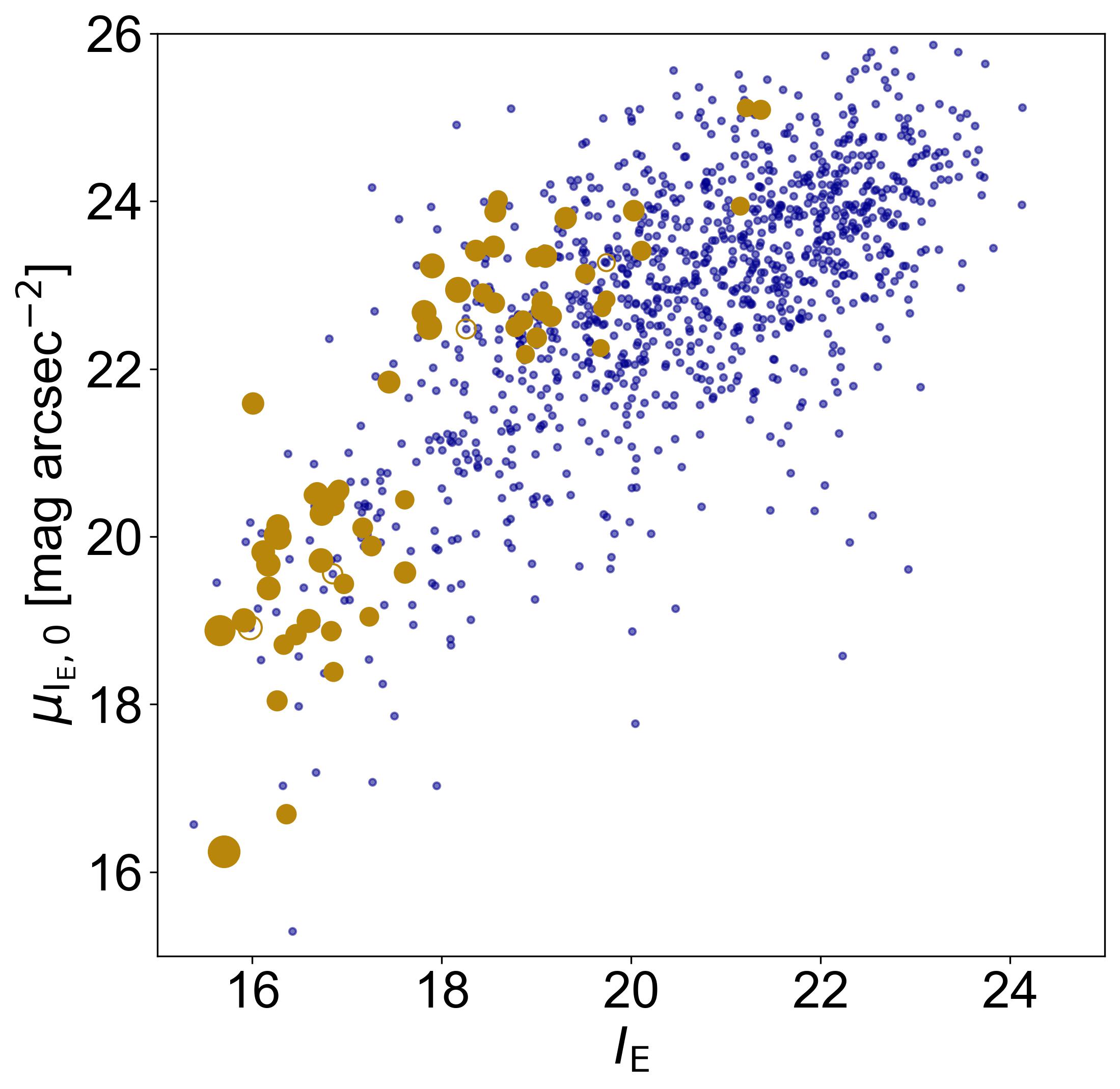}
}
\caption{Same as in Fig, \ref{fig:visual_gcrich}, except that GC-richness is determined through automatic GC detection. Symbols size scales for bright galaxies (grey symbols in the upper right panel) and for the dwarf population (orange symbols in all panels) according to the total population of identified GC candidates, $N_{\rm GC}$, adopting different scaling factors for bright and dwarf galaxies for clarity.} 
\label{fig:ngc_gcrich}
\end{figure*}

As a second step, we investigate possible differences between the structural properties of the nucleated and non-nucleated dwarfs. Recent results strongly indicate that nuclear star clusters (NSCs) in low-mass galaxies grow primarily via dissipationless merging of pre-existing massive star clusters \citep{Sanchez-Janssen2019,Fahrion2022}. In this scenario, it is expected that the structural parameters of the host galaxy play a relevant role in both setting the efficiency of dynamical friction and the strength of the cluster mass loss as it spirals inwards \citep{Leaman2022}. For example, \citet{denBrok2014} find that Coma dwarfs that are more concentrated and rounder tend to host more luminous star clusters. In addition, it has been shown that nucleated cluster dwarfs are intrinsically rounder, at all luminosities, than their non-nucleated counterparts \citep{Sanchez-Janssen2019b,Lisker2007,Venhola2019}.

In Fig.\,\ref{fig:nucprop}, we show the scaling relations between \IE\ and the effective radius $R_{\rm e}$, the S\'ersic index $n$, and the axis ratio of the nucleated and non-nucleated dwarfs in Perseus. Given the low statistics of the dI sample, we focus here on the dEs only. For each relation, we also include histograms of the property distributions within the same magnitude range $17.3\leq \IE<23.6$. We limit the range of S\'ersic index values to $n\geq0.5$ because smaller values indicate a 3D luminosity distribution that is deficient at the centre, which is unphysical \citep{Trujillo2001}. From the distributions of \IE, we see that the nucleated dwarfs tend to be brighter than the non-nucleated ones. Looking at the structural properties at similar magnitudes, we observe similar ranges of properties. However, the distributions do not peak at the same value, such that the nucleated distributions are shifted towards larger $R_{\rm e}$, $n$, and axis ratio. This means that we observe less nuclei in smaller, high ellipticity dwarfs, as well as in galaxies with flatter surface brightness profiles. These results are consistent with the observations in other galaxy clusters and lower density environments (e.g., \citealt{Sanchez-Janssen2019,Neumayer2020,Poulain2021}). 

\subsection{Dwarf nucleated substructures}

During the visual review of the dwarf candidates, the classifiers were allowed to identify multiple nuclei in the dwarfs. A galaxy is visually defined as multinucleated when at least two compact sources of similar luminosity are located close to the photometric centre and are brighter than other surrounding compact sources. The presence of multiple nuclei can be linked to dwarf-dwarf mergers \citep{Pak2016}, where the NSCs of both interacting galaxies are migrating towards the potential well of the newly formed galaxy. Another possibility is the observation of a nucleus with a stellar disc, similar to the double nucleus of M31 \citep{Lauer1993}. Given that the dwarf nuclei can have similar photometric and structural properties to massive GCs (e.g., \citealt{Poulain2021,Hoyer2023}), we can also be observing the infall of massive GCs towards the galaxy centre due to dynamical friction \citep{lotz2001}. Finally, we cannot exclude the possibility that in some cases, the candidate nucleus or nuclei might be a background galaxy or foreground star, seen in projection on top of the galaxy. We show in Fig.\,\ref{fig:multinuclei} examples of four multinucleated dwarfs. 

Together with multiple nuclei, reviewers also noted the presence of complex nuclei in a few tens of objects. These NSCs exhibit some tidal features such as stellar tails, suggesting an ongoing interaction at the centres of the galaxies. These types of structures can be identified only with the depth and high spatial resolution of the VIS observations, since such structures are not resolved in the other bands. Four such nuclei with complex structure are visible in Fig.\,\ref{fig:complexnuclei}. Further investigations are needed to assess the nature of these substructures, which is beyond the scope of this paper.

\subsection{Globular clusters}
\label{sec:GCauto}

To investigate the GC population in galaxies within our field, we adopted two distinct approaches. First, we examined galaxies visually identified as GC-rich, comparing their properties (such as position in the cluster, S\'ersic parameters, etc.) with those of GC-poor galaxies in the sample.
Then, we adopted a more quantitative approach to estimate the total GC population in each galaxy, and inspected the overall characteristics of the GC systems within these galaxies. Showing the results for both visual and automatic detection methods for GC-rich dwarfs  serves two key purposes: maintaining coherence with other sections, which rely on visual inspection, and providing validation for the visual approach.

\subsubsection{GC richness from visual inspection}

As described in Sect.~\ref{sec:morpho_gc_nsc}, during the visual inspection of dwarf galaxies in the Perseus field, classifiers were asked to verify the presence of GC candidates hosted within the galaxies. GC candidates were visually identified as relatively compact and faint sources with a surface distribution centrally concentrated around the galaxy. In this context, a dwarf galaxy was categorized as GC-rich if it contained a minimum of two GC candidates. Part of the reason we chose this low value was to separate the dwarfs with potential GC candidates from the vast majority of dwarfs with zero ($\pm 1$) GC candidates.

With this (somewhat arbitrary) definition of GC richness in mind, Fig.\,\ref{fig:visual_gcrich} compares the spatial, photometric and structural properties of visually GC-rich galaxies with those of the non-GC-rich elements in the catalogue. For the sake of clarity, only dwarfs with GC scores $\geq$\,0.85 are indicated, corresponding to those receiving 6 votes out of 7.

One possible source of confusion in visually inspecting GC richness could arise from the proximity, at least in projection, of a dwarf candidate to a bright, massive companion or to regions of high intracluster GC population density \citep{EROPerseusICL}.
Figure\,\ref{fig:visual_gcrich} (upper left panel) shows the positions of dwarf and bright galaxies in the Perseus field. Here, grey filled circles represent galaxies brighter than ${\IE}=16$ [absolute $M(\IE)=-18.3$], with symbol size scaled with the galaxy effective radius. The full sample of dwarf galaxies is reported with violet-blue dots. GC-rich candidates are denoted by filled or empty orange circles. Filled orange symbols are assigned to dwarfs projected far from any bright galaxy, while empty orange circles are  used for dwarfs within 5 effective radii of a bright companion.

A first observation is the absence of any obvious GC richness enhancement near the bright/massive galaxies in the field. This suggests a real over-density of GCs associated with the dwarfs rather than contamination from the intracluster GCs or the GC system of a nearby massive companion. Furthermore, we do not observe significant evidence of any other trends apart from the clustering of dwarfs around the massive galaxies. Indeed, a clear east-west stretch is evident both for the dwarfs and the giant galaxies.

The remaining panels of Fig.\,\ref{fig:visual_gcrich} show the properties of the dwarf galaxy sample, using the same colour and symbol coding as in the upper left panel, except for the bright galaxies which are not shown. By considering as `consensus sample' the dwarfs with at least 6 out of 7 votes in favour of GC richness, in Fig.\,\ref{fig:visual_gcrich} we observe the following. 

\begin{itemize}
\item No GC-rich and uncontaminated candidate has $R_{\rm e} < 2\farcs0$ ($\sim$800\,pc) or is fainter than $\IE\sim$22 (lower left panel).
\\
\item Two GC-rich objects exhibit $\IE\gtrsim$\,21 [fainter than the absolute magnitude $M(\IE)$\,$\sim\,-13.3$]. One of them is located near the denser cluster area, close to the cluster core, suggesting the possibility of intracluster GCs or bright galaxy GC contamination. The other dwarf (EDwC-0066) appears as a faint diffuse object with a handful of point-like sources within one effective radius (lower left panel).
\\
\item Within fixed $\IE$ intervals, and for galaxies brighter than $\IE\sim20$, the GC-rich candidates appear to have, on average, slightly smaller $n$ values compared to the GC-poor sample. This trend appears to be weak but consistent over the range of $16 \leq \IE \leq 20$. For example, in the 18--19 magnitude bin, we find $\langle n \rangle = 1.0 \pm 0.1$ for the 19 GC-rich dwarfs, compared to $\langle n \rangle = 1.4 \pm 0.3$ for the 28 GC-poor dwarfs (only including those with a score $\leq 1$ out of 7 votes; lower middle panel). 
\\
\item For $\IE \leq20$, there appears to be a tendency for the central surface brightness of GC-rich dwarfs to be slightly fainter than that of the entire sample of dwarfs in the same magnitude range (lower right panel).
\\
\item The S\'ersic exponent, $n$, shows a correlation with the host galaxy central surface brightness. This behaviour has been previously observed \citep[e.g.,][]{Ferrarese2006}; here, we find that it appears to be independent of the GC richness (upper middle panel).
\\
\item Approximately 40 GC-rich uncontaminated dwarfs (orange filled circles in the figure) have $R_{\rm e} \geq 1.5$\,kpc; among them, about ten have central surface brightness $\mu_{\rm {\IE,0}}\,\gtrsim\,23.2$\,mag\,arcsec$^{-2}$ (equivalent to $\mu_{\rm {g,0}}\,\gtrsim\,24.0$\,mag\,arcsec$^{-2}$), hence are possible GC-rich UDG candidates (upper right panel).
\end{itemize}

\begin{figure*}[ht!]
\centerline{
\includegraphics[width=\linewidth]{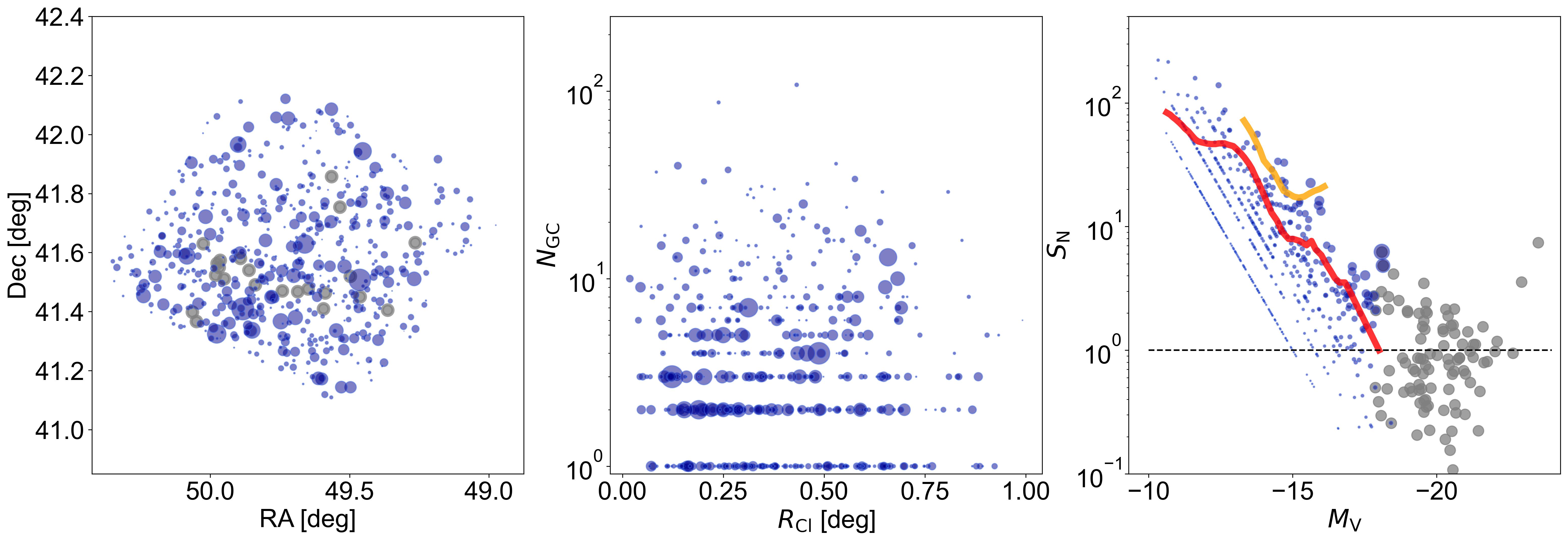}
}
\caption{Galaxy positions and GC content. {\it Left panel}: The {violet-blue circles} indicate the positions of dwarf galaxies, with symbol size scaled to the specific frequency, $S_{{N}}$. {Gray circles} mark the brightest galaxies in the field, with magnitudes $8\leq \IE\leq 13$.
{\it Middle panel}: total population of GC candidates in dwarf galaxies, $N_{\rm GC}$ (log scale), plotted against the cluster-centric distance, $R_{\rm Cl}$, from the BCG. Symbol sizes correspond to those in the {left panel}. {\it Right Panel}: Specific frequency versus magnitude plot, colour and symbols are as in {left panel}, except that the symbol size is scaled to $N_{\rm GC}$. The {red curve} shows the relation from \citet{Lim2020} for UDGs in Virgo, while the {orange curve} is for UDGs in Coma \citep{Lim2018}. The {horizontal dashed line} shows the $S_N=1$ level.
\label{fig:sn_positions}}
\end{figure*}

\begin{figure*}[htbp!]
\begin{center}
\includegraphics[trim={0 0 0 0},angle=0,width=0.9\linewidth]{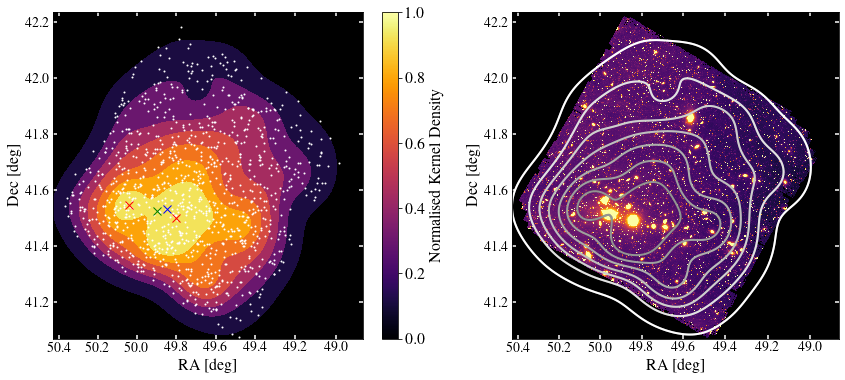}
\includegraphics[trim={0 0 0 0},angle=0,width=0.9\linewidth]{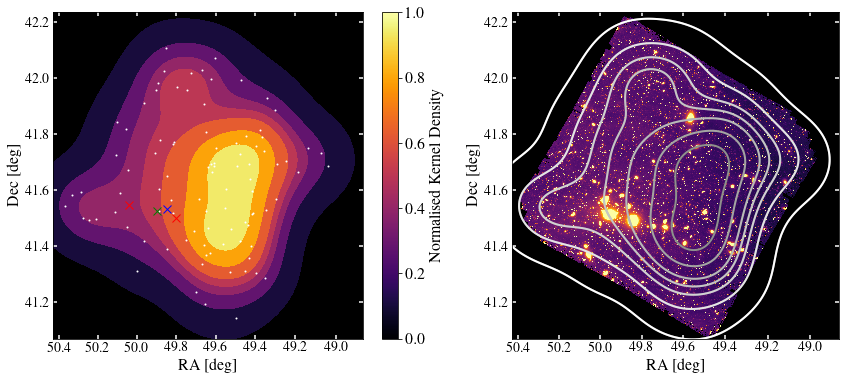}
\includegraphics[trim={0 0 0 0},angle=0,width=0.9\linewidth]{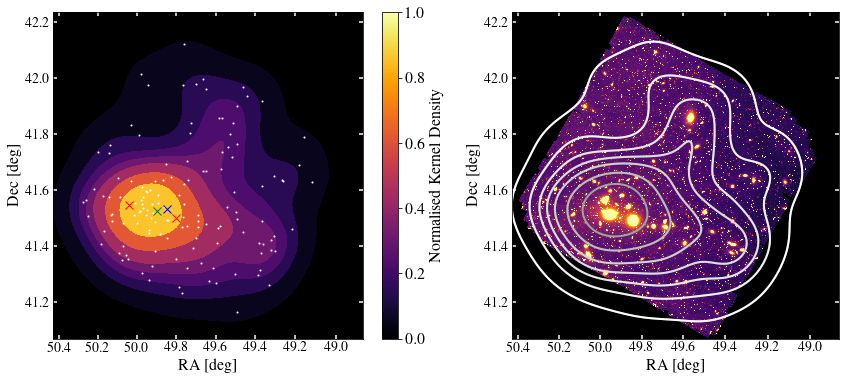}
\vspace{-8pt}
\caption{{\it Top left:} Projected distribution ({white dots}) and density map of the 1100 dwarf candidates within the \Euclid ERO FoV of the Perseus cluster. {\it Top right:} The density map ({white contours}) of the dwarf candidates are overlaid on the \Euclid ERO $\IE$ image of the cluster.  {\it Middle left and right:} Same as above but for the ultra diffuse galaxies identified as cluster members. {\it Bottom left and right:} Same as above but for the bright (non-dwarf) galaxies identified as cluster members. The iso-density centres of the dwarfs, as well as the iso-density centre of the GCs and the isophotal centre of the ICL at semimajor-axis radius 320\,kpc, are indicated with the {two red}, {green}, and {blue} crosses, respectively.}
\label{fig:densitymap}
\end{center}
\end{figure*}

\subsubsection{GC richness from automatic detection}
\label{section:GC_auto_detection}

Using the photometric catalogue of \citet{EROPerseusICL} we conducted a comprehensive analysis of detected sources to identify and characterise GC candidates associated with each galaxy in our footprint of the Perseus cluster (Table\,\ref{appendix:struct-param1}). To estimate the total GC population, we made several assumptions and adopted the procedures already used in the literature \citep[e.g.][]{cantiello2020} and outlined below.

At the distance of Perseus, the typical size of a GC, assuming a half-light radius of 3 pc based on Milky Way GCs \citep{Harris1996}, is about one order of magnitude smaller than the typical FWHM of the data in question. Therefore, these sources can be considered point-like in the ERO Perseus dataset.

Initially, we extracted galaxy characteristics (position, effective radius $R_{\rm e}$, etc.; Table\,\ref{appendix:struct-param1}) to identify all sources in the photometric catalogue within 1.5\,$R_{\rm e}$ of each galaxy \citep{Lim2018}. We adopted 3.7\,$R_{\rm e}$ for bright galaxies \citep{Forbes2017}. Additionally, sources within 3 background annuli with inner and outer radii equal to 5 and 10\,$R_{\rm e}$, 10 and 13.25\,$R_{\rm e}$, and 13.25 and 15.9 $R_{\rm e}$, were identified around each galaxy. We then estimated three over-density values of sources on-galaxy (one value for each background annulus) by subtracting the off-galaxy density from the on-galaxy density. The number of extra sources on-galaxy was then obtained by multiplying the over-density by the effective area of the galaxy. We chose not to apply a further factor of 2 due to spatial coverage, based on a series of empirical tests derived from estimates of $N_{\rm GC}$ within 4 $R_{\rm e}$. By comparing the total GC population estimated within 1.5 and 4.0 $R_{\rm e}$, we find that not doubling the population within 1.5 $R_{\rm e}$ provides a better match with the results within 4 $R_{\rm e}$ (which are spatially more complete). We adopted as reference the central value of the three estimates derived using all background annuli. The numbers and their uncertainties are reported in Table\,\ref{appendix:struct-param1}. Using the same sample of selected GC candidates on-galaxy, we also counted the number of objects within $R_{\rm e}/3$ of the galaxy photocentre, which are potentially nuclear star clusters. The number of these candidates, $N_{\rm NSC}$, is also reported in Table\,\ref{appendix:struct-param1}. 

Subsequently, we refined the analysis to consider only sources that satisfied certain properties for compactness, such as concentration index, FWHM, and elongation. To further narrow down the GC candidate list, we restricted the analysis to compact sources within the expected turnover magnitude (TOM), $\IE \leq \IE^{\rm TOM}$, at the distance of Perseus. For $\IE^{\rm TOM}$ we assumed the distance modulus $m{-}M=34.3$, and derived an absolute TOM of $M(\IE)^{\rm TOM}\approx -8.0$ \citep[see also][]{EROFornaxGCs}. This corresponds to assuming $M_V^{\rm TOM}=-7.5$ \citep[][and references therein]{Rejkuba2012} adjusted by a 0.5 offset from the $V$ band to the $\IE$ band. Consequently, the expected TOM in $\IE$ is approximately 26.3. For the bright cut-off for GC candidates, we considered using a magnitude of 22.7, estimated assuming a 3\,$\sigma_{{\rm GCLF}}$ cut brighter than the expected TOM, adopting a GC luminosity function (GCLF) width of $\sigma_{\rm GCLF}= 1.2$. Although $\sigma_{\rm GCLF} = 1.0$ or smaller is more in line with expectations for dwarfs \citep{Villegas10}, we opted for a larger value due to the (fractionally small) depth of the cluster and the wide range of galaxy magnitudes we consider.

Finally, adopting the procedure described above for assessing the excess of sources on galaxies, and assuming that the GC candidates sample is basically complete at the peak of the GCLF \citep{EROPerseusICL}, we double the number of selected sources for an estimate of the total population of GC on the galaxy, $N_{\rm GC}$, under the assumption of Gaussian GCLF. The entire procedure adopted is depicted in Fig.\,\ref{fig:GCautodetect}, for the dwarf galaxy EDwC-0791.

As a preliminary test, we compared the GC richness results obtained through visual inspection with the number of GC candidates, $N_{\rm GC}$, extracted using the procedure described above. Figure~\ref{fig:ngc_ie} shows $N_{\rm GC}$ versus $\IE$, colour-coded based on the GC richness score obtained from our initial visual inspection (as labelled). The upper panel shows the full sample on a logarithmic scale, while the bottom panel displays only the sample of dwarf galaxies over a limited range of $N_{\rm GC}$, adopting a linear scale.

An initial notable finding is that, except for one dwarf (EDwC-0055), all 37 galaxies with a visual GC richness score of 100\% are estimated to have a population of $N_{\rm GC}>$~0. The case of EDwC-0055 -- shown in the upper left panel of Fig.\,\ref{fig:multinuclei} -- is intriguing: despite not being in proximity to any apparent source of GC contamination, it exhibits a clear concentration of point-like sources within $R_{\rm e}/2$ and visually appears GC-rich according to our adopted definition. We made several tests on the estimated $N_{\rm GC}$ value for this galaxy and found it to be robust against changes in the background regions used. A key factor influencing the outcome might be the presence of three point-like sources close in projection in the galaxy core, which might impact the algorithms for compactness selection. Furthermore, adopting a fainter magnitude limit for GC selection, the $N_{\rm GC}$ for this galaxy indeed makes it GC-rich. This may suggest that the galaxy lies in the background of the cluster or that its GC population is skewed toward fainter magnitudes due to sample size effects, or a combination of both.

Expanding the comparison to objects with a visual GC score of $\geq$\,0.85 (equivalent to six out of seven total votes in favour of GC richness), out of 70 dwarfs, four exhibit $N_{\rm GC}=1$, while the rest have $N_{\rm GC}>$\,2, thus effectively meeting our visual definition of GC richness. A final note regarding the comparison between visual and automated richness assessments is that, in all cases with a score of $\geq$\,0.85, the two estimates are consistent within the $N_{\rm GC}$ uncertainties reported in Table\,\ref{appendix:struct-param1}, which were estimated from the Poisson scatter of the three background regions.

In Fig.\,\ref{fig:ngc_gcrich}, we present the same structural, photometric and positional properties of the galaxy sample as shown in Fig.\,\ref{fig:visual_gcrich}, in this case using the estimated $N_{\rm GC}$ values rather than visual classification. To identify GC-rich dwarfs quantitatively, we first determined the median population of GCs for dwarf galaxies with $N_{\rm GC} > 1$, and found $\langle N_{\rm GC} \rangle = 4$. Then, we further constrained the GC-richness by measuring its robustness against our estimated uncertainty. Specifically, we required $\Delta N_{\rm GC} / N_{\rm GC} \leq 0.5$, i.e. imposing that the fractional error on $N_{\rm GC}$ must be less than or equal to 50\%. In the figure we adopt for bright, dwarf and GC-rich (contaminated and uncontaminated) galaxies the same symbol and colour coding as Fig.\,\ref{fig:visual_gcrich}. Furthermore, the symbol sizes of bright (gray dots in the upper left panel) and GC-rich dwarfs (orange circles in all panels) scale with the total population of GC candidates, adopting different scaling factors for bright and dwarf galaxies for clarity. The overall consistency of the results shown in Fig.\,\ref{fig:visual_gcrich} and Fig.\,\ref{fig:ngc_gcrich} provides evidence of the validity of the visual inspection approach adopted, with the additional evidence that brighter galaxies tend to host a larger GC population.

In the upper panel of Fig.\,\ref{fig:ngc_ie}, we also present the $N_{\rm GC}$ estimates for the bright galaxies in the field. The approach adopted here, as described above, is probably not optimal for these bright galaxies. This is because the background correction to $N_{\rm GC}$ in the more contaminated inner cluster regions, when using our criteria, might not be optimal and would need to be fine-tuned for each single bright/extended galaxy. For the purposes of this work, however, we maintain the current procedure. 

Despite this limitation, the upper panel of Fig.\,\ref{fig:ngc_ie} shows the expected correlation between $N_{\rm GC}$ and $\IE$, i.e., between the (mass of the) total GC population and the galaxy mass \citep{Spitler2009,Burkert2020}, spanning approximately 10 magnitudes and about three orders of magnitude in $N_{\rm GC}$, including the bright and massive galaxies in our sample. In this panel, we plot the linear fit equation to $\logten N_{\rm GC}$ versus $\IE$, adopting different $\IE$ magnitude intervals (see figure caption) and including all galaxies with $N_{\rm GC}>4$. Although the scatter is notably large -- especially considering that both quantities are on a log scale -- the known correlation is well identified. A linear fit to these data for the range of $10\leq \IE\leq21$ appears to align with the entire data set of magnitudes and $\logten N_{\rm GC}$ values (Pearson correlation coefficient $r=-0.8$). However, we find that the slope of the $N_{\rm GC}$-magnitude relation changes for galaxies fainter than $\IE=17$. Therefore, our results seem to support similar analyses from the literature about the flattening of the $N_{\rm GC}$-magnitude relation for faint galaxies \citep{Harris2013, Burkert2020, Jones2023}. Further refinements are in progress, with a more specific selection of GCs around bright galaxies and possibly involving a colour selection in the GC identification procedure using the NISP data. 

The left panel of Fig.\,\ref{fig:sn_positions} shows the positions of the dwarfs in the field (violet-blue circles), with symbol size scaled to the GCs specific frequency ($S_N=N_{\rm GC}\times 10^{0.4(M_V+15)}$). In the panel, we also show the positions of galaxies in the five brightest magnitude ranges ($8\leq \IE \leq 13$), indicated by gray circles. While we find that galaxies with the largest $N_{\rm GC}$ are scattered around the field, inspecting $S_{{N}}$ it appears that galaxies located around the east-west strip, where most of the brightest cluster members are found, tend to exhibit larger $S_{{N}}$ values on average, especially south of the BCG region.

In the middle panel of Fig.\,\ref{fig:sn_positions}, we plot the $N_{\rm GC}$ values versus the cluster-centric radius, $R_{\rm Cl}$, measured with respect to the BCG. In general, there seems to be no compelling evidence of a correlation between $N_{\rm GC}$ or $S_{{N}}$ and $R_{\rm Cl}$.

As a final test for the GC populations in dwarf and bright galaxies in the field, we compared the $S_{{N}}$ versus $V$ band magnitude relation for our entire sample with the mean relations derived for UDGs in the Coma cluster \citep{Lim2018} and in Virgo \citep{Lim2020}, all shown in the right panel of Fig.\,\ref{fig:sn_positions}. Ignoring the elements with $N_{\rm GC}<4$, where the population is typically consistent with zero within uncertainties, our $S_{{N}}$ estimates for dwarfs appear to range between the Virgo and Coma mean relations. Previous studies have shown that low-mass galaxies in denser environments can have higher $S_{\rm N}$ 
\citep{Peng2008,Mistani2016,Lim2018}. Therefore, our results possibly indicate that Perseus is dynamically intermediate: not evolved and relaxed like Coma, not as active and still in evolution as Virgo. For the bright galaxy component, i.e., all galaxies with $\IE<16.3$, we estimate a median $\langle S_N \rangle=0.9$, which appears slightly smaller than the average for galaxies of this magnitude \citep{Harris2017}. However, the scatter is quite large, and, as explained above, the procedure adopted here to estimate $N_{\rm GC}$ for bright targets will probably require further refinements, which will be subject of a future dedicated paper.

As a final note, we encourage the reader to approach the GC selection presented in this section with due consideration. It is based on morphometric and photometric selection from single-band data, along with statistical background decontamination. However, potential residual contamination on the galaxy or in the background estimation regions, coupled with spatially changing background (particularly relevant for brighter galaxies and around bright companions), represents a significant limitation that may affect our estimates of $N_{\rm GC}$. To enhance our analysis, future observations with spatial resolution and depth similar to VIS, especially in the near-IR (e.g., from Roman or JWST), are required. These observations will provide the necessary colour constraints for a more robust cleaning of the sample from interlopers.

\subsection{2D spatial distribution} 

In Fig.\,\ref{fig:densitymap} we show the projected distribution and the density map and contours of the Perseus dwarf galaxies and UDGs across the ERO FoV. The projected density of the bright galaxy sample from \citet{EROPerseusOverview} is also shown in the same figure. 

The dwarf distribution has two main iso-density centres located on each side of the bright galaxy distribution. The main iso-density centre is located approximately at RA\,$=$\,$3^{\rm h}\,19^{\rm m}\,12^{\rm s}.0$ and Dec\,$=\,\ang{41;30;0.0}$ and the second one, located to the left of the main iso-density centre, is centred approximately at RA\,$=$\,$3^{\rm h}\,20^{\rm m}\,8^{\rm s}.4$, Dec\,$=\,\ang{41;32;42.0}$. The double iso-density centre could be the result of recent merger activity reported for this cluster. It is worth noting that the bright galaxies were not removed from the image when the data were visually inspected for the identification of dwarfs in the field via the annotation tool. Therefore, we may be missing a handful of dwarfs that are within the halo of the massive (bright) galaxies.

Figure\,\ref{fig:densitymap} also shows the iso-density centre of the GCs and the isophotal centre of the ICL at semimajor-axis radius 320\,kpc measured by \citet{EROPerseusICL}. The ICL isophotal centre is measured at ($3^{\rm h}\,19^{\rm m}\,22^{\rm s}.253$,\,\ang{41;31;58.902}) and the GC iso-density centre at ($3^{\rm h}\,19^{\rm m}\,35^{\rm s}.031$,\,\ang{41;31;22.108}). The ICL isophotal centre is located westward of the BCG core by 60\,kpc \citep{EROPerseusICL}. The comparison of the location of the dwarfs, GCs, and ICL in the ERO field indicates that all three distributions appear to have a main centre displaced to the west of the galaxy light distribution, with the dwarf main centre being the most displaced (about $110$\,kpc) from the BCG. This agreement likely indicates that these shifts are real and not driven by some bias introduced in the data reduction and analysis. The brightest X-ray emission is centred on the BCG and only on larger scales is offset to the east, i.e., opposite to the ICL \citep{EROPerseusICL}. 

The 2D projected distribution of UDGs appears strongly asymmetric with respect to the cluster centre, showing a large offset to the west of the iso-density centres of the dwarf distribution. An asymmetry of the UDGs with respect to the cluster centre has also been reported for the Hydra I cluster \citep{LaMarca2022}. For that cluster, roughly half of the UDGs are concentrated close to the cluster core, in the same direction as the GCs and ICL centres, and around a subgroup of galaxies to the north, while the other half are found uniformly distributed at larger clustercentric distances. We also note a similar split in the distribution for the UDGs in Perseus, with approximately half concentrated in the west over-density and the remaining half found more uniformly distributed at larger clustercentric distances. There is a hint that the spatial distribution of the UDGs may be associated with over-densities of bright galaxies in the cluster. This would be consistent with the ﬁndings for other galaxy clusters where over-densities of UDGs are found close to subgroups of galaxies \citep{Janssens2019}. If the UDGs are associated with the grouping of bright galaxies, this may suggest that the UDGs joined the cluster via accretion of subgroups. On the other hand, we cannot rule out that they could also be native to the cluster \citep{Sales2020}.

\section{\label{sc:Conclusion} Conclusions}

As part of the \Euclid ERO series of papers, we have used \Euclid imaging of the Perseus cluster to demonstrate the capabilities of the telescope for dwarf galaxy science. The depth, spatial resolution, and FoV have allowed us to detect and characterize 1100 dwarf galaxy candidates in a 0.7\,deg$^2$ field, slightly offset from the BCG, NGC\,1275. We visually classified their morphologies, extracted their photometric and structural properties, quantified their GC populations, and mapped their spatial distribution across the cluster. Some of our key findings are highlighted below:

\begin{itemize}

\item With 1100 dwarf candidates, this catalogue more than doubles the number of dwarfs associated with the Perseus cluster in the literature. The detections appear to be robust, and the dwarfs follow known scaling relations in the literature, although the diffuse end of the dwarf parameter relations are more densely populated than in previous studies. Morphologically, the sample can be separated into:  96\% [4\%] dEs [dIs]; 53\% [47\%] nucleated [non-nucleated];  74\% [26\%] GC-poor [GC-rich], and 94\% [6\%] not morphologically disturbed [morphologically disturbed]. Furthermore, 8\% of the dwarfs can be classified as UDGs. The UCDs were not considered during the visual examination of the images due to their small sizes and relatively high surface brightnesses.
\\
\item The nucleated dwarfs in the Perseus cluster follow the same trends found in the literature, with nuclei typically found in brighter, larger, and rounder dwarfs with flatter surface brightness profiles. In this work, however, we also illustrate the potential for \Euclid to identify complex nuclear structures; it appears that several nuclei are associated with tidal features, suggesting that we are witnessing formation or transformation events. With the large number of dwarfs expected from the EWS, we may be able to place observational constraints on the frequency of these events.
\\
\item The combination of the surface brightness sensitivity, spatial resolution and wide field of \Euclid allowed us to detect and characterise the GC systems of the dwarf galaxies. We automated the GC detection, and the resulting correlation between $N_{\rm GC}$ and \IE, including the flattening below $\IE \sim 17$, agrees well with the literature. In terms of distribution, the dwarfs with the largest GC counts can be found scattered throughout the cluster, but those with the largest $S_N$ values are typically concentrated south of the BCG, potentially another indication of a recent merger. Our $S_{N}$ estimates for the Perseus dwarfs appear to range between the mean relations of those in the Virgo and Coma clusters, suggesting that Perseus is dynamically intermediate, neither as active as the Virgo cluster nor as relaxed as the Perseus cluster.
\\
\item We compared the 2D projected distribution of the dwarfs in the cluster with those of both the GCs and ICL distributions \citep{EROPerseusICL}. The dwarf distribution is characterized by a double iso-density centre, perhaps due to a recent merger event. The main iso-density centre of this distribution is shifted to the west of the BCG, in agreement with the iso-density centre of the GCs and the isophotal centre of the ICL at semimajor-axis radius 320\,kpc. The UDGs show a more asymmetric distribution than the dwarf population as a whole; roughly half of the UDGs are located westward of the cluster centre, in the same direction as the GC iso-density centre and the ICL isophotal centre, while the other half can be found more uniformly distributed at larger clustercentric distances. There is a hint that the spatial distribution of the UDGs may be associated with over-densities of bright galaxies in the cluster, which could indicate their recent accretion via groups. However, we cannot rule out that they may also be native to the cluster.

\end{itemize}

The ERO VIS images of the Perseus cluster have demonstrated the capability of \Euclid to not only detect new LSB galaxies, including those close to bright stars, but more importantly, to identify them as genuine dwarf galaxies. The pristine PSF and high spatial resolution of \Euclid provide the ability to distinguish dwarfs from other sources, such as background galaxies, without the need of follow-up observations. Furthermore, we showed that \Euclid offers the added benefit of being able to also provide a census of the GC systems and nuclei associated with the dwarfs. In particular, looking for high concentrations of GCs in the \Euclid surveys shows great potential for the detection of GC-rich dwarfs and UDGs. 

The upcoming EWS will have a sky coverage $\sim$\,14\,000\,deg$^{2}$. Although it is expected to be about 0.75\,mag arcsec$^{-2}$ less deep than the data presented in this work, it will transform our ability to characterise dwarf galaxies across a range of stellar masses and located in a variety of environments. The analysis of such a large data set will require the development of automatic detection methods, which are currently underway. The dwarf catalogue presented here should prove to be a useful benchmark for testing present and future machine learning techniques.

\begin{acknowledgements}
\AckERO \AckEC

This work has made use of the Early Release Observation (ERO) data from the Euclid mission of the European Space Agency (ESA), 2024, https://doi.org/10.57780/esa-qmocze3. O.\,Marchal would like to acknowledge Thomas Boch and Matthieu Baumann at the Observatoire de Strasbourg. M.\,Poulain and A.\,Venhola are both supported by the Academy of Finland grant number 347089. A.\,Ferr\'e-Mateu acknowledges support from RYC2021-031099-I and PID2021-123313NA-I00 of MICIN/AEI/10.13039/501100011033/FEDER,UE,NextGenerationEU/PRT. L.\,Hunt, R.\,Scaramella, M.\,Cantiello, and R.\,Habas acknowledge funding from the Italian INAF Large Grant 12-2022.

\end{acknowledgements}

\bibliography{paper, Euclid, EROplus}

\begin{appendix}
\onecolumn

\section{Properties of the dwarf galaxy candidates}
\label{AppendixA}

The dwarf galaxy candidates are classified by visual inspection with the methodology described in Sect.\,\ref{sc:Methods}. The final sample contains 1100 dwarf galaxies of which 96\% are classified as dE, and 4\% as dI. The dE galaxies are further classified as nucleated (53\%), GC-rich (26\%), and with disturbed morphology (6\%). \\\\
Table\,\ref{appendix:final-dwarf-catalogue} lists the entire sample of dwarf galaxy candidates, ordered by increasing RA. The columns are as follows: a unique identifier (ID), RA in degrees, Dec in degrees, morphology (either dE or dI), GC-rich flag (recall that in the visual classification we defined dwarfs with $N_{\rm GC} > 2$\,GCs as GC-rich), a flag for the presence of a nucleus (Nucleated), and a disturbed morphology flag (Disturbed). Both dE and dI galaxies could be classified as disturbed. \\\\
Table\,\ref{appendix:struct-param1} lists the photometric and structural parameters of the dwarf galaxy candidates (ordered by increasing RA) obtained with the methodology described in Sect.\,\ref{sc:Phot-Str}. The column definitions are as follows: a unique identifier (ID), RA in degrees, Dec in degrees, apparent magnitude in the \IE\ filter (\IE), effective radius in arcsec ($R_{\rm e}$), S\'ersic index (n), axis ratio (AR), position angle in radians (PA), central surface brightness ($\mu_{\IE, \rm 0}$, mag~arcsec$^{-2}$), surface brightness at $R_{\rm e}$ ($\mu_{\IE, \rm e}$, mag~arcsec$^{-2}$), surface brightness within $R_{\rm e}$ ($\langle \mu_{\IE,\rm e} \rangle$, mag~arcsec$^{-2}$), the number of associated globular clusters ($N_{\rm GC}$), and the number of nuclear star clusters ($N_{\rm NSC}$).\\\\
Table\,\ref{appendix:struct-param2} lists the aperture magnitudes and extinction corrections (EC) described in Sect.\,\ref{sc:colours} and \ref{sec:extcorr}, respectively. The columns are: a unique identifier (ID), the apparent magnitude in the \IE\ filter, the extinction correction in \IE, the apparent magnitude in the \YE\ filter, the extinction correction in \YE, apparent magnitude in the \JE\ filter, the extinction correction in \JE, the apparent magnitude in the \HE\ filter, and the extinction in \HE. The dwarf galaxy candidates are ordered by increasing RA. \\\\
Finally, images of the 1100 dwarf galaxy candidates in the final sample of this work are shown in Fig.\,\ref{fig:cutouts-all-dwarfs}. The cutouts were created from the \IE image. For better visibility, the dimensions of each cutout have been scaled to twice the area of the original annotations and an arcsinh stretch has been applied. The dwarf candidates are ordered by decreasing surface brightness $\langle \mu_{\IE,\rm e} \rangle$. \\\\

\begin{table*}[ht!]
\caption{Table of the visual properties of the dwarfs galaxy candidates. The full table will be available as supplementary material at the Strasbourg astronomical Data Center (CDS; \url{https://www.aanda.org/for-authors/latex-issues/tables}).}
\label{appendix:final-dwarf-catalogue}

\centering
\begin{tabular}{lllcccc}
\hline
\hline
\noalign{\smallskip}
\omit\hfil ID \hfil & \omit\hfil RA \hfil & \omit\hfil Dec \hfil & \omit\hfil Morphology \hfil & \omit\hfil \phantom{00}GC-rich\phantom{00} \hfil & \omit\hfil Nucleated \hfil & \omit\hfil \phantom{00}Disturbed \hfil \\
   & \omit\hfil [deg] \hfil & \omit\hfil [deg] \hfil & \omit\hfil & \omit\hfil & \omit\hfil & \omit\hfil \\
\hline
\noalign{\smallskip}
EDwC-0001\tablefootmark{*} & 48.975450 & 41.692800 & dE & Yes & Yes & Yes \\
EDwC-0002 & 49.021117 & 41.783423 & dE & No & No & No \\
EDwC-0003 & 49.033821 & 41.683320 & dE & Yes & Yes & No \\
EDwC-0004 & 49.047517 & 41.621950 & dE & No & Yes & No \\
EDwC-0005 & 49.064550 & 41.624950 & dE & No & No & Yes \\
\vdots & \vdots & \vdots & \vdots & \vdots & \vdots & \vdots \\
EDwC-1095 & 50.325621 & 41.517185 & dE & No & No & No \\
EDwC-1096 & 50.329711 & 41.521662 & dE & No & Yes & No \\
EDwC-1097 & 50.335171 & 41.549552 & dE & Yes & Yes & No \\
EDwC-1098 & 50.344780 & 41.503080 & dE & No & Yes & No \\
EDwC-1099 & 50.349776 & 41.570365 & dE & No & No & No \\
EDwC-1100\tablefootmark{*} & 50.360450 & 41.540150 & dE & No & No & No \\
\hline 
\end{tabular}
\tablefoot{
\tablefoottext{*}{Denotes dwarf candidates that fall just outside of the \Euclid footprint, and that were identified in an earlier data product with a slightly larger FoV. They are included here for completeness.}
}
\end{table*}

\newpage

\begin{sidewaystable}[htbp!]

\small
\centering
\newcommand{\pd}{\phantom{1}}
\setlength{\tabcolsep}{3.25pt}
\caption{Photometric and structural properties of the dwarf galaxy candidates. The full table will be available as supplementary material at the Strasbourg astronomical Data Center (CDS) (\url{https://www.aanda.org/for-authors/latex-issues/tables}).}
\label{appendix:struct-param1}
\smallskip
\label{table:1}
\smallskip

\resizebox{\textwidth}{!}{%
\begin{tabular}{lllrrrrrrrrrr}
\hline
\hline

&&&&&&&&&&&&\\[-7pt]
\omit\hfil ID \hfil & \omit\hfil RA \hfil & \omit\hfil Dec \hfil & \omit\hfil \IE\ \hfil & \omit\hfil $R_{\rm e}$ \hfil & \omit\hfil $n$ \hfil & \omit\hfil AR \hfil & \omit\hfil PA \hfil & \omit\hfil $\mu_{\IE,\rm{0}}$ \hfil & \omit\hfil $\mu_{\IE, \rm {e}}$ \hfil & \omit\hfil $\langle \mu_{\IE, \rm {e}} \rangle$ \hfil & \omit\hfil $N_{\rm GC}$ \hfil & \omit\hfil $N_{\rm NSC}$ \hfil\\
\omit\hfil  & \omit\hfil [deg] \hfil & \omit\hfil [deg] \hfil & \omit\hfil [mag] \hfil & \omit\hfil [arcsec] \hfil & \omit\hfil & \omit\hfil & \omit\hfil [rad] \hfil & \omit\hfil [mag arcsec$^{-2}$] \hfil & \omit\hfil [mag arcsec$^{-2}$] \hfil & \omit\hfil [mag arcsec$^{-2}$] \hfil & \omit\hfil  & \omit\hfil \\
&&&&&&&&&&&&\\[-8pt]
\hline
&&&&&&&&&&&&\\[-8pt]
EDwC-0001\tablefootmark{*} & 48.975450 & 41.692800 & 15.64 $\pm$ 0.31 & 15.00 $\pm$ 1.72 & 3.82 $\pm$ 0.17 & 0.89 $\pm$ 0.01 & 1.95 $\pm$ 0.04 & 16.83 $\pm$ 0.01 & 24.76 $\pm$ 0.24 & 19.16 $\pm$ 0.00 & 6.0 $\pm$ 17.0 & 1 \\
EDwC-0002 & 49.021117 & 41.783423 & 22.82 $\pm$ 0.26 & 1.25 $\pm$ 0.07 & 0.50 $\pm$ 0.21 & 0.69 $\pm$ 0.06 & 2.61 $\pm$ 0.12 & 24.54 $\pm$ 0.61 & 25.29 $\pm$ 0.62 & 24.93 $\pm$ 0.36 & 0.0 $\pm$ 0.0 & 0 \\
EDwC-0003 & 49.033821 & 41.683320 & 17.99 $\pm$ 0.00 & 7.57 $\pm$ 0.03 & 0.35 $\pm$ 0.00 & 0.50 $\pm$ 0.00 & 2.76 $\pm$ 0.00 & 23.49 $\pm$ 0.00 & 23.93 $\pm$ 0.01 & 22.15 $\pm$ 0.00 & 5.0 $\pm$ 5.0 & 3 \\
EDwC-0004 & 49.047517 & 41.621950 & 19.08 $\pm$ 0.28 & 3.50 $\pm$ 0.34 & 1.47 $\pm$ 0.17 & 0.41 $\pm$ 0.01 & 1.26 $\pm$ 0.01 & 20.87 $\pm$ 0.05 & 23.71 $\pm$ 0.13 & 21.90 $\pm$ 0.01 & 0.0 $\pm$ 2.0 & 1 \\
EDwC-0005 & 49.064550 & 41.624950 & 19.25 $\pm$ 0.02 & 4.24 $\pm$ 0.09 & 1.65 $\pm$ 0.03 & 0.46 $\pm$ 0.00 & 0.51 $\pm$ 0.01 & 21.25 $\pm$ 0.00 & 24.48 $\pm$ 0.26 & 22.21 $\pm$ 0.00 & 1.0 $\pm$ 3.0 & 1 \\
EDwC-0006 & 49.066602 & 41.735277 & 22.75 $\pm$ 0.25 & 1.70 $\pm$ 0.12 & 1.13 $\pm$ 0.23 & 0.54 $\pm$ 0.04 & 0.12 $\pm$ 0.06 & 23.87 $\pm$ 0.16 & 25.98 $\pm$ 0.04 & 24.92 $\pm$ 0.06 & $-$1.0 $\pm$ 0.0 & 0 \\
EDwC-0007 & 49.067834 & 41.780197 & 21.94 $\pm$ 0.03 & 2.68 $\pm$ 0.09 & 1.18 $\pm$ 0.04 & 0.45 $\pm$ 0.01 & 1.78 $\pm$ 0.01 & 23.77 $\pm$ 0.01 & 25.98 $\pm$ 0.14 & 24.59 $\pm$ 0.00 & 1.0 $\pm$ 2.0 & 0 \\
EDwC-0008 & 49.069858 & 41.731817 & 20.57 $\pm$ 0.07 & 5.88 $\pm$ 0.37 & 1.40 $\pm$ 0.07 & 0.82 $\pm$ 0.01 & 1.61 $\pm$ 0.06 & 24.37 $\pm$ 0.01 & 27.05 $\pm$ 0.01 & 23.98 $\pm$ 0.00 & $-$1.0 $\pm$ 4.0 & 2 \\
EDwC-0009 & 49.070800 & 41.719650 & 19.74 $\pm$ 0.00 & 3.15 $\pm$ 0.01 & 0.32 $\pm$ 0.00 & 0.32 $\pm$ 0.00 & 2.02 $\pm$ 0.00 & 22.91 $\pm$ 0.00 & 23.29 $\pm$ 0.06 & 22.97 $\pm$ 0.00 & 5.0 $\pm$ 3.0 & 0 \\
\vdots & \vdots & \vdots & \vdots & \vdots & \vdots & \vdots & \vdots & \vdots & \vdots & \vdots & \vdots & \vdots \\
EDwC-1095 & 50.325621 & 41.517185 & 22.06 $\pm$ 0.02 & 2.00 $\pm$ 0.06 & 0.68 $\pm$ 0.03 & 0.48 $\pm$ 0.01 & 0.97 $\pm$ 0.02 & 24.16 $\pm$ 0.02 & 25.30 $\pm$ 0.06 & 24.61 $\pm$ 0.02 & $-$1.0 $\pm$ 0.0 & 0 \\
EDwC-1096 & 50.329711 & 41.521662 & 21.62 $\pm$ 0.23 & 2.30 $\pm$ 0.17 & 1.71 $\pm$ 0.23 & 0.41 $\pm$ 0.02 & 3.04 $\pm$ 0.03 & 22.05 $\pm$ 0.06 & 25.40 $\pm$ 0.13 & 23.89 $\pm$ 0.01 & $-$1.0 $\pm$ 0.0 & 0 \\
EDwC-1097 & 50.335171 & 41.549552 & 20.38 $\pm$ 0.02 & 3.63 $\pm$ 0.08 & 1.04 $\pm$ 0.02 & 0.68 $\pm$ 0.01 & 1.48 $\pm$ 0.02 & 23.58 $\pm$ 0.01 & 25.48 $\pm$ 0.05 & 23.43 $\pm$ 0.00 & 6.0 $\pm$ 4.0 & 2 \\
EDwC-1098 & 50.344780 & 41.503080 & 17.24 $\pm$ 0.01 & 5.98 $\pm$ 0.04 & 2.01 $\pm$ 0.01 & 0.94 $\pm$ 0.00 & 1.26 $\pm$ 0.02 & 20.09 $\pm$ 0.00 & 24.10 $\pm$ 0.00 & 20.43 $\pm$ 0.02 & 6.0 $\pm$ 5.0 & 0 \\
EDwC-1099 & 50.349776 & 41.570365 & 21.11 $\pm$ 0.02 & 2.88 $\pm$ 0.08 & 0.51 $\pm$ 0.03 & 0.71 $\pm$ 0.01 & 1.60 $\pm$ 0.05 & 24.67 $\pm$ 0.02 & 25.45 $\pm$ 0.06 & 24.13 $\pm$ 0.00 & 1.0 $\pm$ 2.0 & 0 \\
EDwC-1100\tablefootmark{*} & 50.360450 & 41.540150 & 19.24 $\pm$ 0.03 & 5.92 $\pm$ 0.05 & 0.36 $\pm$ 0.01 & 0.50 $\pm$ 0.01 & 2.06 $\pm$ 0.01 & 24.17 $\pm$ 0.01 & 24.63 $\pm$ 0.03 & 23.10 $\pm$ 0.00 & $-$2.0 $\pm$ 3.0 & 0 \\
\hline
\end{tabular}}
\tablefoot{
\tablefoottext{*}{Denotes dwarf candidates that fall just outside of the \Euclid footprint, and that were identified in an earlier data product with a slightly larger FoV. They are included here for completeness.}
}
\smallskip
\end{sidewaystable}

\begin{table}[hp!]

\centering
\newcommand{\pd}{\phantom{1}}
\setlength{\tabcolsep}{3.25pt}
\caption{Aperture magnitudes and extinction correction (EC) in the \Euclid \YE, \JE, and \HE\ bands for each dwarf galaxy candidate. The full table will be available as supplementary material at the Strasbourg astronomical Data Center (CDS) (\url{https://www.aanda.org/for-authors/latex-issues/tables}).}
\label{appendix:struct-param2}
\smallskip
\label{table:2}
\smallskip
\resizebox{\textwidth}{!}{%
\begin{tabular}{lrrrrrrrr}
\hline
\hline

&&&&&&&&\\[-7pt]
\omit\hfil ID \hfil & \omit\hfil \IE \hfil & \omit\hfil EC (\IE) \hfil & \omit\hfil \YE\ \hfil & \omit\hfil EC (\YE) \hfil & \omit\hfil \JE\ \hfil & \omit\hfil EC (\JE) \hfil & \omit\hfil \HE\ \hfil & \omit\hfil EC (\HE) \hfil\\
&&&&&&&&\\[-8pt]
\hline
&&&&&&&&\\[-8pt]
EDwC-0001\tablefootmark{*} & 16.63 $\pm$ 0.00 & 0.26 $\pm$ 0.01 & \omit\hfil \dots \hfil & 0.13 $\pm$ 0.01 & \omit\hfil \dots \hfil & 0.09 $\pm$ 0.00 & \omit\hfil \dots \hfil & 0.06 $\pm$ 0.00 \\
EDwC-0002 & 23.72 $\pm$ 0.06 & 0.29 $\pm$ 0.01 & 23.17 $\pm$ 0.05 & 0.15 $\pm$ 0.01 & 23.33 $\pm$ 0.05 & 0.10 $\pm$ 0.00 & 23.50 $\pm$ 0.05 & 0.07 $\pm$ 0.00 \\
EDwC-0003 & 16.80 $\pm$ 0.00 & 0.26 $\pm$ 0.01 & 16.41 $\pm$ 0.00 & 0.13 $\pm$ 0.01 & 16.42 $\pm$ 0.00 & 0.09 $\pm$ 0.00 & 16.48 $\pm$ 0.00 & 0.06 $\pm$ 0.00 \\
EDwC-0004 & 19.54 $\pm$ 0.01 & 0.26 $\pm$ 0.02 & 18.85 $\pm$ 0.01 & 0.13 $\pm$ 0.01 & 18.92 $\pm$ 0.01 & 0.09 $\pm$ 0.01 & 18.47 $\pm$ 0.01 & 0.06 $\pm$ 0.00 \\
EDwC-0005 & 19.83 $\pm$ 0.01 & 0.26 $\pm$ 0.02 & 19.32 $\pm$ 0.01 & 0.13 $\pm$ 0.01 & 19.34 $\pm$ 0.01 & 0.09 $\pm$ 0.01 & 19.23 $\pm$ 0.01 & 0.06 $\pm$ 0.00 \\
EDwC-0006 & 23.63 $\pm$ 0.05 & 0.28 $\pm$ 0.01 & 23.57 $\pm$ 0.06 & 0.14 $\pm$ 0.01 & 23.28 $\pm$ 0.05 & 0.10 $\pm$ 0.00 & 23.23 $\pm$ 0.05 & 0.06 $\pm$ 0.00 \\
EDwC-0007 & 22.52 $\pm$ 0.03 & 0.30 $\pm$ 0.01 & 22.08 $\pm$ 0.03 & 0.15 $\pm$ 0.01 & 22.07 $\pm$ 0.03 & 0.10 $\pm$ 0.00 & 21.97 $\pm$ 0.03 & 0.07 $\pm$ 0.00 \\
EDwC-0008 & 20.86 $\pm$ 0.02 & 0.28 $\pm$ 0.01 & 20.05 $\pm$ 0.01 & 0.14 $\pm$ 0.01 & 19.83 $\pm$ 0.01 & 0.10 $\pm$ 0.00 & 19.56 $\pm$ 0.01 & 0.06 $\pm$ 0.00 \\
EDwC-0009 & 20.11 $\pm$ 0.01 & 0.28 $\pm$ 0.01 & 19.52 $\pm$ 0.01 & 0.14 $\pm$ 0.01 & 19.47 $\pm$ 0.01 & 0.09 $\pm$ 0.00 & 19.39 $\pm$ 0.01 & 0.06 $\pm$ 0.00 \\
\vdots & \vdots & \vdots & \vdots & \vdots & \vdots & \vdots & \vdots & \vdots \\
EDwC-1095 & 22.59 $\pm$ 0.03 & 0.33 $\pm$ 0.02 & 22.21 $\pm$ 0.03 & 0.17 $\pm$ 0.01 & 22.19 $\pm$ 0.03 & 0.11 $\pm$ 0.01 & 22.15 $\pm$ 0.03 & 0.07 $\pm$ 0.00 \\
EDwC-1096 & 22.26 $\pm$ 0.03 & 0.33 $\pm$ 0.02 & 21.95 $\pm$ 0.03 & 0.17 $\pm$ 0.01 & 21.85 $\pm$ 0.03 & 0.11 $\pm$ 0.01 & 21.82 $\pm$ 0.03 & 0.07 $\pm$ 0.00 \\
EDwC-1097 & 21.00 $\pm$ 0.02 & 0.32 $\pm$ 0.02 & 20.70 $\pm$ 0.01 & 0.16 $\pm$ 0.01 & 20.63 $\pm$ 0.01 & 0.11 $\pm$ 0.01 & 20.63 $\pm$ 0.01 & 0.07 $\pm$ 0.00 \\
EDwC-1098 & 18.00 $\pm$ 0.00 & 0.35 $\pm$ 0.02 & 17.42 $\pm$ 0.00 & 0.17 $\pm$ 0.01 & 17.31 $\pm$ 0.00 & 0.12 $\pm$ 0.01 & 17.22 $\pm$ 0.00 & 0.08 $\pm$ 0.00 \\
EDwC-1099 & 21.73 $\pm$ 0.02 & 0.32 $\pm$ 0.02 & 21.25 $\pm$ 0.02 & 0.16 $\pm$ 0.01 & 21.25 $\pm$ 0.02 & 0.11 $\pm$ 0.01 & 21.23 $\pm$ 0.02 & 0.07 $\pm$ 0.00 \\
EDwC-1100\tablefootmark{*} & 19.94 $\pm$ 0.01 & 0.34 $\pm$ 0.02 & 19.37 $\pm$ 0.01 & 0.17 $\pm$ 0.01 & 19.29 $\pm$ 0.01 & 0.11 $\pm$ 0.01 & 19.34 $\pm$ 0.01 & 0.07 $\pm$ 0.00 \\
\hline
\end{tabular}}
\tablefoot{
\tablefoottext{*}{Denotes dwarf candidates that fall just outside of the \Euclid footprint, and that were identified in an earlier data product with a slightly larger FoV. They are included here for completeness.}
}
\smallskip
\end{table}

\clearpage
\begin{figure*}[ht!]
\centerline{\includegraphics[width=\linewidth]{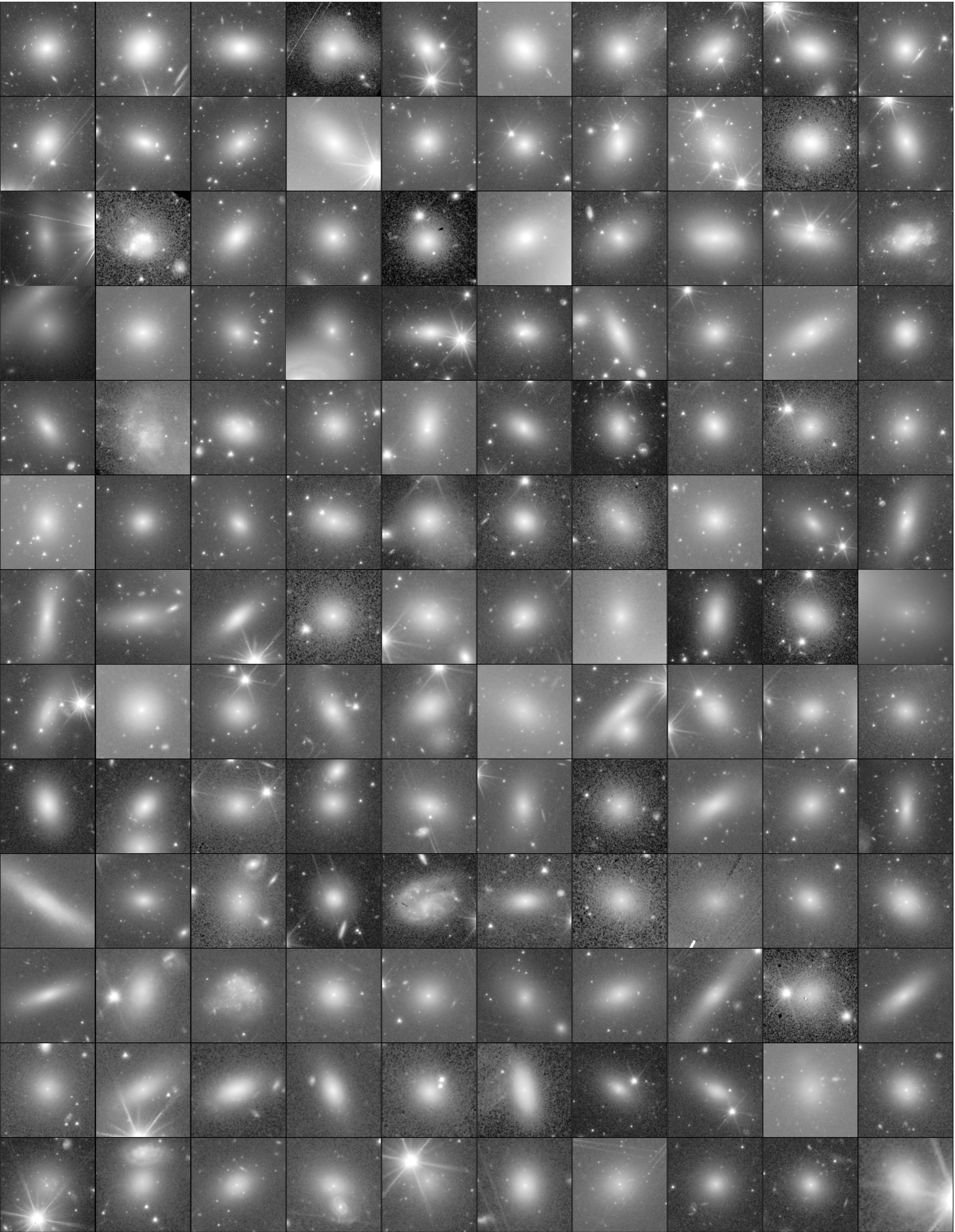}}
\caption{Cutouts of the $\IE$ image with arcsinh stretching of all our dwarf candidates, ordered by decreasing surface brightness $\langle \mu_{\IE,\rm e} \rangle$. The size of the cutouts are proportional to twice the area determined from the annotation of the classifiers, with north up and east to the left. The figure continues on the next pages.}
\label{fig:cutouts-all-dwarfs}
\end{figure*}

\begin{figure*}[ht!]\ContinuedFloat
\centerline{\includegraphics[width=\linewidth]{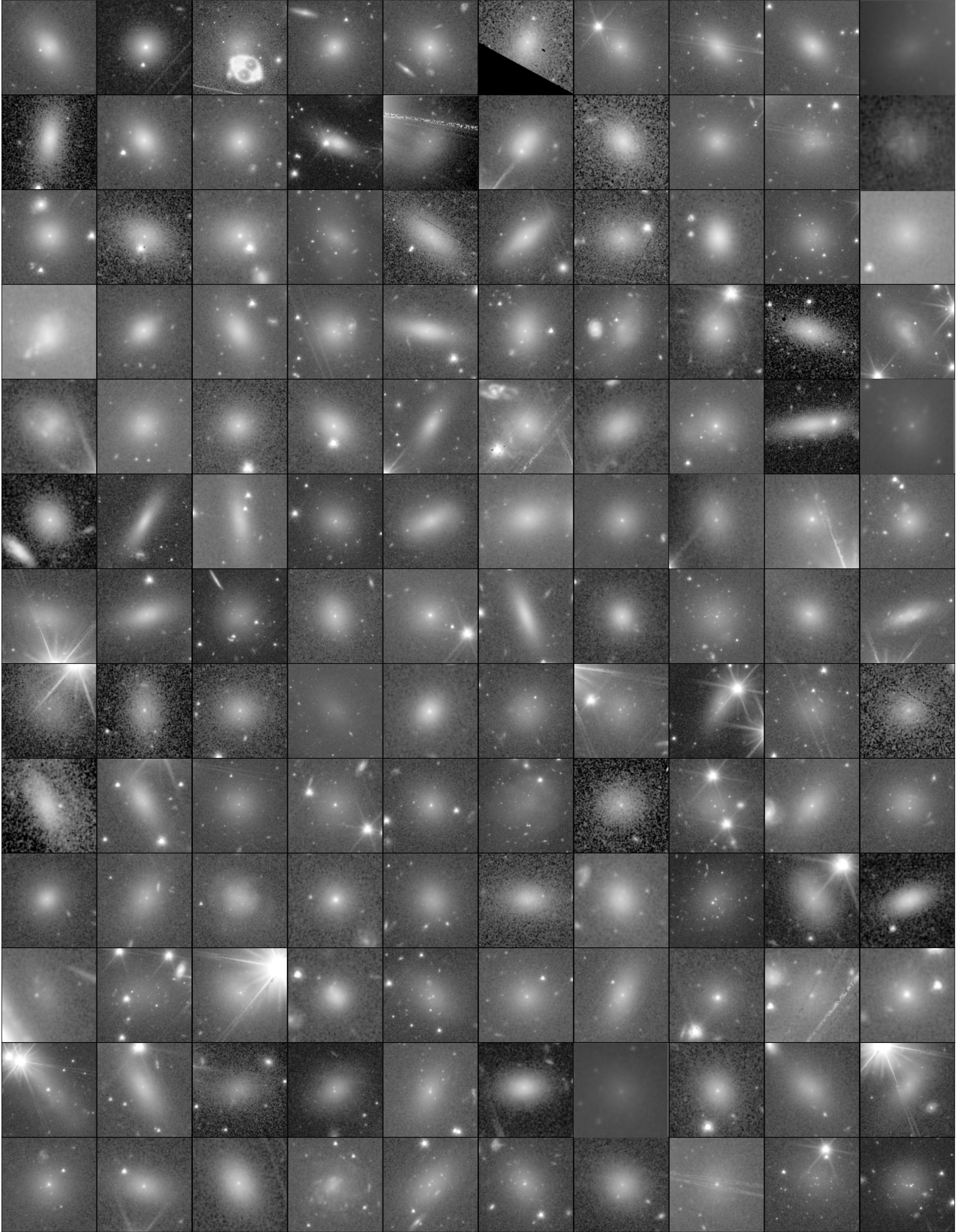}} 
\caption{Continued.}
\label{fig:cutouts-all-dwarfs2}
\end{figure*}

\begin{figure*}[ht!]\ContinuedFloat
\centerline{\includegraphics[width=\linewidth]{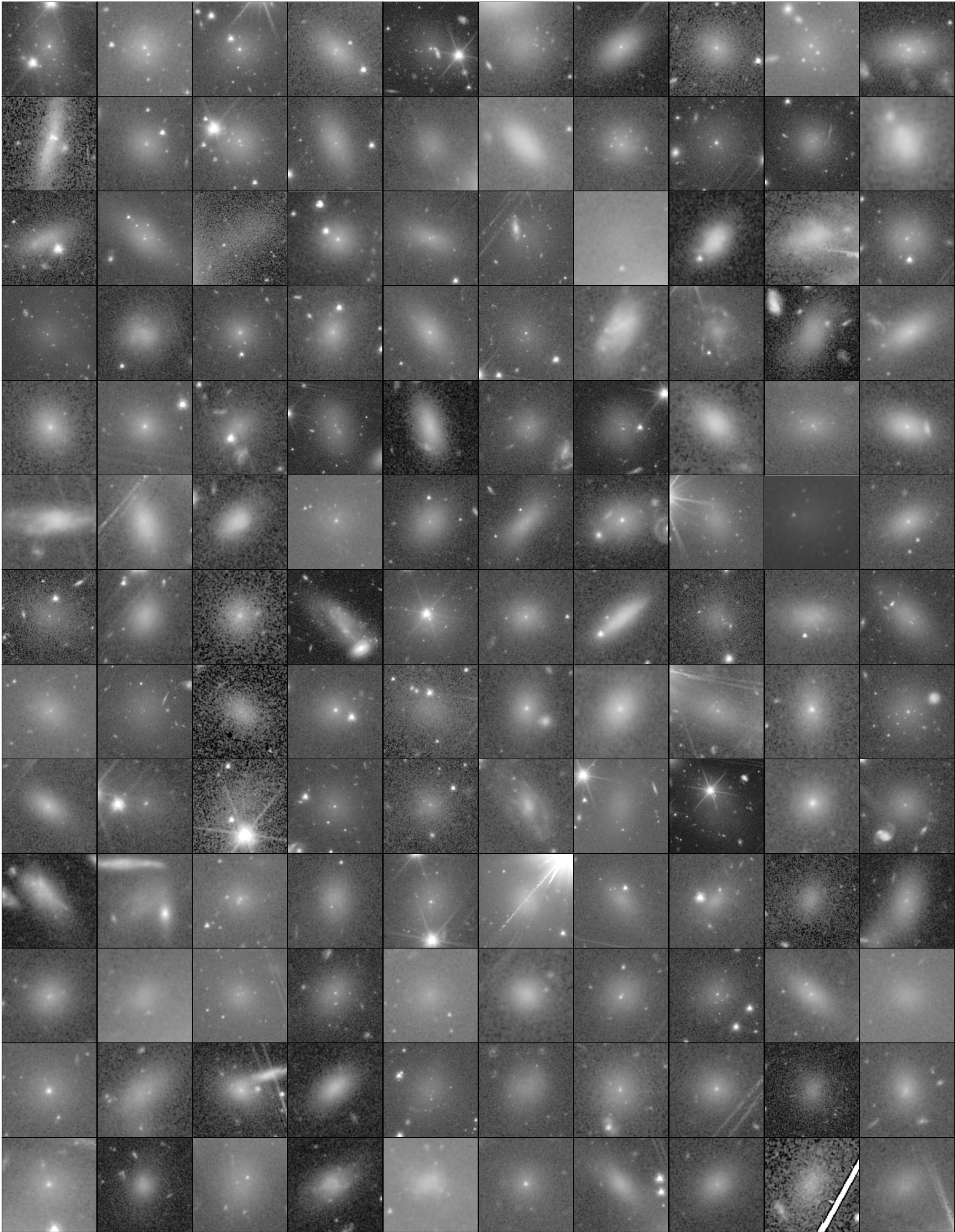}} 
\caption{Continued.}
\label{fig:cutouts-all-dwarfs3}
\end{figure*}

\begin{figure*}[ht!]\ContinuedFloat
\centerline{\includegraphics[width=\linewidth]{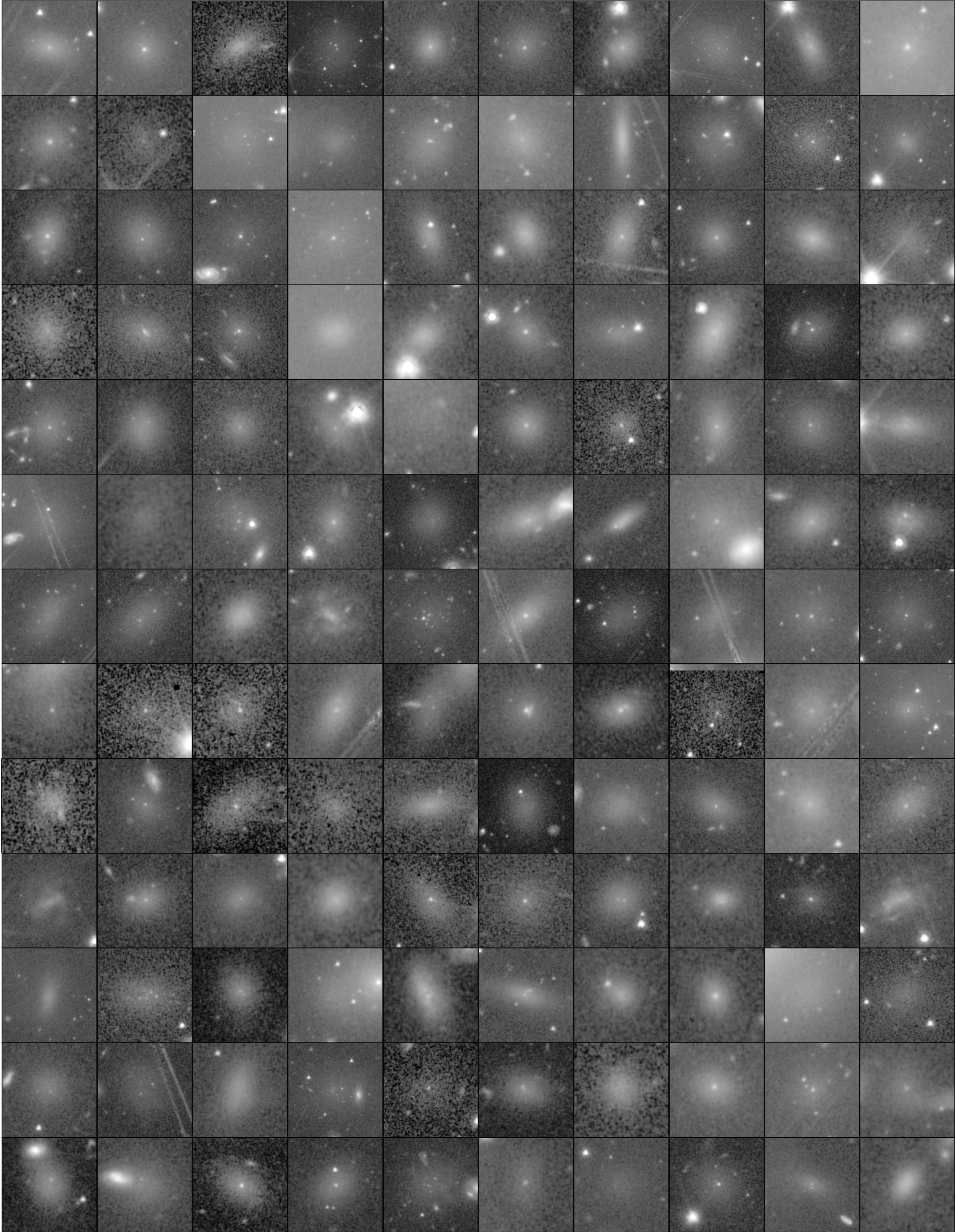}}
\caption{Continued.}
\label{fig:cutouts-all-dwarfs4}
\end{figure*}

\begin{figure*}[ht!]\ContinuedFloat
\centerline{\includegraphics[width=\linewidth]{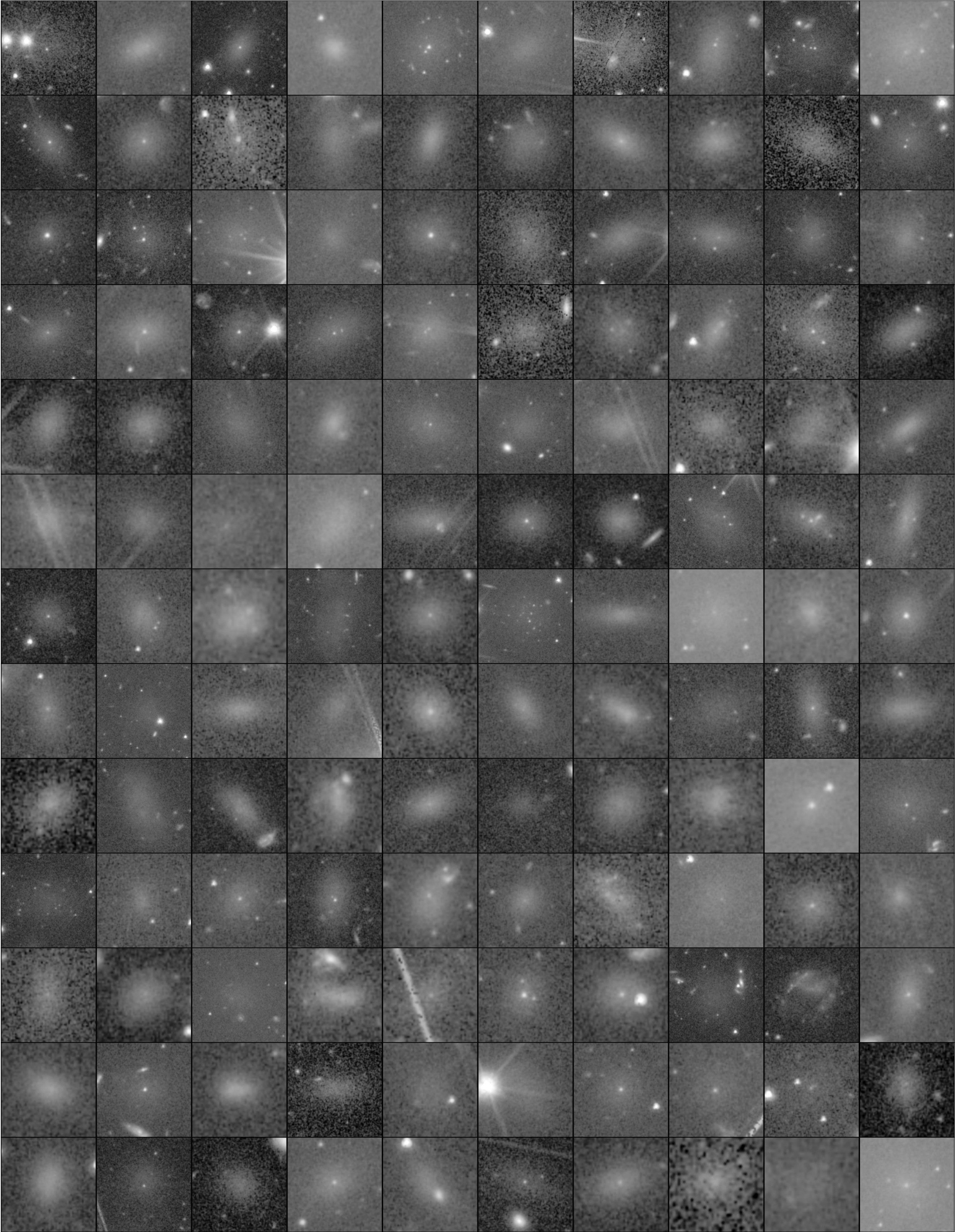}}
\caption{Continued.}
\label{fig:cutouts-all-dwarfs5}
\end{figure*}

\begin{figure*}[ht!]\ContinuedFloat
\centerline{\includegraphics[width=\linewidth]{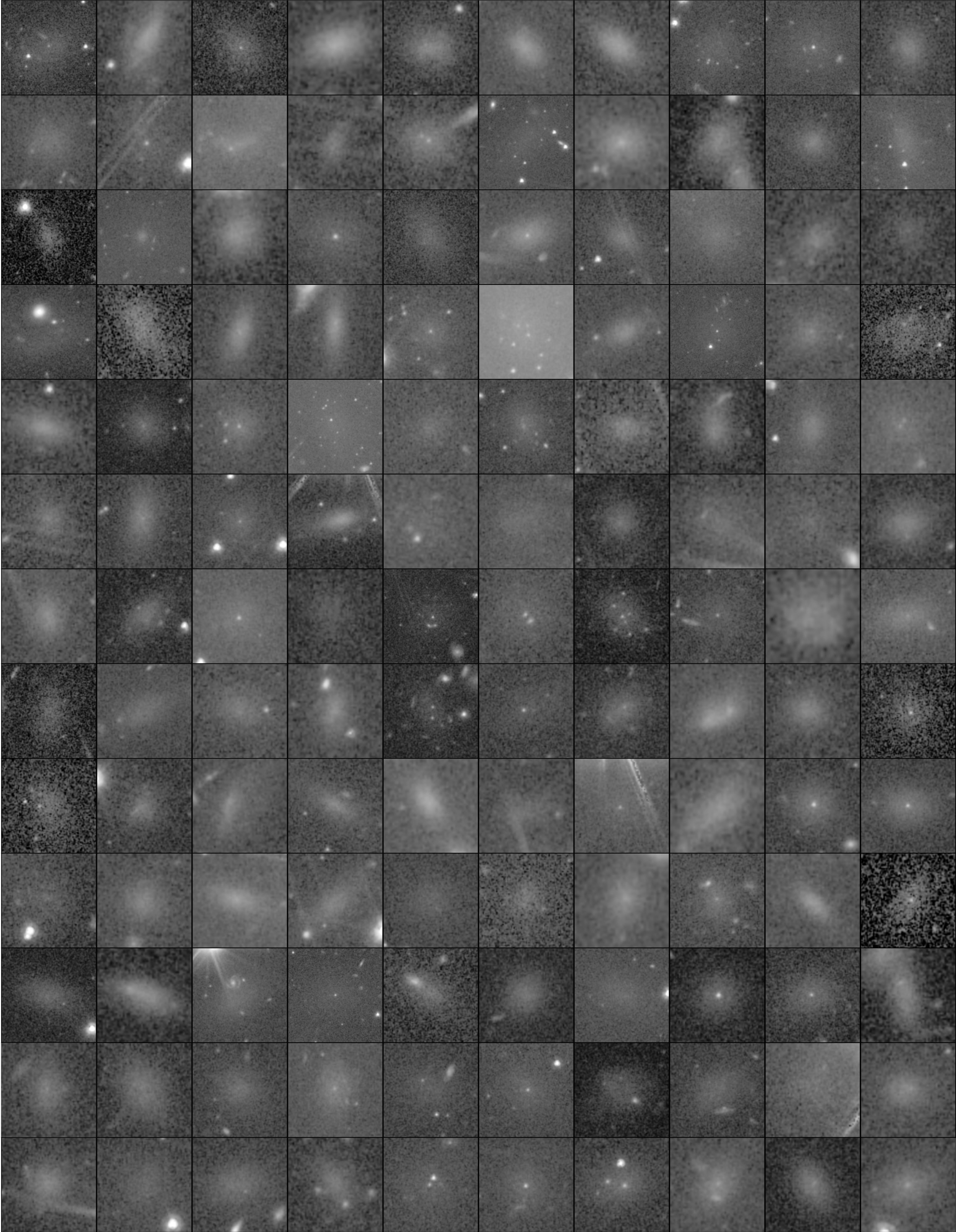}}
\caption{Continued.}
\label{fig:cutouts-all-dwarfs6}
\end{figure*}

\begin{figure*}[ht!]\ContinuedFloat
\centerline{\includegraphics[width=\linewidth]{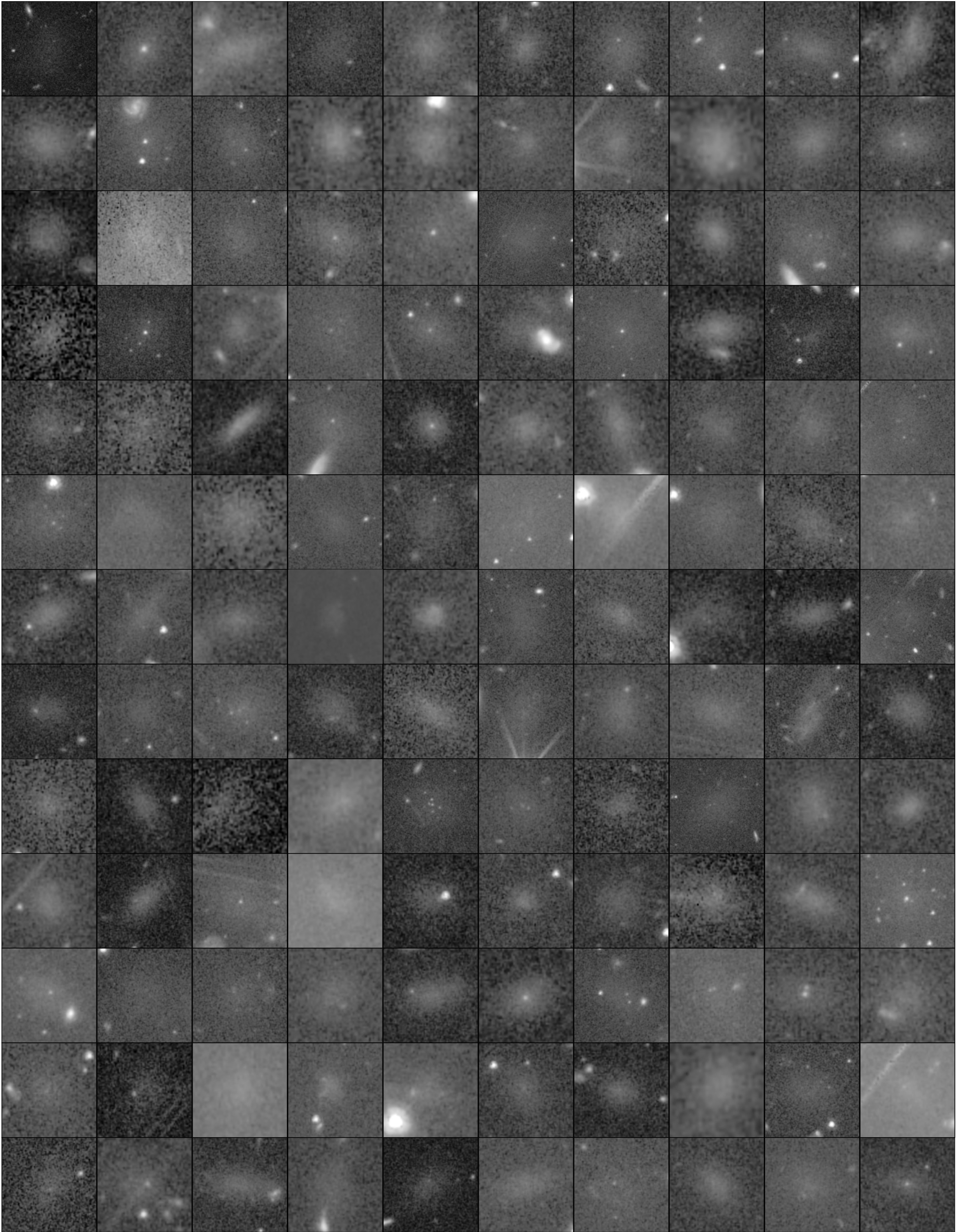}}
\caption{Continued.}
\label{fig:cutouts-all-dwarfs7}
\end{figure*}

\begin{figure*}[ht!]\ContinuedFloat
\centerline{\includegraphics[width=\linewidth]{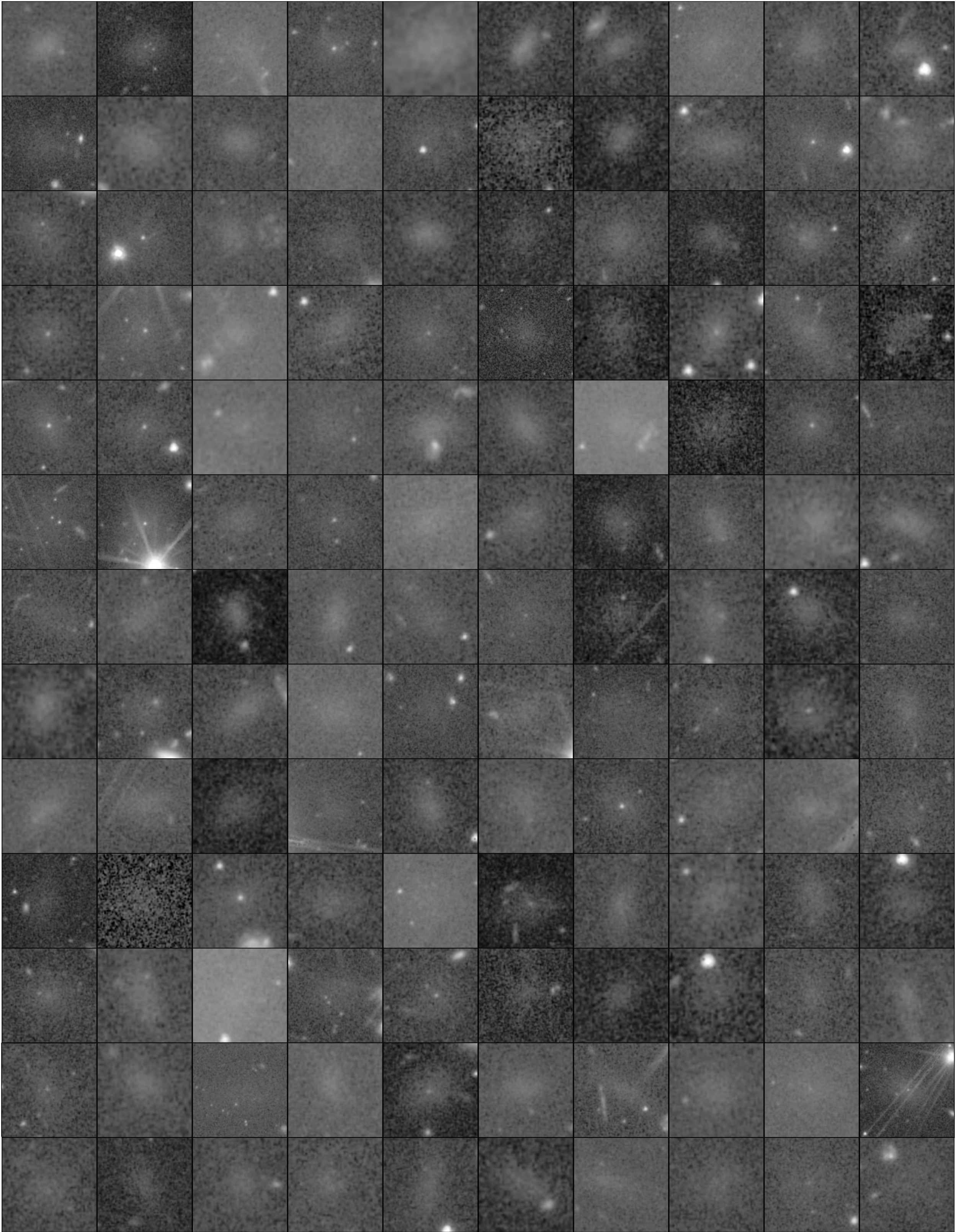}}
\caption{Continued.}
\label{fig:cutouts-all-dwarfs8}
\end{figure*}

\begin{figure*}[ht!]\ContinuedFloat
\centerline{\includegraphics[width=\linewidth]{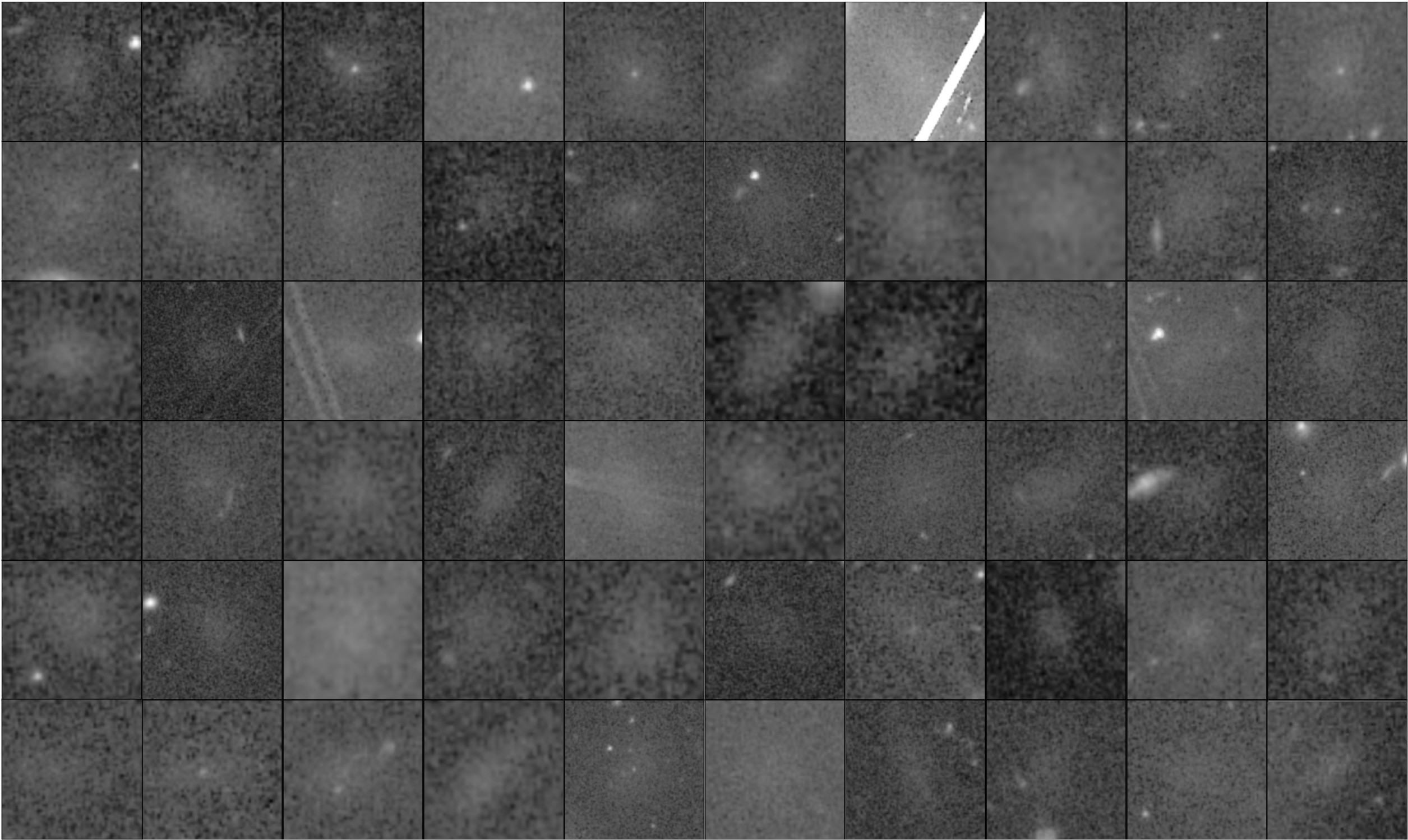}}
\caption{Continued.}
\label{fig:cutouts-all-dwarfs9}
\end{figure*}

\clearpage

\section{Visual properties of the galaxies not included in the final dwarf catalogue}
\label{AppendixB}

The final catalogue of dwarf galaxies presented in Appendix\,\ref{AppendixA} is based on the voting system presented in Sect.~\ref{sc:Methods}. A score of 0.7 (corresponding to the agreement of 5/7 classifiers) was the threshold we adopted to keep or reject a galaxy from the final catalogue. Some 160 objects from our initial list of 1260 objects fall below this threshold and were subsequently removed from the sample. Nevertheless, these are objects that were initially flagged as potential dwarf candidates by at least two team members, and may include some legitimate dwarf galaxies; hence, we include their properties here. Table\,\ref{appendix:table-rejected-dwarfs} presents the visual properties of the candidates, ordered by increasing RA. The columns include: a unique identifier (ID), RA in degrees, Dec in degrees, morphology (either dE or dI), a GC-rich flag (where dwarfs with $N_{\rm GC} > 2$\,GCs were considered GC-rich), a flag denoting the presence of a nucleus (Nucleated), and a flag for galaxies with disturbed morphologies (Disturbed).\\\\
Images of these rejected candidates are shown in Fig.\,\ref{fig:cutouts-rejected1}. The cutouts were created from the \IE image. For better visibility, the dimensions of each cutout have been scaled to twice the area of the original annotations and an arcsinh stretch has been applied. The dwarf candidates are ordered by increasing RA.\\\\

\begin{table}[ht!]
\centering
\caption{Visual properties of the dwarf galaxy candidates not included in the final catalogue. The full table will be available as supplementary material at the Strasbourg astronomical Data Center (CDS; \url{https://www.aanda.org/for-authors/latex-issues/tables).}}
\label{appendix:table-rejected-dwarfs}

\vspace{0.5cm}

\begin{tabular}{lllcccc}
\hline 
\hline

\noalign{\smallskip}
\omit\hfil ID \hfil & \omit\hfil RA \hfil & \omit\hfil Dec \hfil & \omit\hfil Morphology \hfil & \omit\hfil \phantom{00}GC-rich\phantom{00} \hfil & \omit\hfil Nucleated \hfil & \omit\hfil \phantom{00}Disturbed \hfil \\
   & \omit\hfil [deg] \hfil & \omit\hfil [deg] \hfil & \omit\hfil & \omit\hfil & \omit\hfil & \omit\hfil \\
\hline
\noalign{\smallskip}
1101 & 49.003850 & 41.815750 & dE & No & No & No \\
1102 & 49.025050 & 41.822250 & dE & No & Yes & Yes \\
1103 & 49.040055 & 41.820055 & dE & No & No & Yes \\
1104 & 49.055566 & 41.797613 & dE & No & No & No \\
1105 & 49.071610 & 41.690478 & dE & No & No & No \\
\vdots & \vdots & \vdots & \vdots & \vdots & \vdots & \vdots  \\
1256 & 50.314450 & 41.437250 & dE & No & No & No \\
1257 & 50.317150 & 41.477000 & dE & No & No & No \\
1258 & 50.319350 & 41.467400 & dE & No & No & No \\
1259 & 50.326200 & 41.478567 & dE & No & No & No \\
1260 & 50.344600 & 41.583450 & dE & No & No & No \\
\hline 
\end{tabular}
\end{table}

\clearpage
\begin{figure*}[ht!]
\centerline{\includegraphics[width=\linewidth]{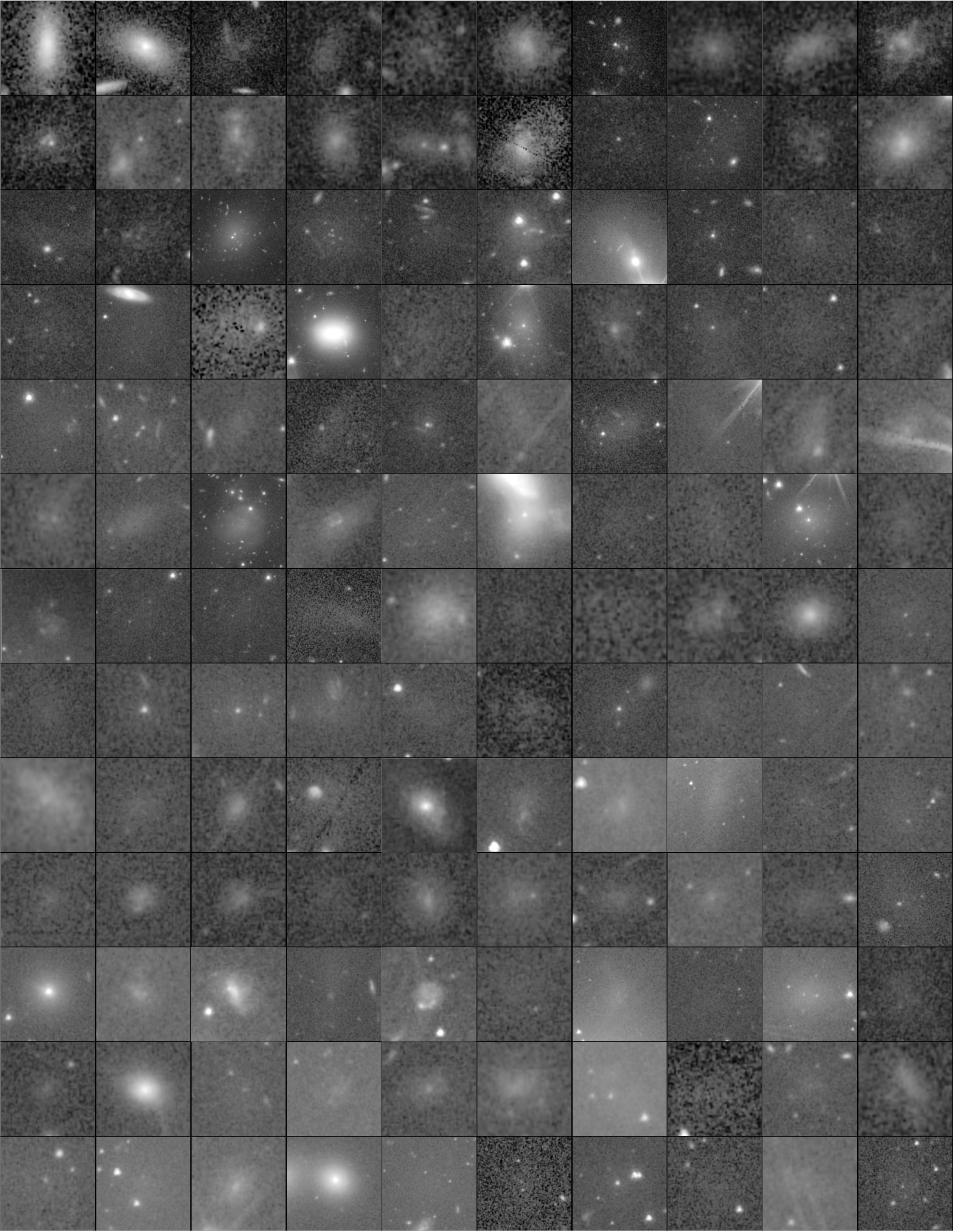}}
\caption{Cutouts of the $\IE$ image with arcsinh stretching of the 160 objects rejected as dwarf candidates, ordered by increasing right ascension. The size of the cutouts are proportional to twice the area determined from the annotation of the classifiers, with north up and east to the left. The figure continues on the next page.}
\label{fig:cutouts-rejected1}
\end{figure*}

\begin{figure*}[ht!]\ContinuedFloat
\centerline{\includegraphics[width=\linewidth]{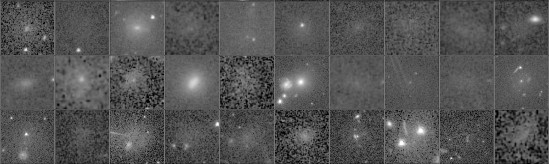}}
\caption{Continued.}
\label{fig:cutouts-rejected2}
\end{figure*}
\clearpage

\vspace{5cm}

\newpage

\section{\Euclid NISP-NISP colours as a function of VIS magnitude \texorpdfstring{\IE}{IE}}
\label{AppendixC}
\begin{figure*}[ht!]
\centerline{
\includegraphics[width=0.7\linewidth]{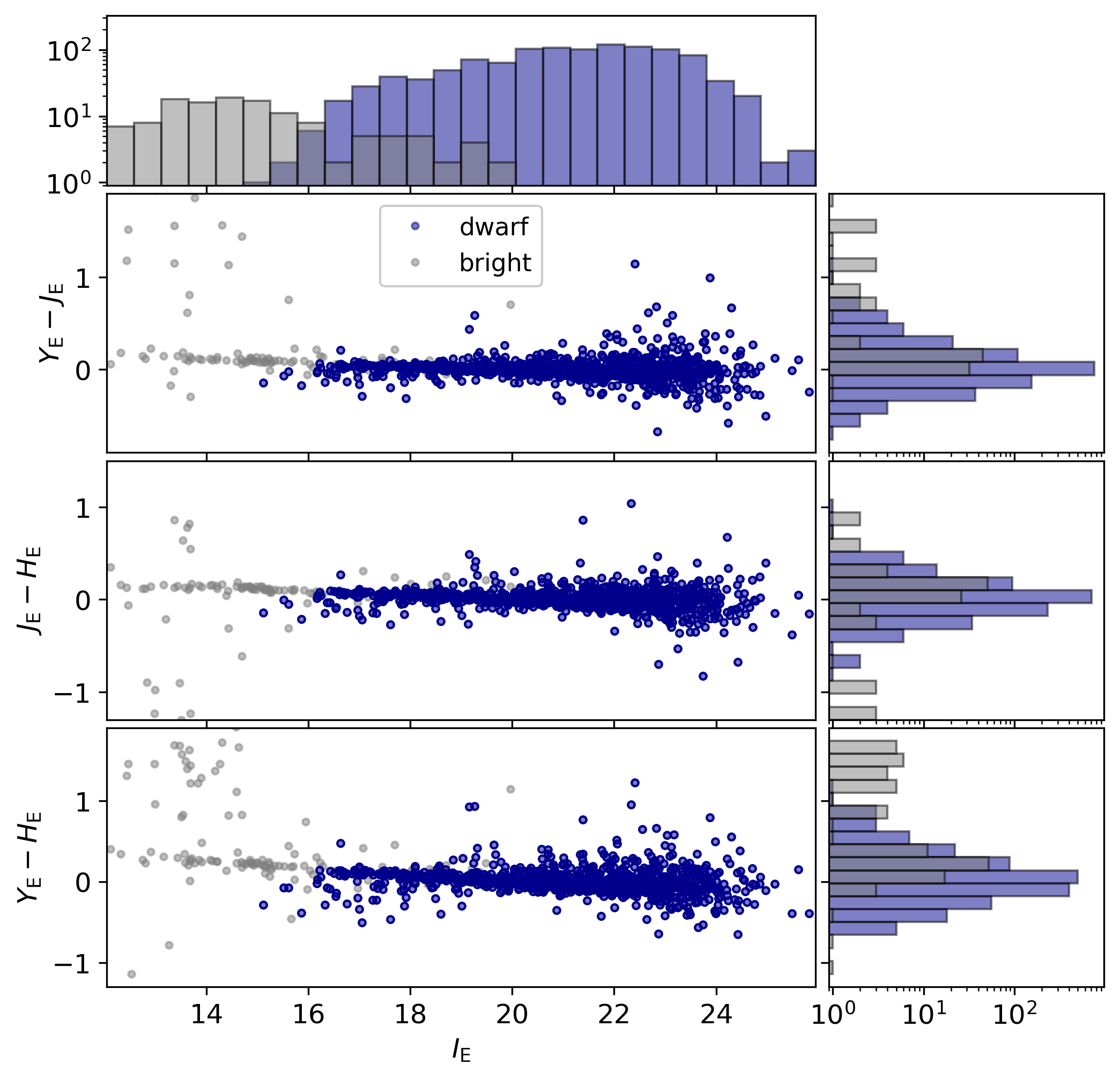}
}
\caption{\Euclid NISP-NISP colours as a function of \IE\ measured using aperture photometry within 1\,$R_{\rm e}$ of the 1100 dwarf galaxy candidates and the bright galaxy sample in the ERO Perseus field from \citet{EROPerseusOverview} shown in {violet-blue} and {light-grey colours}, respectively. The magnitudes were corrected for extinction before the colours were computed, as described in Sect.\,\ref{sec:extcorr}. 
\label{fig:colour-colour}}
\end{figure*}

Following the discussion of the colours in Sect.\,\ref{sec:extcorr}, we also include the \Euclid NISP-NISP colours in Fig.\,\ref{fig:colour-colour}.

\clearpage

\end{appendix}

\end{document}